\documentclass[useAMS,usenatbib,psfig]{mn2e}
\usepackage{graphicx,rotating,times}
\usepackage{grffile}
\usepackage{longtable,supertabular,multirow}
\usepackage{amsmath,amssymb,rotating}
\usepackage{lscape,xcolor}
\newcommand{\nhthree}{\mbox{NH$_3$}}
\newcommand{\nhone}{\mbox{NH$_3$\,(1,1)}}
\newcommand{\nhtwo}{\mbox{NH$_3$\,(2,2)}}

\newcommand{\water}{\mbox{H$_2$O}}

\newcommand{\kms}{\mbox{kms$^{-1}$}}
\def \kms{\mbox{kms$^{-1}$}}

\def \cm2{\mbox{cm$^{-2}$}}
\def \cm3{\mbox{cm$^{-3}$}}
\def \hydrogen{\mbox{H$_2$}}

\title[HOPS $\nhthree$ source properties]{$\water$ Southern Galactic
  Plane Survey (HOPS): Paper III -- Properties of Dense Molecular Gas
  across the Inner Milky Way}

\author[S.N.Longmore et al.]{S. N. Longmore$^{1,2,3}$\thanks{E-mail:s.n.longmore@ljmu.ac.uk},
 A. J. Walsh$^{4}$, C. R. Purcell$^{5,6}$, D. J. Burke$^{3}$, J. Henshaw$^{1}$, D. Walker$^{1}$, \newauthor
J. Urquhart$^{7}$, A. T. Barnes$^{1}$, M. Whiting$^{8}$, M. G. Burton$^{9}$, S. L. Breen$^{5, 8}$, T. Britton$^{8, 6}$,   \newauthor
K. J. Brooks$^{8, 10}$, M. R. Cunningham$^9$, J. A. Green$^8$, L. Harvey-Smith$^8$, L. Hindson$^{9}$,  \newauthor
M. G. Hoare$^{12}$, B. Indermuehle$^8$, P. A. Jones$^9$, N. Lo$^{11}$, V. Lowe$^{7, 6, 14}$, T. J. T. Moore$^{1}$,   \newauthor
M. A. Thompson$^{11}$ and M. A. Voronkov$^8$\\
$^{1}$ Astrophysics Research Institute, Liverpool John Moores University, 146 Brownlow Hill, Liverpool L3 5RF, UK;\\
$^{2}$European Southern Observatory, Karl-Schwarzchild-Str. 2, Garching bei M\"{u}nchen, Germany;\\
$^{3}$Harvard-Smithsonian Center For Astrophysics, 60 Garden Street, Cambridge, MA, 02138, USA;\\
$^{4}$ICRAR, Curtin University, Australia;\\
$^{5}$IfA, Sydney University, Australia;\\
$^{6}$Dept. of Physics \& Astronomy, Macquarie University, NSW 2109, Australia;\\
$^{7}$Centre for Astrophysics and Planetary Science, University of Kent, Canterbury, CT2\,7NH, UK;\\
$^{8}$CSIRO Astronomy and Space Science, Australia Telescope National Facility, PO BOX 76, Epping, NSW 1710, Australia;\\
$^{9}$School of Physics, University of New South Wales, Sydney, NSW 2052, Australia; \\
$^{10}$Murdoch University, 90 South Street, Murdoch, WA 6150, Australia; \\
$^{11}$University of Hertfordshire, UK;\\
$^{12}$University of Leeds, UK\\
$^{13}$Universidad de Chile, Santiago, Chile;\\
$^{14}$ CBA, 11 Harbour Street, Sydney, NSW 2000, Australia;\\
} 
\begin{document}
\hyphenation{kruijs-sen}
 
\date{in prep.}
\pagerange{\pageref{firstpage}--\pageref{lastpage}} \pubyear{2010}
\maketitle

\begin{abstract}

The {\bf H}$_2${\bf O} Southern Galactic {\bf P}lane {\bf S}urvey (HOPS) has mapped 100 square degrees of the Galactic plane, from $l=290^{\circ}$ through to $l=30^{\circ}$ and $b\pm0.5^{\circ}$, for water masers and thermal molecular line emission using the 22-m Mopra telescope. We describe the automated spectral-line fitting pipelines used to determine the properties of spectral-line emission detected in HOPS datacubes, and use these to derive the physical and kinematic properties of gas in the survey.  A combination of the angular resolution, sensitivity, velocity resolution and high critical density of lines targeted make the HOPS data cubes ideally suited to finding precursor clouds to the most massive and dense stellar clusters in the Galaxy. We compile a list of the most massive HOPS ammonia ($\nhthree$) regions and investigate whether any may be young massive cluster (YMC) progenitor gas clouds. HOPS is also ideally suited to trace the flows of dense gas in the Galactic Centre. We find the kinematic structure of gas within the inner 500\,pc of the Galaxy is consistent with recent predictions for the dynamical evolution of gas flows in the centre of the Milky Way. We confirm a recent finding that the dense gas in the inner 100\,pc has an oscillatory kinematic structure with characteristic length scale of $\sim$20\,pc, and also identify similar oscillatory kinematic structure in the $\nhthree$ data for the gas at radii larger than 100\,pc. We discuss the potential origin and implications for these oscillations. Finally, we make all of the above fits and the remaining HOPS data cubes (including C$_2$S, HC$_3$N, HC$_5$N and several CH$_3$OH, radio recombination line and  metastable $\nhthree$ transitions) across the 100 square degrees of the survey available to the community through the HOPS website.

\end{abstract}


%------------------------------------------------------------
\section{Introduction}

The majority of stars in the Universe form in the highest density regions of giant molecular clouds. Finding and characterising the properties of this dense gas as a function of environment is a primary goal  for star formation studies and has been the focus of many recent observational surveys  \citep[see the review by][]{molinari14}. 

Among these surveys, the {\bf H}$_2${\bf O} Southern Galactic {\bf P}lane {\bf S}urvey (HOPS) is unique. It is the only blind, large-area, spectral line survey of the Galaxy combining molecular line transitions probing gas with a higher volume density than CO-bright gas, with water maser emission -- a well-known tracer of star formation activity. The  HOPS spectral line sensitivity of 0.2\,K per 0.4\,$\kms$ channel corresponds to 5$\sigma$ molecular cloud mass completeness limits of 400, 5000, and 3$\times10^4$\,M$_\odot$ for clouds at distances of 3.2\,kpc, 8.5\,kpc (Galactic Centre distance) and 18\,kpc (far side of the Galaxy) \citep{purcell2012}. HOPS is therefore less sensitive to low column density gas or lower mass star formation regions outside the solar neighbourhood than mm-continuum surveys with comparable survey area. However, the 2$\arcmin$ angular resolution, sensitivity and high critical density of lines targeted make the HOPS data cubes ideally suited to finding gas clouds with radii of several pc, masses greater than several thousand M$_\odot$ and average number densities of $\geq$10$^{3-4}$\,cm$^{-3}$ throughout the Galaxy. These are exactly the properties predicted for gas clouds expected to form the most massive and dense stellar clusters in the disc of the Galaxy, and the gas clouds within the inner few hundred pc of the Galactic Centre.

The additional kinematic information of the HOPS data confers several additional advantages over continuum-only surveys. As well as making it trivial to identify sources with spuriously high column density due to confusion from multiple physically-distinct clouds along the line of sight, it becomes possible to investigate the dynamical structure of the massive and dense gas clouds detected in the survey.

In this paper we seek to unlock this kinematic dimension and derive the physical and kinematic properties of the HOPS sources. In $\S$~\ref{sec:spectral-line_fitting} we first describe the automated spectral-line fitting routines that were developed in order to do this, and then apply the pipelines on the HOPS ammonia ($\nhthree$) data cubes. In $\S$~\ref{sec:peak_spectra_fit_results} we use the $\nhone$ and $\nhtwo$ fit results to derive the characteristic dense molecular gas properties for the HOPS dense molecular gas clouds detected in the disk of the Galaxy. We then use these fits to identify a sample of clouds that may be precursors to the next generation of the most massive and dense clusters in the Galaxy. In $\S$~\ref{sec:galactic_centre} we use the fits to the $\nhone$, (2,2), (3,3) and (6,6) data cubes to investigate the kinematics of dense gas in the inner few hundred pc of the Galaxy and compare to recent theoretical predictions. Finally, in the Appendix we make the remaining HOPS data cubes available to the community through the HOPS website.

%------------------------------------------------------------
\section{Spectral-line fitting of HOPS datacubes}
\label{sec:spectral-line_fitting}

HOPS is a survey with the Mopra radio telescope designed to simultaneously map spectral-line emission along the Southern Galactic plane across the 12-mm band (frequencies of 18.5 to 27.5\,GHz). HOPS has mapped 100 square degrees of the Galactic plane, from $l=290^\circ$ to $l=30^\circ$ and  $|b|\leq 0.5^\circ$ \citep{walsh2008}, aiming to provide an untargeted census of 22.235\,GHz $\water$ (6$_{16}-$5$_{23}$) masers and thermal line emission towards the inner Galaxy.

The survey properties, observing parameters, data reduction and $\water$ maser catalogue are described in \citet[][hereafter Paper~I]{walsh2011}. The general methodology used to find spectral-line emission in HOPS datacubes and its specific application to ammonia [\nhthree]\,(J,K) = (1,1) and (2,2)  data are described in \citet[][hereafter Paper~II]{purcell2012}. The accurate water maser positions are published in \citet{walsh14}. Below we describe the automated spectral-line fitting routines and apply them to the $\nhone$ and $\nhtwo$ data cubes. 

With a large survey area (100 deg$^2$) and a wide (8\,GHz) bandwidth including many bright transitions, HOPS detected a large number of spectral line emission regions. The properties of the emission varies widely, containing transitions with resolved hyperfine structure, line widths ranging from marginally resolved ($<$1\,$\kms$) to several tens of $\kms$ (e.g. pressure-broadened radio recombination lines and molecular lines towards the Galactic Centre) and regions with emission from multiple velocity components. To handle these data, an automated emission finding pipeline and spectral-line fitting pipeline were developed. The emission finding pipeline, described in Paper~II, searched the full dataset, identified regions with emission and extracted sub-cubes that encompass the spatial and velocity extent of the emission, taking into account hyperfine structure where applicable. The output from the emission finding pipeline was then fed into the spectral-line fitting pipeline.

The $\nhone$ and $\nhtwo$ transitions were chosen to test the automated fitting pipeline for several reasons. Firstly, a large number of sources were detected in both transitions with a large range in brightness temperature and linewidth. This allowed the robustness of the fitting routines to be tested over a wide range of parameter space. Secondly, the $\nhone$ hyperfine structure is easily-detectable and the main/inner satellites can be spectrally resolved for sources with narrow line widths (i.e. less than a few $\kms$). At the same time the main/inner satellites are sufficiently close in frequency that they overlap in velocity for sources with large linewidth ($>5\,\kms$). These properties make $\nhone$ emission ideal to test how the pipeline deals with simultaneously fitting multiple lines and line-blending. Conversely, the $\nhtwo$ satellite lines are much weaker and offset in velocity from the main line making a single Gaussian component fit a good approximation. Thirdly, the emission from the $\nhone$ and (2,2) transitions is expected to come from very similar gas, so the line centre velocity (V$_{\rm LSR}$) and line width ($\Delta$V) from the same source should be similar for the two transitions, allowing further checks on the robustness of the fitting results. Finally, these lines are well known as robust probes of the physical conditions (especially the temperature) in dense molecular gas  \citep[e.g.][]{ho_townes1983}. In this way, at the same time as testing the automated spectral line fitting routines, we can identify and determine the broad physical properties of regions in the Galaxy where the earliest stages of high mass star formation are taking place.

The first step in the automated fitting process was to remove sources flagged in Paper~II as potentially spurious. These included artefacts at the extreme ends of the spectral bandpass caused by the baseline fitting (flag labels `vpF' and `vmF' in Paper~II) and those identified by eye as spurious (flag label `artifactF' in Paper~II). For the purposes of testing the spectral-line fitting pipelines, we also removed sources identified with more than one velocity component along the line of sight (flag label `multiF' in Paper~II). This left 653 emission regions. 

A large number of these regions exhibit asymmetric intensity profiles in the $\nhone$ hyperfine structure. For these sources, the usual assumption of equal excitation temperature for all the hyperfine levels does not hold, potentially leading to errors in the derived optical depth and other gas properties (e.g. temperature and column density). To ensure this did not affect the fit results, two versions of the pipeline were written: one making the standard `equal excitation' assumption ($\S$~\ref{sub:class_fitting}), and the other where the intensity of the $\nhone$ outer/inner satellite lines were left as free parameters ($\S$~\ref{sub:sherpa_fitting}). To further check the fitting robustness over the large range in expected line profile shapes, signal-to-noise etc, the two versions of the pipeline were written using independent software packages. The two versions of the pipeline are described in detail below.

%------------------------------
\subsection{Spectral-line fitting of $\nhone$ and (2,2) emission with Sherpa}
\label{sub:sherpa_fitting}

The first of the spectral-line fitting pipelines is based on {\sc  Sherpa}\footnote{Part of the CIAO~4.1 package \citep{freeman2001,refsdal2007}: http://cxc.harvard.edu/sherpa/ } routines. The input to the spectral-fitting pipeline are the emission sub-cubes from the source-finding pipeline described in Paper~II. This source-finding pipeline also provides masks identifying which pixels in the sub-cubes contain real emission. The spectral-line fitting pipeline then fits Gaussian profiles (with multiple components for transitions with hyperfine structure) to the velocity axis of each of these pixels, determines the robustness of each fit, and, for regions with at least one good fit, derives characteristic gas properties for the pixel with peak emission. For regions with enough good-fit pixels (see $\S$~\ref{sec:peak_spectra_fit_results}), the pipeline produces maps of the fit results. 

This version of the fitting pipeline does not assume equal excitation temperatures for the $\nhone$ hyperfine components. Instead, the $\nhone$ emission was fit with five Gaussian components -- one for each of the $F_1 = 1 \rightarrow 0$ and $F_1 = 0 \rightarrow 1$ outer satellite components, $F_1 = 1 \rightarrow 2$ and $F_1 = 2 \rightarrow 1$ inner satellite components and the $F_1\rightarrow F_1$ main component. In the fitting process, these were all fixed to have the same line width, with velocity separations given by their predicted frequency offsets \citep{rydbeck1977}. $\chi^2$ minimization was used to find the best-fit to the following free parameters: (1) the V$_{\rm LSR}$ of the $F_1\rightarrow F_1$ main-line component, (2) a single line width for the 5 Gaussian components and (3) the intensity of each of the five Gaussian components. The $\nhtwo$ emission was fit in a similar way to the $\nhone$ emission but only using a single Gaussian component as the $\nhtwo$ outer/inner satellite lines are much weaker and typically not detected. In all cases the initial guess for the V$_{\rm LSR}$ was taken from the velocity of peak $\nhone$ emission identified by the source-finding algorithm in Paper~II. If the fit converged, it returned the best-fit peak intensity (T$_{\rm peak}$) for each Gaussian component, the linewidth ($\Delta$V), emission velocity (V$_{\rm LSR}$) and the residuals (the r.m.s. variation in the spectra with the best-fit Gaussian component(s) subtracted).

Several quality control steps were then implemented. Firstly, a non-detection was defined where one or more of the following was true: i) the fit did not converge, ii) T$_{\rm peak}<5\sigma$, where $\sigma$ was calculated as the RMS in the residuals of those pixels which could be robustly fit, iii) $\Delta$V$<$0.05$\kms$ (i.e. linewidths smaller than the channel width), iv) $\Delta$V$>$10$\kms$. Criteria iv) was important for discarding spurious fits across most of the survey area and  highlighting sources for which the \nhone\ satellite and main components can no longer be spectrally resolved (limiting the ability to derive robust opacities). This also had the effect of flagging emission from clouds close to the Galactic centre where the linewidth is generally $>10\,\kms$. However, the complicated velocity structure and extreme gas excitation conditions mean a full analysis of this emission is beyond the scope of this pipeline. These data are discussed in $\S$~\ref{sub:CMZ_fitting}, and have been analysed in more detail elsewhere \citep{longmore13}. 

Spatial pixels which passed criteria i) to iv) were defined to have reliable fits. For regions with at least one $\nhone$ detection, the `peak pixel' was defined as that with the highest $\nhone$ main component brightness temperature. Following further quality control steps outlined below, the fits to the $\nhthree$ emission at this peak pixel were used to determine characteristic gas properties of the region.

Gas properties were derived for each pixel depending on the reliability of the $\nhone$ and (2,2) detections. For pixels with reliable $\nhone$ fits, the $\nhone$ main-line optical depth, $\tau_{\rm m}^{11}$, was derived from the ratio of the main and satellite component peak brightness temperatures \citep[see   e.g.][]{ho_townes1983}. To mitigate uncertainties in the optical depth caused by hyperfine-structure intensity anomalies, an average of the outer/inner satellite line intensities was used to determine the intensity ratio of these components to the main line. With an optical depth for the transition, the $\nhone$ excitation temperature, T$^{11}_{\rm ex}$, was then derived through the detection equation, assuming a background temperature of 2.73\,K and a beam-filling factor of 1. For pixels which also had reliable $\nhtwo$ fits, the gas rotational temperature (T$_{\rm rot}$) and total $\nhthree$ column density (N$_{NH3}^{TOT}$) were derived following \citet{ungerechts1986}. The gas kinetic temperature, T$_{\rm kin}$, was calculated from T$_{\rm rot}$ following \citet[][]{tafalla2004} who used the collisional rates of $\nhthree$ with $\hydrogen$ determined by \citet{danby1988}. Finally, the non-thermal contribution to the $\nhone$ linewidth, $\Delta$V$_{\rm NT}^{11}$, was calculated by subtracting the thermal contribution, $\Delta$V$_{\rm TH}^{11}$ (determined from T$_{\rm   kin}$), in quadrature from the measured $\nhone$ linewidth, $\Delta$V$^{11}$.

%------------------------------
\subsection{Spectral-line fitting of $\nhone$ and (2,2) emission with {\sc class}}
\label{sub:class_fitting}

{\sc class}\footnote{{\sc class} is part of the {\sc GILDAS} data reduction package: http://www.iram.fr/IRAMFR/GILDAS/} was used as a second, independent method of fitting the spectral-line data. For each region identified as containing emission, the source-finding pipeline output a spectrum at the pixel of peak emission. These peak spectra were then read into {\sc class} and a zeroeth-order baseline was subtracted from line-free channels. The line-free channels were selected as those offset from the velocity of peak emission, V$^{\rm peak}_{\rm LSR}$, by more than V$^{\rm peak}_{\rm LSR} \pm 25\,\kms$. The $\nhone$ and (2,2) baseline-subtracted spectra were then fit with the ``nh3(1,1)'' (assuming equal excitation for the hyperfine components) and ``gauss'' methods within {\sc class}, respectively. Gas properties were derived in the same way as for the {\sc Sherpa} routine.
 
%------------------------------
\subsection{Spectral-line fitting of $\nhthree$ emission in the Galactic Centre}
\label{sub:CMZ_fitting}

As mentioned above, the $\nhthree$ emission from the inner Galaxy ($|l| \lesssim 5^\circ$ -- corresponding to the inner few hundred parsecs from the Galactic Centre), is very different in nature from the $\nhthree$ emission throughout the rest of the disk due to the very different physical conditions there \citep{KL13}. The line widths are an order of magnitude larger and the velocity structure is complex, with multiple components along many lines of sight and a velocity gradient of around $200~\kms$ within $|l| \lesssim 2^\circ$. The fitting pipelines developed in $\S$~\ref{sub:class_fitting} and \ref{sub:sherpa_fitting} are not intended to work with such complicated data cubes.

Instead, we used the \textbf{S}emi-automated multi-\textbf{CO}mponent \textbf{U}niversal \textbf{S}pectral-line fitting \textbf{E}ngine ({\sc scouse}) of \citet{henshaw16}. {\sc scouse} is a line-fitting algorithm designed to analyse large volumes of spectral-line data in an efficient and systematic way. It finds a balance between user-interaction and efficiency by: i) systematically excluding regions of low significance from the analysis; ii) breaking up the input map into smaller regions, and requiring the user to fit \emph{only} the spectra that have been spatially-averaged over these regions; iii) using the user-provided values (such as the number of spectral components and parameter estimates) as \emph{non-restrictive guides} in the fitting process. Provided therefore, that the line profile of a spatially averaged spectrum is an adequate representation of its composite spectra, {\sc scouse} provides an efficient and systematic approach to fitting large quantities of spectral-line data. 

We provide {\sc scouse} with the same input parameters as in \citet{henshaw16}, since these have already been (empirically) optimised for emission from gas within the inner few hundred pc of the Galactic Centre (the Central Molecular Zone, CMZ). We refer the reader to that paper for full details of the fitting process, but provide the values of key input parameters used to fit the Galactic Centre data in Table~\ref{Table:global_stats} of \citet{henshaw16}. At present, {\sc scouse} fits all spectra with (multiple) independent Gaussian components, so does not explicitly attempt to fit the NH$_{3}$ hyperfine structure. However, at the size scales probed by the resolution of the HOPS observations, the velocity dispersion of the Galactic Centre gas is comparable to, or greater than, the velocity separation of the hyperfine components for the low $\nhthree$(J,K) transitions \citep{townes_schawlow1955, ho_townes1983, shetty12}. We therefore expect the satellite lines to be blended with the main component for these low $\nhthree$(J,K) transitions. In the unexpected case that the intrinsic velocity dispersion is low enough to resolve the hyperfine components, the known offset and relative intensity of the main, inner and outer satellite components \citep{townes_schawlow1955, ho_townes1983} would make the hyperfine structure trivial to spot. We note that in the case of moderate to high opacity in the low $\nhthree$(J,K) transitions, the brightness temperature of the satellite lines can become comparable to the main line, leading to opacity-broadening when fitting the emission with a single Gaussian component. In this case, without being able to resolve the hyperfine structure, it is not possible to determine the extent of this effect. For that reason, the measured velocity dispersion of the low $\nhthree$(J,K) transitions will only represent an upper limit to the intrinsic velocity dispersion. The expected intensity of the inner and outer satellite hyperfine components drops exponentially with increasing $\nhthree$(J,K) transition so we do not expect to detect the satellite lines for the high $\nhthree$(J,K) transitions, and the above problems should not be an issue. 

\begin{figure}
\includegraphics[width=0.45\textwidth]{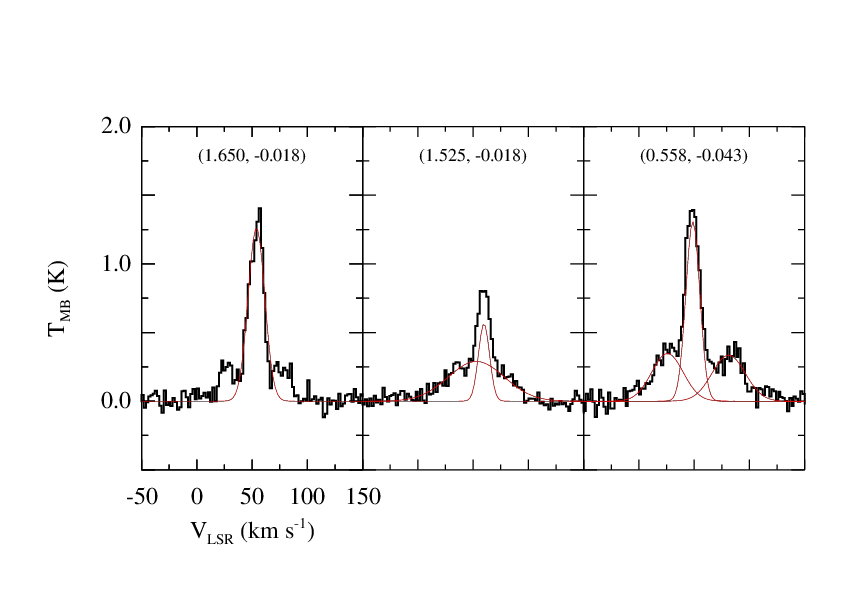}
\caption{Example $\nhthree$\,(3,3) spectra that proved difficult to interpret. The coordinates in parentheses given the position in Galactic longitude and latitude.}
\label{fig:nhthree_33_spectra}
\end{figure}

Visual inspection of the output shows {\sc scouse} generally does a good job of robustly fitting the complex spectral profiles of the $\nhone$, (2,2), (3,3) and (6,6) emission across the full inner 5$^\circ$ longitude range and that hyperfine structure does not result in the ``detection" of spurious additional velocity components. The only exception to this are that some of the $\nhthree$(3,3) spectra showed unusual profiles which proved difficult to interpret. Figure~\ref{fig:nhthree_33_spectra} shows examples of such spectra. The common feature linking these problematic spectra is a bright, narrow velocity component superimposed on the broader, weaker velocity components which are more typical of gas in the Galactic Centre. The left panel of Figure~\ref{fig:nhthree_33_spectra} shows one example where the bright, narrow component lies at a velocity half way in between two velocity components with similar, lower peak brightness temperatures. Such a profile is qualitatively characteristic of the $\nhthree$ hyperfine structure. However, the relative intensity and velocity separation of the lines does not match that predicted quantum mechanically \citep{townes_schawlow1955}.

The middle and right panels of Figure~\ref{fig:nhthree_33_spectra} show two more examples of typical $\nhthree$(3,3) spectra that proved difficult to interpret, in which the bright, narrow component either lay on top of another broader component, or at some intermediate velocity in between two (or more) components with different brightness temperatures that are also asymmetric around the central component. Given that $\nhthree$(3,3) is well known as a maser transition in regions with conditions similar to that characteristic of gas in the Galactic Centre, and many maser species appear widespread throughout the region, we conclude that the most likely explanation for the narrow spectral $\nhthree$(3,3) component is maser emission. As we are interested primarily in the variation in kinematic properties of the thermal gas component, we ignore them in the subsequent discussions.

\begin{table}
	
	\caption{{\sc scouse}: Global statistics. The input parameters provided to {\sc scouse} for the fitting of each transition were identical to those used by \citet{henshaw16}. } 
	\centering  
	\tabcolsep=0.15cm \normalsize{
	\begin{tabular}{ l  c  c  c  c }
	\hline	
 	Statistic &  NH$_{3}$ (1, 1) &  NH$_{3}$ (2, 2) &  NH$_{3}$ (3, 3) &  NH$_{3}$ (6, 6) \\ 
	\hline  
	
	$N_{\rm tot}$ & 95159 & 95159 & 95159 & 95159 \\ [1.0ex] 
	$N_{\rm tot, SAA }$ & 23742 & 18444 & 16740 & 2700 \\ [1.0ex]
	$N_{\rm SAA }$ & 430 & 322 & 296 & 34 \\ [1.0ex]
	$N_{\rm fit}$ & 13621 & 9887 & 13163 & 1494 \\ [1.0ex]
	$N_{\rm comp}$ & 16373 & 11649 & 18516 & 1584 \\ [1.0ex]
	$N_{\rm comp}$/$N_{\rm fit}$ & 1.2 & 1.2 & 1.7 & 1.2 \\ [1.0ex]
	$\overline{\sigma}_{\rm resid/rms}$ & 1.1 & 1.1 & 1.2 & 1.1 \\ [1.0ex]
	
	\hline
	\end{tabular}
	\vspace{0.5cm}
	
\begin{minipage}{0.5\textwidth}\footnotesize{
	\centering  
	\tabcolsep=0.1cm
	\begin{tabular}{ l  l }
	$N_{\rm tot}$ & Total number of positions in the mapped area.\\   
	$N_{\rm tot, SAA}$ & Total number of positions included in the coverage.\\
	$N_{\rm SAA}$ & Total number of spectral averaging areas.\\
	$N_{\rm fit}$ & Total number of positions fitted using the {\sc scouse} routine. \\ 
	$N_{\rm comp}$ & Total number of components fitted using the {\sc scouse} routine. \\
	$N_{\rm comp}$/$N_{\rm fit}$ & Fractional number of Gaussian components per position. \\ 
	$\overline{\sigma}_{\rm resid/rms}$ & Mean $\sigma_{\rm resid}/\sigma_{\rm rms}$. \\ 
	\end{tabular}
}
\end{minipage}
}
\label{Table:global_stats}
\end{table}

%----------------
%----------------

In $\S$~\ref{sec:galactic_centre} we describe the output of the fitting and discuss the implications of this for the dense gas kinematics in the Galactic Centre. The fit results themselves are available through the HOPS website (http://www.hops.org.au).

%============================================================
%Tables

\begin{table*} 
\footnotesize
\caption{Example output from the automated spectral line fitting procedure described in $\S$~\ref{sec:spectral-line_fitting}. Columns 1 and 2 give the Galactic longitude and latitude of the region taken from the source name in Paper~II. Columns 3 to 11 list the best-fit $\nhone$ and $\nhtwo$ parameters at the peak pixel towards each region. For each transition, T$_{\rm B}$ is the peak brightness temperature of the main component, V$_{\rm LSR}$ is the line-centre velocity and $\Delta V$ is the linewidth.  Columns 6 and 7 show the opacity and excitation temperature of the $\nhone$ transition. The subscript `m' denotes the main hyperfine component. Column 8 shows the $\nhtwo$ integrated intensity. A dash denotes no reliable fit was obtained for that parameter. Only the first 5 sources are shown. The full catalogue is available in Appendix 2. }
\begin{center} 
\begin{tabular}{|c|c|c|c|c|c|c|c|c|c|c|} 
\hline \hline 
 l             & b     & T$_{\rm B,m}^{11}$ & V$_{\rm LSR}^{11}$ & $\Delta$V$^{11}$  & $\tau_{m}^{11}$ & T$_{\rm ex}^{11}$ & $\int$ T$_{\rm B}^{22}$dV & V$_{\rm LSR}^{22}$ & $\Delta$V$^{22}$  & T$_{\rm B}^{22}$ \\ 
 ($^\circ$) & ($^\circ$) & (K) & (kms$^{-1}$) & (kms$^{-1}$) & & (K) & (K.kms$^{-1}$) & (kms$^{-1}$) & (kms$^{-1}$ &  (K)  \\ \hline 
0.223 & -0.473 & 2.87 $\pm$ 0.39 & 16.90 $\pm$ 0.07 & 1.96 $\pm$ 0.15 & 1.65 $\pm$ 0.49 & 4.46 & 0.65 $\pm$ 0.20 & 16.71 $\pm$ 0.17 & 1.08 $\pm$ 0.40 & 0.57 \\  
0.238 & -0.189 & 0.74 $\pm$ 0.61 & 76.60 $\pm$ 1.55 & 6.75 $\pm$ 3.78 & 3.37 $\pm$ 3.81 & 2.94 & 2.05 $\pm$ 0.45 & 79.13 $\pm$ 0.75 & 6.34 $\pm$ 1.39 & 0.30 \\  
0.507 & -0.321 & 0.31 $\pm$ 0.07 & -64.00 $\pm$ 0.85 & 6.69 $\pm$ 1.40 & 0.10 $\pm$ 2.74 & 5.83 & 2.47 $\pm$ 0.69 & -64.48 $\pm$ 3.40 & 24.64 $\pm$ 7.92 & 0.09 \\  
0.658 & -0.213 & 0.51 $\pm$ 0.37 & -50.00 $\pm$ 0.67 & 2.60 $\pm$ 1.28 & 1.27 $\pm$ 3.13 & 3.13 & 0.44 $\pm$ 0.16 & -59.87 $\pm$ 0.22 & 1.20 $\pm$ 0.47 & 0.35 \\  
0.860 & -0.067 & 11.30 $\pm$ 1.13 & 12.80 $\pm$ 0.26 & 15.00 $\pm$ 0.90 & 1.92 $\pm$ 0.25 & 8.61 & 126.40 $\pm$ 2.11 & 12.67 $\pm$ 0.25 & 31.51 $\pm$ 0.64 & 3.77 \\  \hline

\end{tabular} 
\end{center} 
\label{tmp} 
\end{table*} 

%------------------------------------------------------------

\begin{table*}
\caption{Example derived gas properties for sources with robust $\nhone$ and $\nhtwo$ detections (see $\S$~\ref{sec:peak_spectra_fit_results}). The rotational temperature, total $\nhthree$ column density,  kinetic temperature and the non-thermal and thermal contributions to the line width are listed in columns 3 to 7, respectively, for the peak $\nhone$ pixel. Only the first 5 sources are shown. The full table is given in Appendix 2. 
}
\begin{tabular}{|c|c|c|c|c|c|c|}

\hline \hline
  \multicolumn{1}{|c|}{l} &
  \multicolumn{1}{c|}{b} &
  \multicolumn{1}{c|}{T$_{\rm rot}$} &
  \multicolumn{1}{c|}{N$_{\rm NH3}^{\rm tot}$} &
  \multicolumn{1}{c|}{T$_{\rm kin}$} &
  \multicolumn{1}{c|}{$\Delta$V$_{\rm nt}$} &
  \multicolumn{1}{c|}{$\Delta$V$_{\rm th}$} \\
  \multicolumn{1}{|c|}{($^\circ$)} &
  \multicolumn{1}{c|}{($^\circ$)} &
  \multicolumn{1}{c|}{(K)} &
  \multicolumn{1}{c|}{(10$^{14}$cm$^{-2}$)} &
  \multicolumn{1}{c|}{(K)} &
  \multicolumn{1}{c|}{($\kms$)} &
  \multicolumn{1}{c|}{($\kms$)} \\
\hline
  29.939 & -0.031 & 19 & 1.69 & 22 & 3.13 & 0.24\\
  27.774 & 0.068 & 21 & 0.69 & 26 & 2.37 & 0.26\\
  27.278 & 0.152 & 42 & 0.45 & 93 & 2.32 & 0.50\\
  24.785 & 0.095 & 17 & 2.22 & 20 & 3.58 & 0.23\\
  24.475 & 0.478 & 34 & 0.33 & 59 & 3.04 & 0.39\\ \hline
\hline\end{tabular}
\label{tab:nh3_derived_vals}
\end{table*}

%============================================================
%Fit results

%------------------------------------------------------------
\section{Peak-spectra fit results}
\label{sec:peak_spectra_fit_results}

The results of peak-spectra fitting are listed in Table~2. Columns 1 and 2 give the Galactic longitude and latitude of the region taken from the source name in Paper~II. Columns 3 to 11 list the best-fit $\nhone$ and $\nhtwo$ parameters at the peak pixel towards each region. For each transition, T$_{\rm B}$ is the peak brightness temperature of the main component, V$_{\rm LSR}$ is the line-centre velocity and $\Delta V$ is the linewidth.  Columns 6 and 7 show the opacity and excitation temperature of the $\nhone$ transition. The subscript `m' denotes the main hyperfine component. Column 8 shows the $\nhtwo$ integrated intensity. A dash denotes no reliable fit was obtained for that parameter. 

For Table~2 we adopted an inclusive approach, where all potential detections are included. While the automated fitting algorithm is designed to remove obviously spurious fits, distinguishing between fits to genuine low signal-to-noise detections and fits to noise involves further work.  In other words, Table~2 is optimal in terms of completeness, but a small number of the low signal-to-noise detections may not be real. To remove such false detections, we used a two-step approach. Firstly we compared the fit results from the two pipelines and the properties of the $\nhone$ and $\nhtwo$ detections, and used these to define appropriate `robustness' criteria. Then we checked the `robust' spectra by eye to confirm the criteria were producing sensible results.

The required robustness of the fits depends on the specific science goals. For our purposes below ($\S$~\ref{sec:nh3_mas}) we wanted to isolate the most robust detections, so defined a set of `highly reliable' $\nhone$ detections as those with: (i) a signal-to-noise in both the {\sc Sherpa} and {\sc Class} pipelines of $>$5; (ii) the {\sc Sherpa} and {\sc Class} V$_{\rm LSR}$ agree to within 3\,$\kms$; (iii) the {\sc Sherpa} and {\sc Class} $\Delta$V agree to within 50\%. 81 of the $\nhone$ detections pass these criteria. Using similar criteria for the $\nhtwo$ detections, which were a subset of the reliable $\nhone$ reliable detections, we identify 64 sources with highly reliable $\nhtwo$ detections.

Table~\ref{tab:nh3_derived_vals} shows the derived gas properties for sources with robust $\nhone$ and $\nhtwo$ detections. The rotational temperature, total $\nhthree$ column density,  kinetic temperature and the non-thermal and thermal contributions to the line width are listed in columns 3 to 7, respectively, for the peak $\nhone$ pixel.

\begin{figure*}
\begin{center}
\begin{tabular}{ccc}
\includegraphics[height=5.0cm, angle=-90, trim=0 0 -5 0]{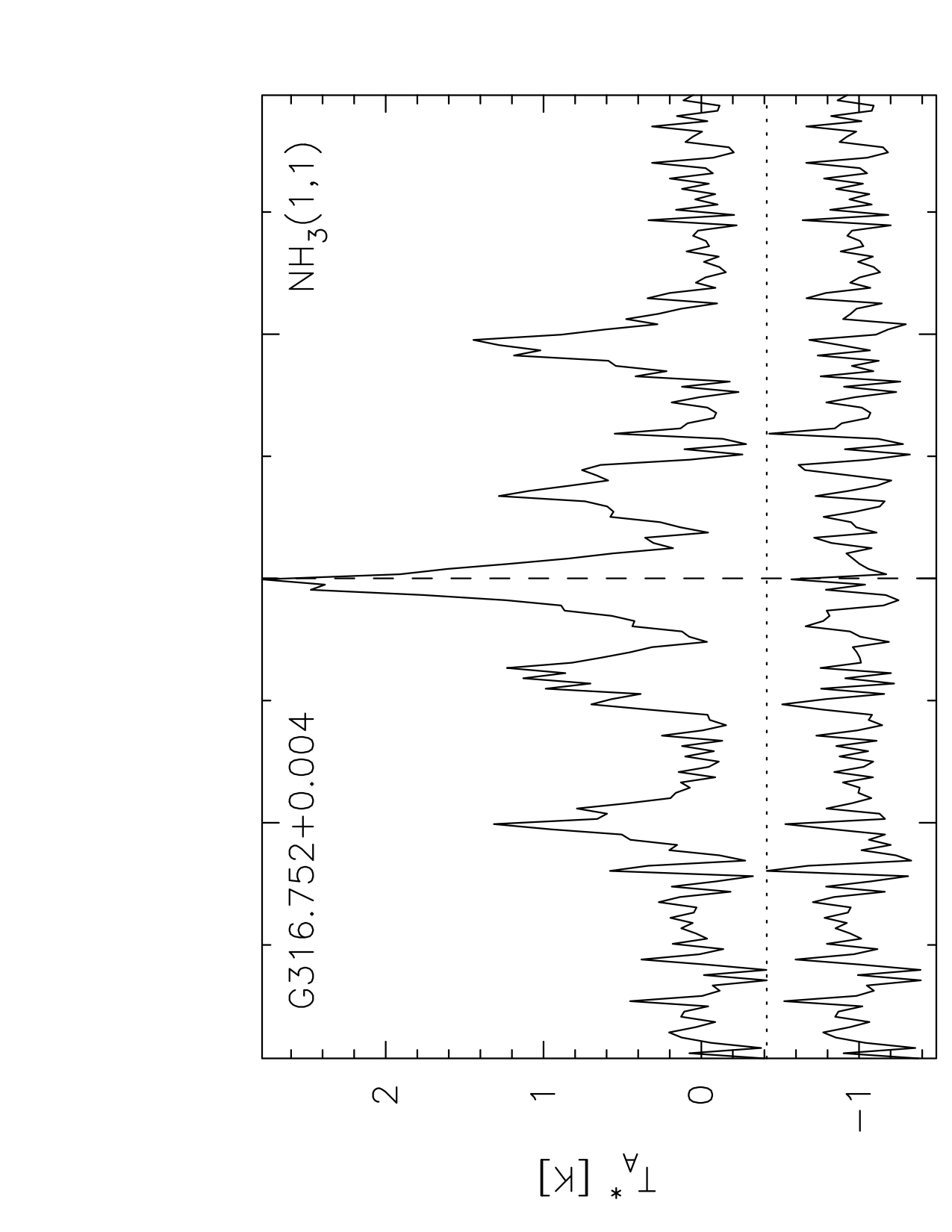} &
\includegraphics[height=5.0cm, angle=-90, trim=0 0 -5 0]{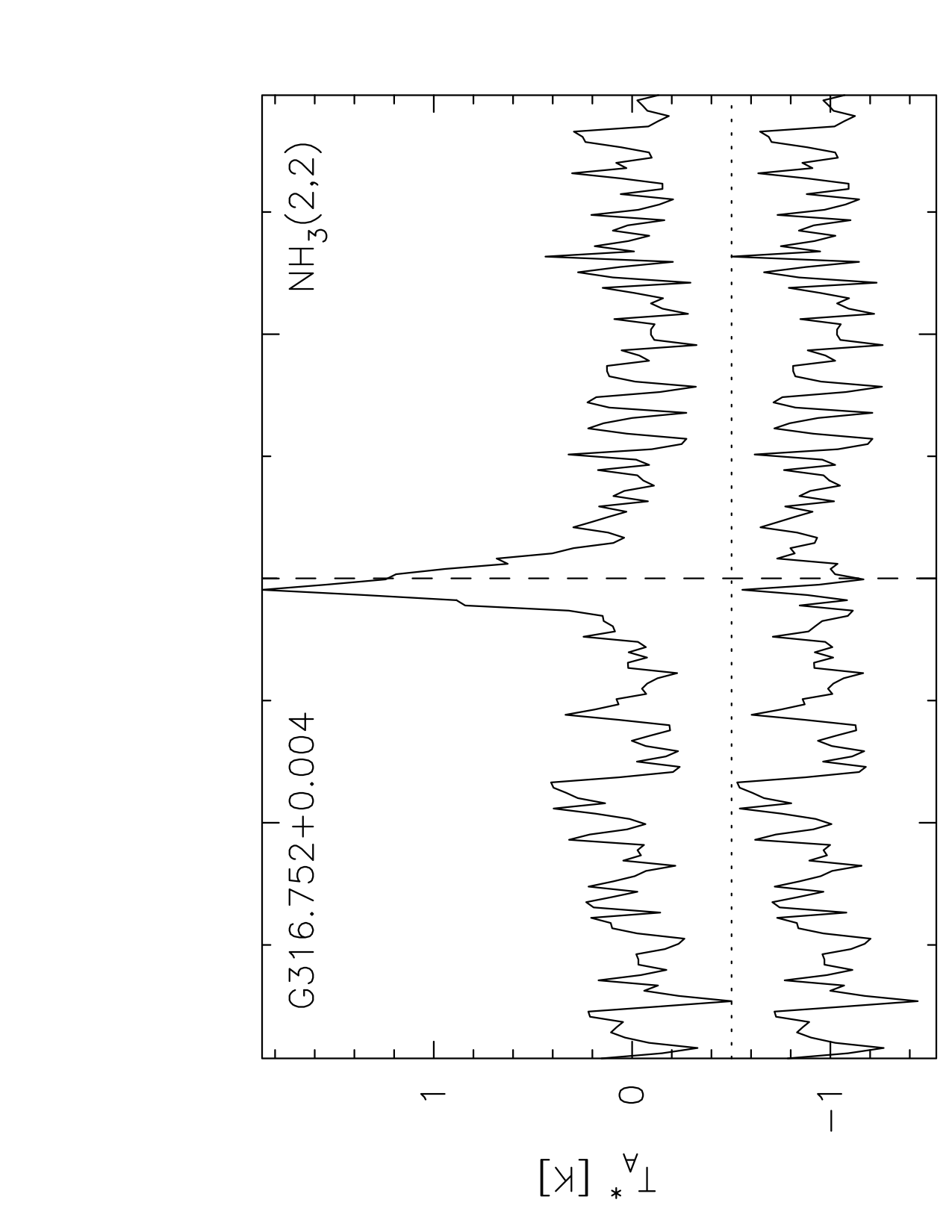} \\
\includegraphics[height=5.0cm, angle=-90, trim=0 0 -5 0]{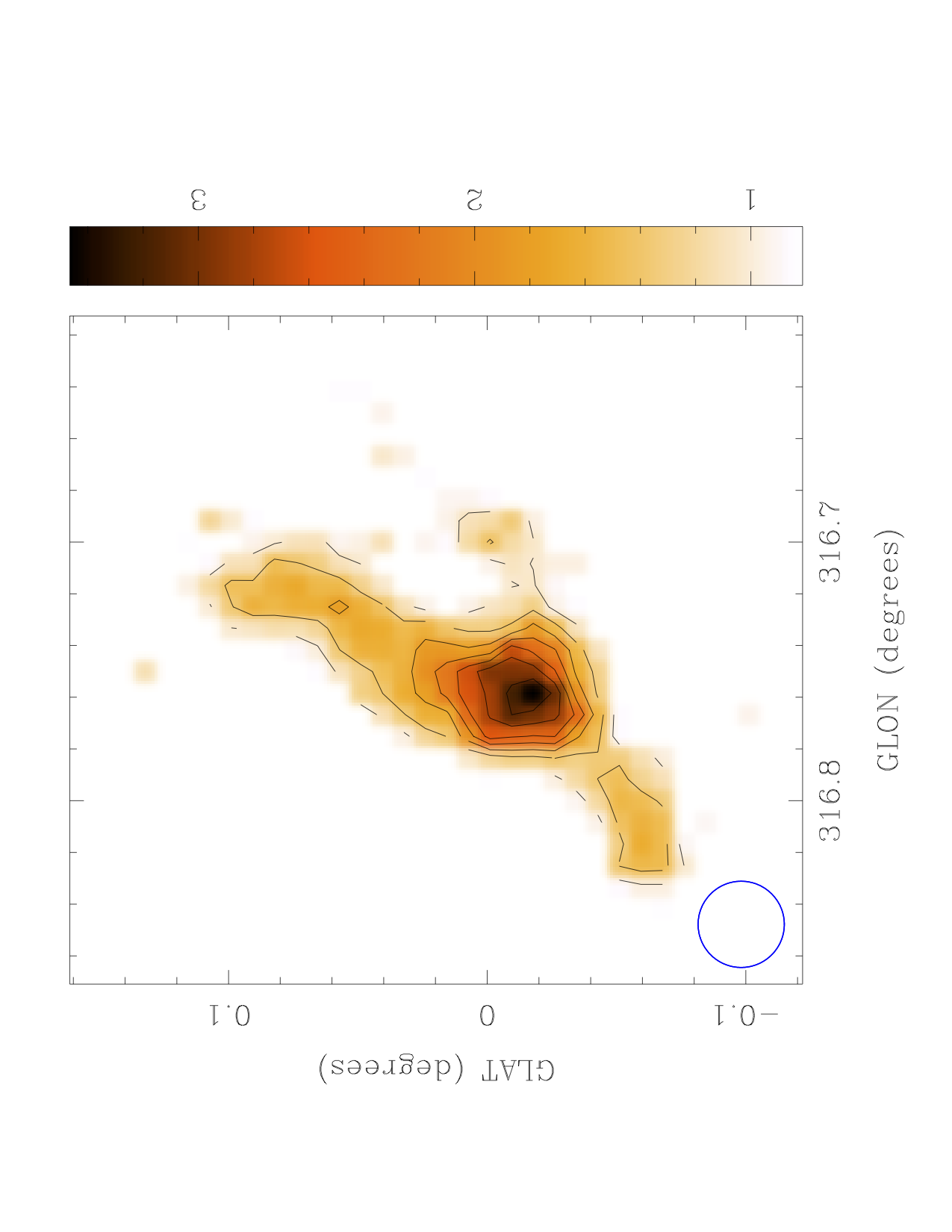} &
\includegraphics[height=5.0cm, angle=-90, trim=0 0 -5 0]{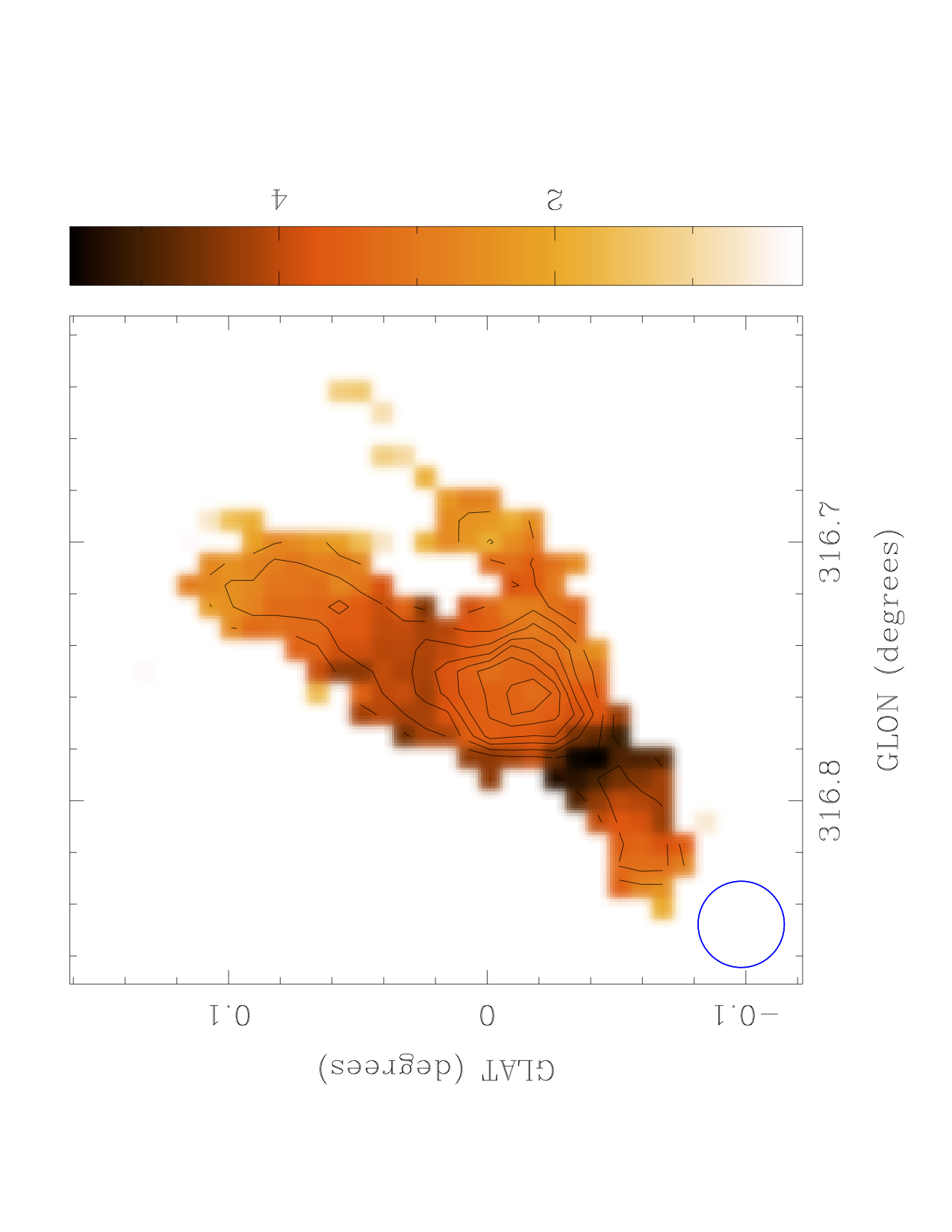} \\
\includegraphics[height=5.0cm, angle=-90, trim=0 0 -5 0]{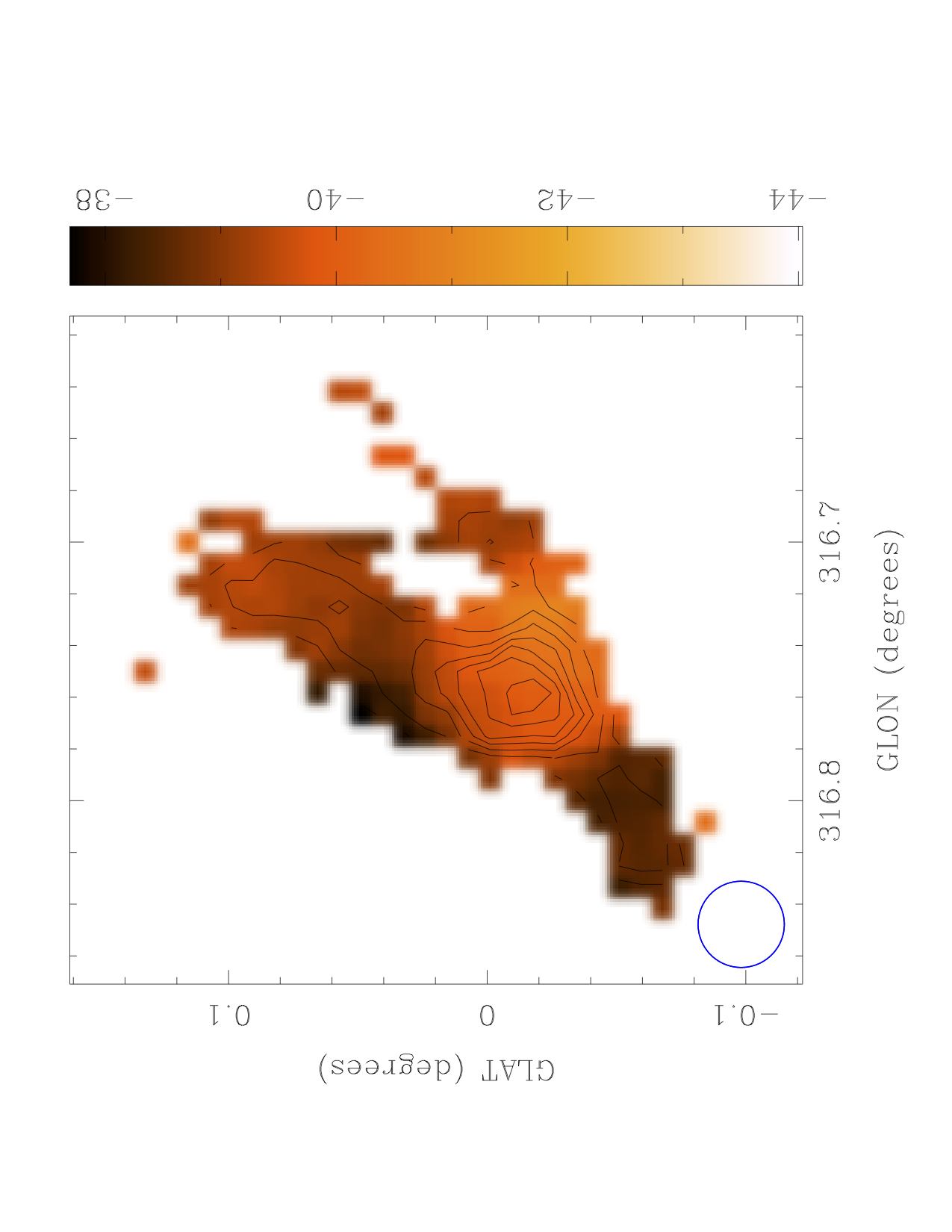} &
\includegraphics[height=5.0cm, angle=-90, trim=0 0 -5 0]{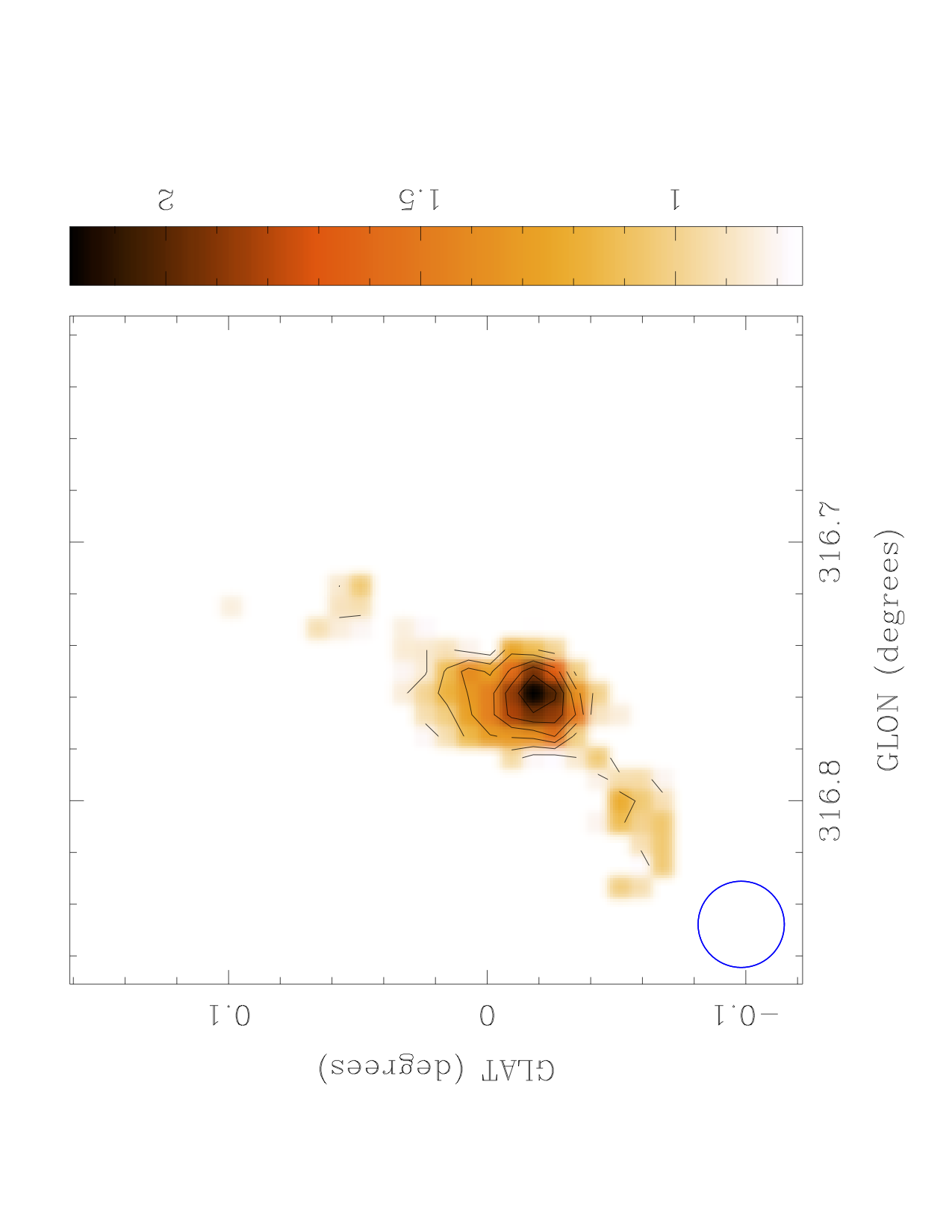} \\
\includegraphics[height=5.0cm, angle=-90, trim=0 0 -5 0]{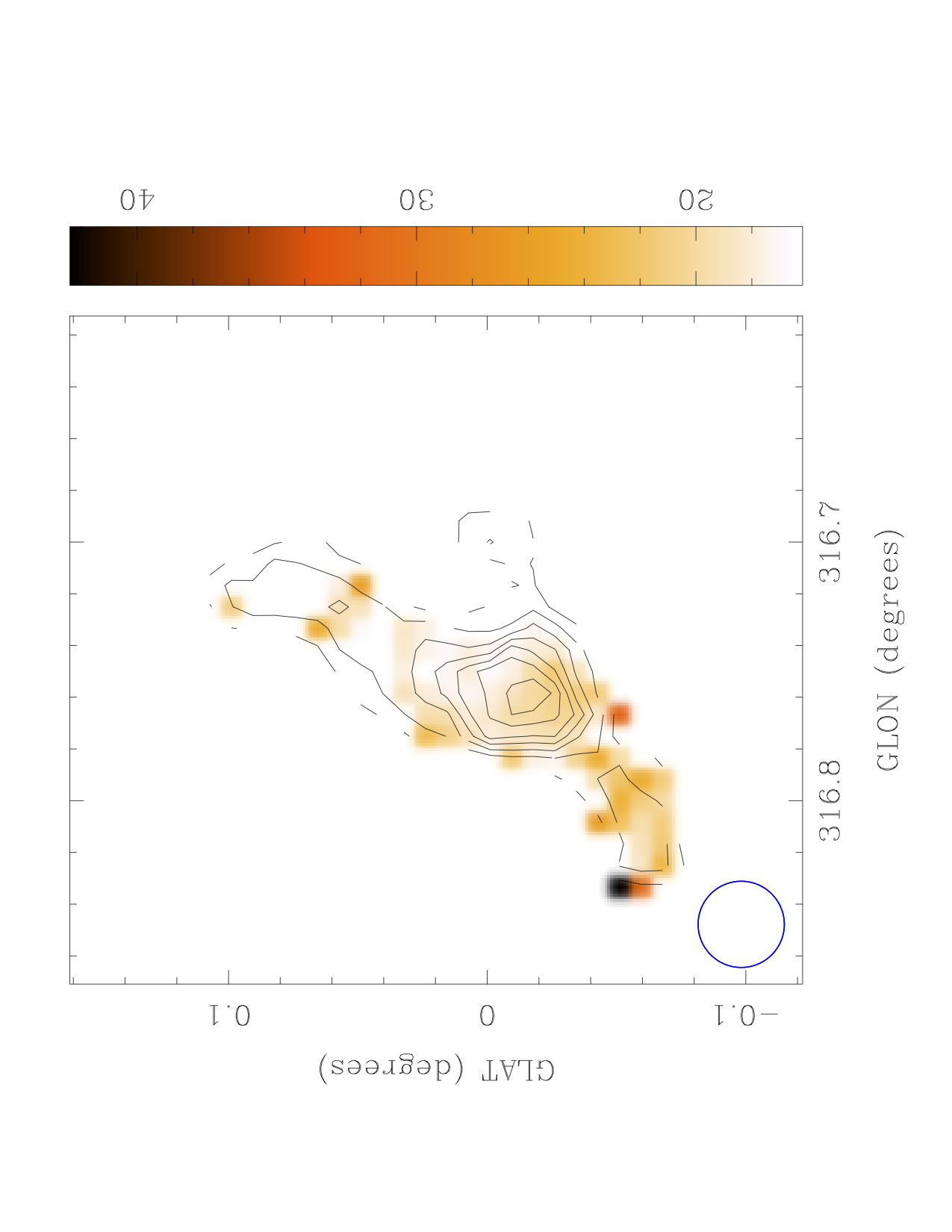} &
\includegraphics[height=5.0cm, angle=-90, trim=0 0 -5 0]{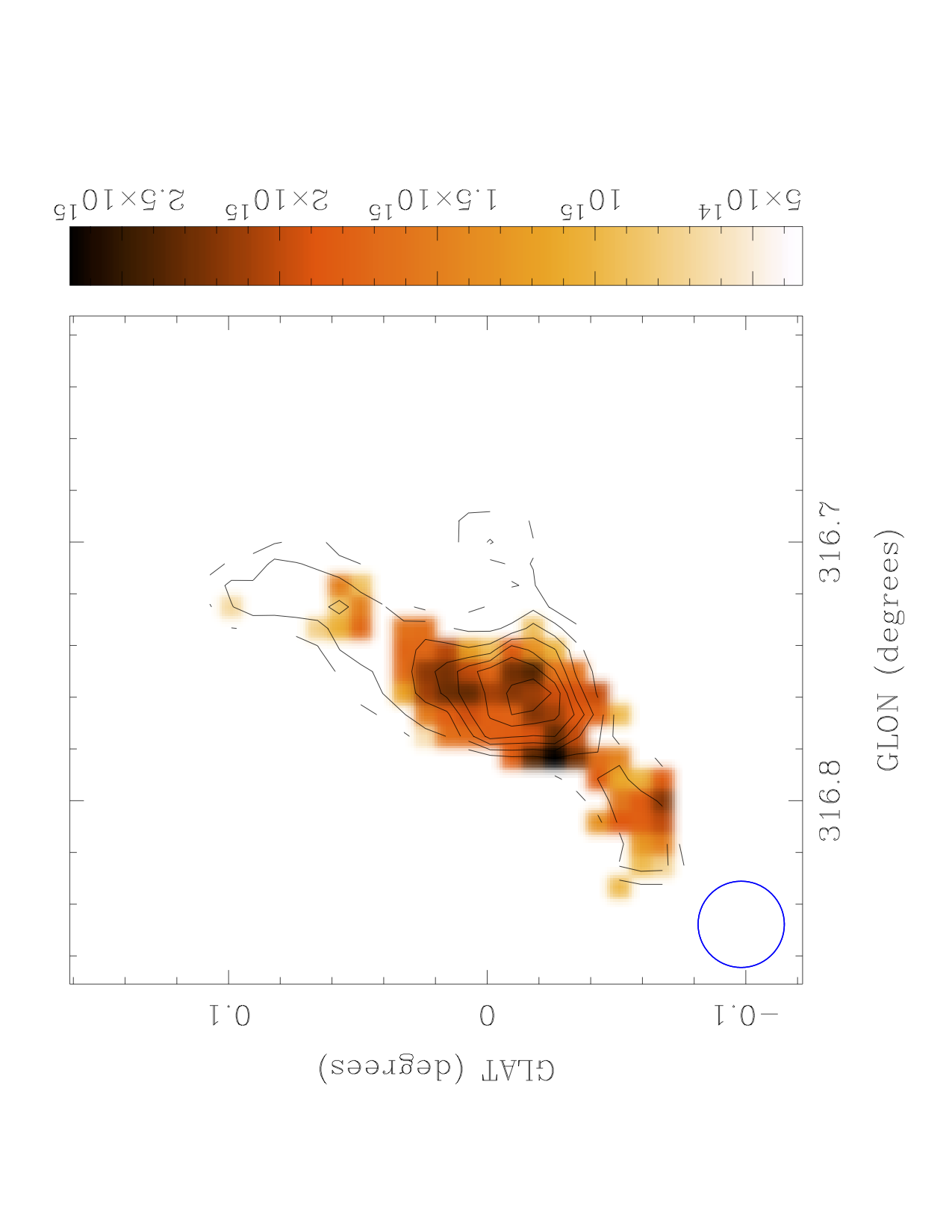} \\
\end{tabular}
\caption{Spatially-resolved gas properties output from the automated spectral line fitting routine for example source G$316.752+0.004-38.2$ (see Paper~II for the naming definition). \nhone\ [top left] and \nhtwo\ [top right] spectra from the pixel with peak \nhone\ emission, with residuals from the fit displayed underneath. The vertical dashed lines show the $\nhone$ V$_{\rm LSR}$. The left panel of the second row shows the brightness temperature (in K) of the \nhone\ emission overlaid with contours in 10\% steps of the peak value, from 90\% down. The right panel of the second row and left panel of the third row show the velocity dispersion and V$_{LSR}$ (in $\kms$) of the \nhone\ emission determined from the fit to each pixel. The contours for these two plots show the same $\nhone$ brightness temperature contours as the left panel of the second row. The right panel on the third row shows the brightness temperature (in K) of \nhtwo\ emission overlaid with contours in 10\% steps of the peak value, from 90\% down. The bottom left and right panels show the kinetic temperature (in K) and total \nhthree\ column density (in cm$^{-2}$) maps, respectively. The contours for these two plots show the same $\nhtwo$ brightness temperature contours as the right panel of the third row. The primary beam is shown in the bottom left corner of each image.}
\label{fig:sherpa_maps}
\end{center}
\end{figure*}

Maps of the fit parameters and derived gas properties were made for those regions with at least 30 pixels (i.e. $>$1.5 times the 2$\arcmin$ beam area) with reliable $\nhone$ fits. An example of these maps is shown in Figure~\ref{fig:sherpa_maps}. The top left and top right panels show $\nhone$ and $\nhtwo$ peak pixel spectra. If a reliable fit was obtained, the residual (the best-fit profiles subtracted from the data) is shown below the observed spectra for that transition, separated by a horizontal dotted line. For regions with a reliable $\nhone$ fit, the $\nhone$ V$_{\rm LSR}$ is shown as a dashed vertical line. For regions where no reliable fits were returned, only the observed spectra at the position of peak emission are shown. The left panel of the second row shows the NH$_3$(1,1) brightness temperature overlayed with contours in 10\% steps of the peak value, from 90\% down. The right panel of the second row and left panel of the third row show the velocity dispersion and V$_{\rm LSR}$ of the NH$_3$(1,1), respectively, determined from the fit. The contours for these two plots show the $\nhone$ brightness temperature. For those regions which also have at least 30 pixels with reliable $\nhtwo$ fits (for example, the region in Figure~\ref{fig:sherpa_maps}), the right panel on the third row shows the NH$_3$(2,2) brightess temperature overlayed with contours in 10\% steps of the peak value, from 90\% down. The bottom left and right panels show the kinetic temperature and total NH$_3$ column density maps, respectively. The contours for these two plots show the $\nhtwo$ brightness temperature. Similar plots for all regions with at least 30 pixels with reliable $\nhone$ fits are available in an online appendix. Plots of $\nhone$ and $\nhtwo$ peak pixel spectra (similar to those in the top row of Figure~\ref{fig:sherpa_maps}) are available for all sources in an online appendix.

In $\S$~\ref{sec:nh3_mas} we use these catalogues to try and determine the relative ages of the $\nhthree$ sources identified in HOPS.

%------------------------------
%Sherpa map figure

%============================================================
\section{Characterising the properties of HOPS NH$_3$ sources in the disk}
\label{sec:nh3_mas}

$\nhthree$ observations have played an important role in deriving the physical properties of gas in high-mass star formation regions and, in combination with other diagnostics, trying to define a self-consistent observational evolutionary sequence to order regions by their relative ages \citep[e.g.][]{pillai06, L07A, hill10, ragan11, dunham11, urquhart11, wienen12, battersby14b, urquhart15}. 

Previous work has used the diagnostic power of $\nhthree$ to determine the gas properties of young high mass star formation regions that have been selected through various methods, for example, infrared-dark clouds \citep[IRDCs:][]{pillai06, ragan11, battersby14a}, mid-IR bright young stellar objects \citep{urquhart11, urquhart15}, and (sub)mm continuum emission \citep{dunham11, wienen12}. These selection criteria are complementary. IRDCs identify regions at early evolutionary stages but require a bright mid-IR background, so preferentially pick out nearby objects. Mid-IR bright regions can be detected across the Galaxy but are more evolved. Longer wavelength (sub)mm continuum data identify both of these evolutionary stages with good completeness, but also select lower mass regions which are unlikely to form high-mass stars. 

As a large-area, blind, molecular line survey, HOPS allows us to take a different approach, by starting with an $\nhthree$-selected sample. HOPS is a relatively shallow survey, so this approach has disadvantages compared to the previous identification methods outlined above. For example, the survey is incomplete for lower mass and lower density molecular clouds. However, $\nhthree$ has an effective critical density of several $\sim$10$^3$\,$\cm3$ \citep{evans1999}, so traces dense molecular gas, and is found in regions cold enough that more common gas tracers, like CO, tend to deplete by freeze out onto dust grains (Bergin et al. 2006). Also, spectral line observations automatically provide the underlying kinematic gas structure, which is crucial for determining the nature of continuum sources (e.g. whether bright sources are in fact multiple regions superimposed along the line of sight). As shown in Paper~II, HOPS is therefore an ideal survey to identify the most massive and dense molecular clouds in the Galaxy. These are particularly interesting as potential precursors to the most massive and dense stellar clusters, dubbed `young massive clusters'  \citep[YMC's;][]{L12,L14PPVI}. We discuss potential YMC precursor clouds in $\S$~\ref{sub:ymc_candidates}.

Before doing this, our first goal is to understand the range of evolutionary stages covered by the HOPS $\nhthree$ sources. For this purpose we chose 22-GHz water maser \citep{walsh14} and 6.7-GHz methanol maser \citep{caswell10,caswell11,green10,green12, breen15} emission as diagnostics of embedded star formation activity within gas clouds. We then divided the HOPS $\nhthree$ detected regions into different groups depending on whether or not the sources contained maser emission within the primary beam (2$\arcmin$). We defined the following four groups: (i) NH$_3$ sources that are not associated with any known 22-GHz water or 6.7-GHz methanol maser; (ii) those that are associated with just 22-GHz water maser emission; (iii) those associated with just 6.7-GHz methanol maser emission, and (iv) those associated with both types of maser. 

Of the 687 HOPS $\nhthree$ sources, there are 490, 54, 70 and 73 in groups (i) to (iv), respectively. The vast majority of HOPS $\nhthree$ sources have no associated maser emission down to the sensitivity limits of the HOPS and Methanol Multi-beam surveys. Taken at face value the dominance of group (i) sources suggests that HOPS preferentially selects young regions which do not have prodigious embedded star formation activity. However, it may be the case that there are lower mass star formation regions in the sample. These would not be expected to have 6.7-GHz methanol maser emission \citep{minier03,breen13}. Also, with a 5$\sigma$ sensitivity of $\sim$1\,Jy, HOPS is a comparatively shallow 22-GHz water maser survey. So while it has the advantage of being complete, weaker water masers will have been missed. Therefore, the $\nhthree$-only group may contain a number of clouds with active low mass star formation. 

Previous studies have found such biases when comparing regions with different selection criteria. For example, \citet{hill10} found that, in their sample, millimetre continuum sources without other signs of star formation (including methanol masers and radio continuum emission) tended to have smaller sizes, masses and velocity dispersions. This was interpreted as either evidence for these sources to be at an earlier stage of star formation, or to be associated with lower mass star formation that does not exhibit methanol masers or radio continuum sources. 

\begin{figure}
\includegraphics[width=0.45\textwidth]{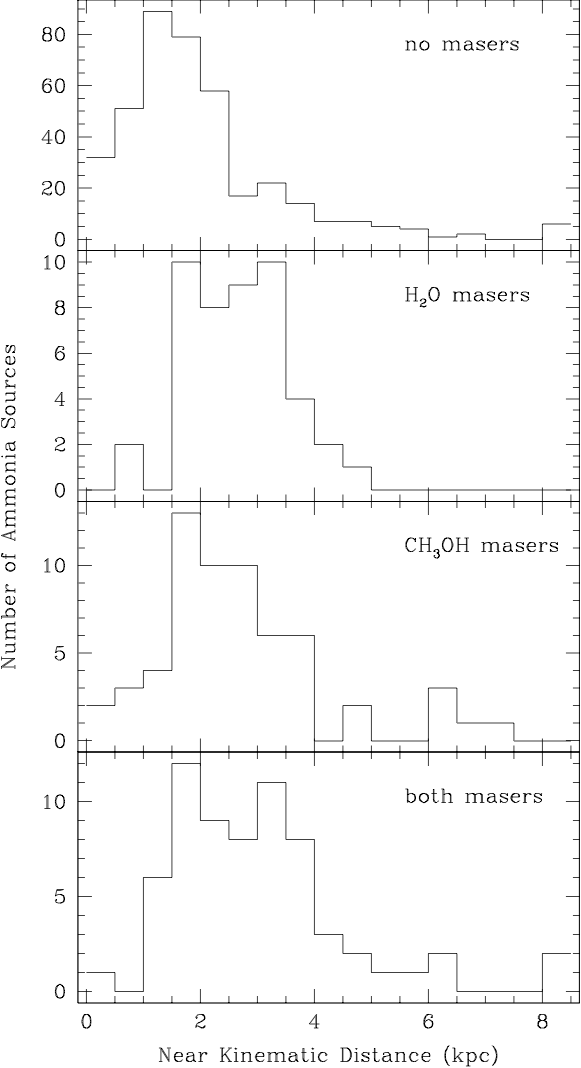}
\caption{Distributions of distances for NH$_3$ sources with no known maser association (top), associations with just water masers (next panel down), associations with just methanol masers (next panel down) and associations with both types of masers (bottom panel). }
\label{smplot1full}
\end{figure}

Given the molecular cloud mass completeness limits of 400\,M$_\odot$ for clouds at distances of 3.2\,kpc \citep{purcell2012}, HOPS should only be able to detect low mass star forming regions out to distances of a few kpc. Figure~\ref{smplot1full} shows the distribution of near kinematic distances derived in Paper~II for sources in the four groups. The $\nhthree$-only sources appear to be systematically closer. In particular, the fraction of sources at distances $>$3\,kpc is reduced compared to the other distributions. There are no clear differences in the distance distributions for sources associated with maser emission. This is consistent with the notion that the $\nhthree$-only group contains more nearby, low mass star formation regions.

\begin{figure}
\includegraphics[width=0.45\textwidth]{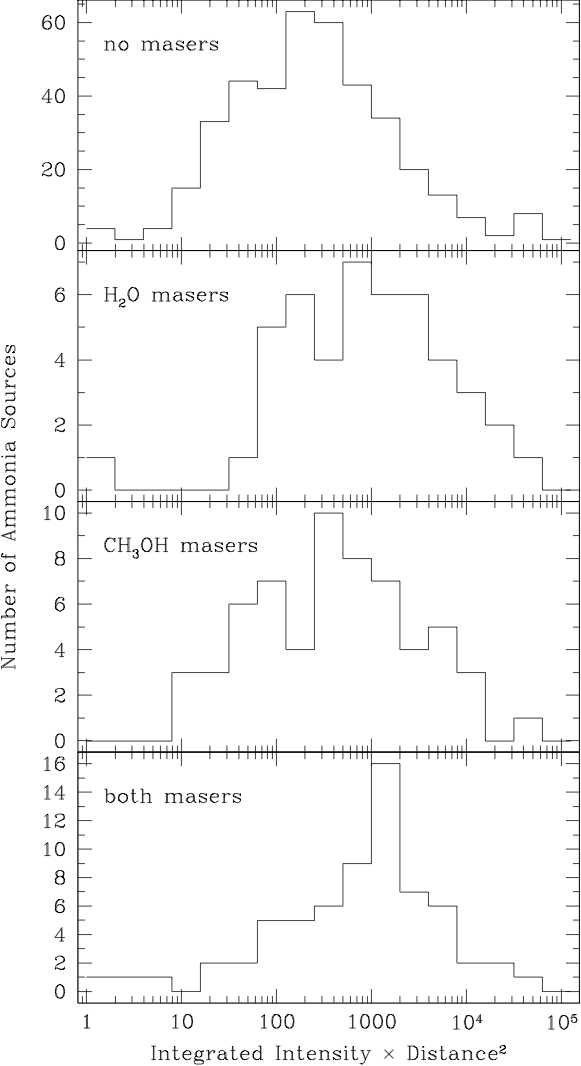}
\caption{Distributions of integrated intensity, multiplied by the square of the distance for all NH$_3$ sources (in units of K\,$\kms$\,kpc$^2$). This quantity is a crude estimation for the relative clump masses. The distributions are shown for NH$_3$ clumps with no known maser association (top), associations with just water masers (next panel down), associations with just methanol masers (next panel down) and associations with both types of masers (bottom panel). }
\label{smplot2full}
\end{figure}

Figure \ref{smplot2full} shows the distributions of $\nhone$ integrated intensity, multiplied by the square of the distance, for the same four sub-samples. This quantity is useful as a measure of the \emph{relative} mass of each region. While far too crude to use as an absolute mass measurement (it does not take into account the effects of gas temperature or variations in the NH$_3$ relative abundance, for example), this has the potential to highlight any major systematic differences in the $\nhthree$-derived mass between the groups. The range of values in all four groups are similar, but the combination of low number of sources in the groups and systematic uncertainties in the measurements mean we cannot determine statistically meaningful differences between the distributions. 

We then focused on the `highly-reliable' sources identified using the criteria outline in $\S$~\ref{sec:peak_spectra_fit_results}. Applying these criteria, there are then 13, 14, 12 and 24 sources in groups (i) to (iv), respectively. The number of $\nhthree$ sources associated with maser emission [groups (ii), (iii) and (iv)] drops by factors of several. However, the number of $\nhthree$-only sources [group (i)] drops by a factor $\sim$40. 

\begin{figure}
\includegraphics[width=0.45\textwidth]{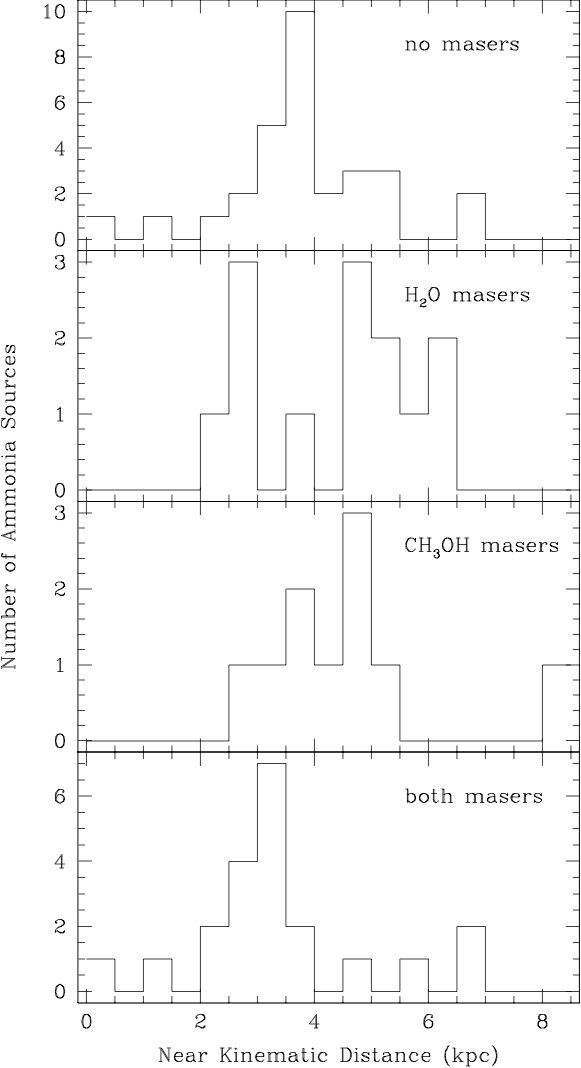}
\caption{Distributions of distances for `highly-reliable' NH$_3$ sources with no known maser association (top), associations with just water masers (next panel down), associations with just methanol masers (next panel down) and associations with both types of masers (bottom panel). There are no systematic differences between the four groups.}
\label{smplot1}
\end{figure}

\begin{figure}
\includegraphics[width=0.45\textwidth]{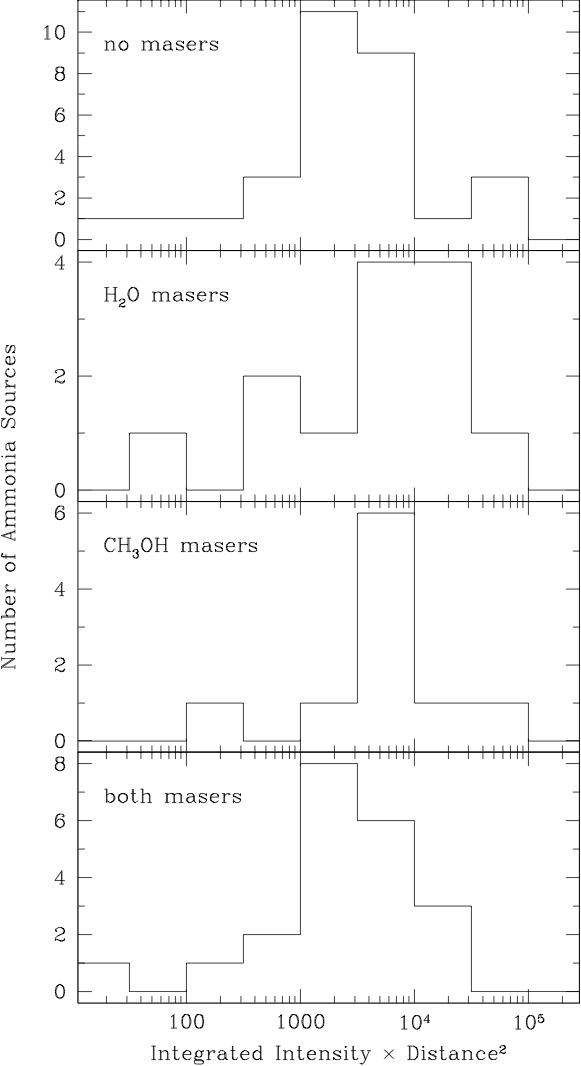}
\caption{Distributions of integrated intensity, multiplied by the square of the distance for `highly-reliable' NH$_3$ sources (in units of K\,$\kms$\,kpc$^2$). This quantity is a crude estimation for the relative clump masses. The distributions are shown for NH$_3$ clumps with no known maser association (top), associations with just water masers (next panel down), associations with just methanol masers (next panel down) and associations with both types of masers (bottom panel). There are no systematic differences between the four groups.}
\label{smplot2}
\end{figure}

\begin{figure}
\includegraphics[width=0.45\textwidth]{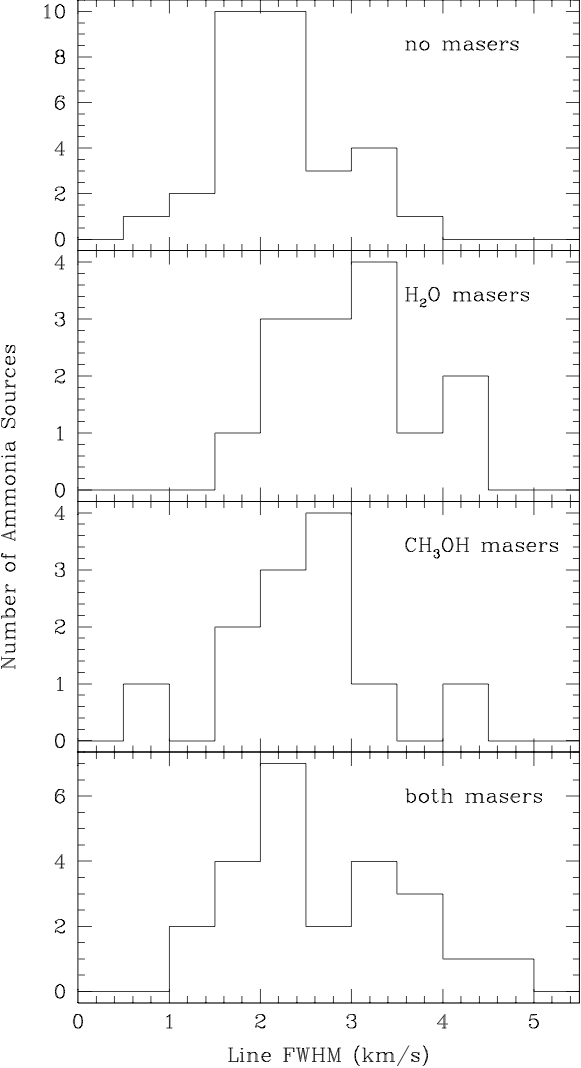}
\caption{Distributions of spectral line FWHM for NH$_3$ sources. The distributions are shown for NH$_3$ clumps with no known maser association (top), associations with just water masers (next panel down), associations with just methanol masers (next panel down) and associations with both types of masers (bottom panel). The top distribution shows typically narrower line widths.}
\label{smplot3}
\end{figure}

\begin{figure}
\includegraphics[width=0.45\textwidth]{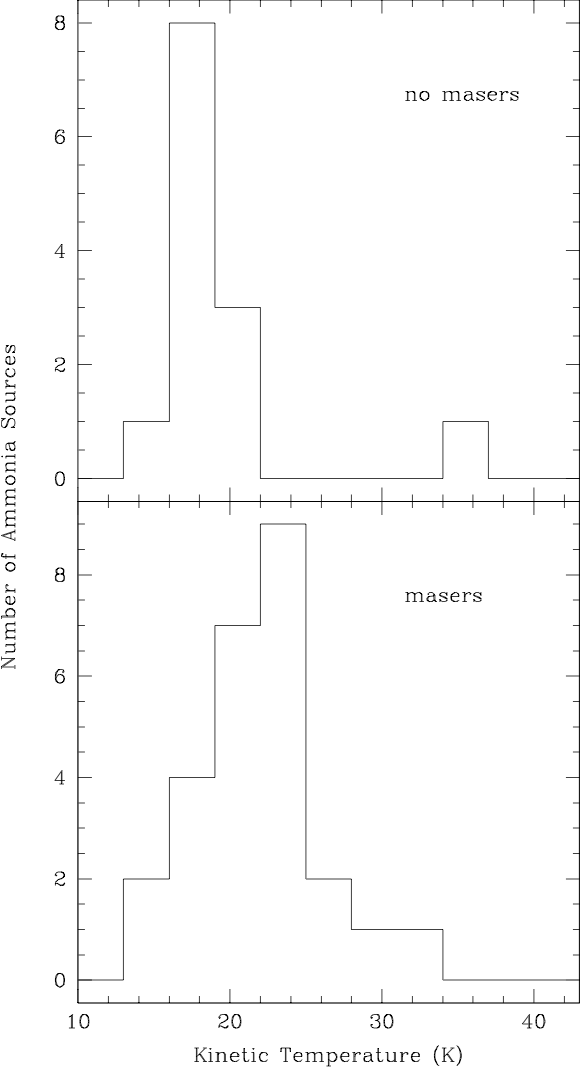}
\caption{Distributions of kinetic temperatures for NH$_3$ sources. The distributions are shown for NH$_3$ clumps with no known maser association (top) and associations either water and/or methanol masers (bottom). The top distribution shows typically lower kinetic temperatures.}
\label{smplot4a}
\end{figure}

Figures~\ref{smplot1}, \ref{smplot2}, \ref{smplot3} and \ref{smplot4a} show the distribution of near kinematic distance, integrated intensity multiplied by the square of the distance, linewidth and temperature, respectively, for the `highly-reliable' sources. There are no clear systematic differences between any of the distributions. There may be a slight trend for $\nhthree$-only sources to have lower line widths and temperatures than sources with masers -- similar to results found by other authors \citep[e.g.][]{L07A, hill10, wienen12} -- but the low numbers of sources in each group and associated systematic uncertainties mean the trend is not statistically significant.

%---------------------

%---------------------

%---------------------

%---------------------

In summary,  the $\nhthree$-only sources in the full catalogue are on average closer and have less reliable detections than $\nhthree$ sources associated with maser emission. The `highly-reliable' $\nhthree$-only sources have similar distance and relative mass distributions as the $\nhthree$ sources associated with maser emission, and may have slightly narrower $\nhone$ line widths and lower gas temperatures as would be expected if they are younger than $\nhthree$ sources associated with masers. 

%-----------------------------------
\subsection{Searching for Young Massive Cluster (YMC) progenitors}
\label{sub:ymc_candidates}

The most massive ($>$10$^4$\,M$_\odot$) and compact (radius $\sim$pc) stellar clusters are powerful probes of star formation across cosmological timescales. Their very high (proto)stellar densities constitute the most extreme known stellar birth environments, providing stringent tests for star and planet formation theories. Being both very bright and compact, the clusters can be readily observed to large distances. The youngest of these are therefore ideal tracers of the recent star formation history of galaxies. Having survived for close to a Hubble time, globular clusters provide a window into star formation when the Universe was still in its infancy, and also trace the subsequent mass assembly of galaxies. Understanding how such extreme clusters form is therefore a key question in many areas of astrophysics.

The properties of young massive clusters (YMCs) have been studied in detail at optical and infrared (IR) wavelengths for decades \citep[see][for a review]{pz10}. One of their defining characteristics are very centrally-concentrated stellar mass profiles, and it is often assumed that the stars formed at similar (or even higher) proto-stellar densities. However, by the time the clusters are visible in the optical and infrared, they are already a few Myr old. As the dynamical time of such dense stellar systems is much shorter than this (of order 0.1\,Myr), the stars very quickly reach dynamical equilibrium, and the imprint of the initial conditions is lost. This makes it very difficult to unambiguously determine if the stars \emph{formed} at such high density, or have dynamically evolved into this state from initially hierarchical substructure as observed for nearby star forming regions and in cluster scale simulations \citep[see e.g.][]{smith09,maschberger10,parker14, wright14}.

A complementary approach to investigate how YMCs form is to search for the likely gas cloud progenitors of these clusters and determine the spatial/kinematic structure before they are visible in the optical/IR and dynamical effects have erased the initial conditions. The capabilities of both current and planned observational facilities mean for the next several decades at least, it will only be possible to detect and resolve individual stars forming in such clouds within our own Galaxy. Finding all the progenitor clouds in the Milky Way is therefore a key step towards understanding the formation mechanism of YMCs. 

The recent completion of many multi-wavelength, sensitive and high-resolution Galactic plane surveys has made it possible to begin systematic searches for YMC precursor clouds in the Milky Way \citep[see][for a review]{L14PPVI}. To date these have focused on finding gas clouds with enough mass ($>$ few $10^4$\,M$_\odot$) in a small enough volume (radius $\sim1$\,pc) that they have the potential to form a stellar system with a density profile similar to that observed for YMCs \citep{bressert12}. These surveys should easily be able to detect any such massive and compact objects through the Galaxy, and a small number of such candidate YMC precursor clouds have been found in both the Galactic disk \citep{ginsburg12,urquhart13, contreras17} and the Galactic Centre \citep{L12, kauffmann13, longmore13b, johnston14, rathborne14a, rathborne14_turb, rathborne15}. 

\citet{walker15,walker16} analysed the radial gas distribution of all known YMC precursor candidates and showed that the gas mass profiles are generally shallower than the stellar mass surface density profiles of the YMCs. Therefore, despite dedicated observational searches which should be complete to such objects throughout the Galaxy, no clouds have been found with a sufficiently high gas mass concentration to form a system with stellar mass as centrally-concentrated as observed for YMCs (i.e. $\sim$10$^4$\,M$_\odot$ within core radii of $\sim$0.5\,pc). The fact that more quiescent, less evolved clouds contain less mass in their central regions than highly star-forming clouds, suggests an evolutionary trend in which YMC progenitor clouds continue to accumulate mass towards their centres after the onset of star formation. However, even if the gas keeps flowing in, the stellar mass of the forming cluster may still dominate the local gravitational potential due to short free-fall times and rapid star formation, leading to central gas exhaustion \citep{kruijssen12, ginsburg16b}.

There are several natural consequences of this ``conveyor-belt" mass accumulation scenario for YMC formation. Firstly, at the earliest evolutionary stages, the mass in the precursor clouds that ultimately ends up in the YMC is spread over a much larger radius. As such, potential YMCs precursor clouds at the earliest evolutionary stages may have been excluded from previous searches looking for clouds with $>$10$^4$\,M$_\odot$ of gas within radii of $<$0.5\,pc. Secondly, the gas at these larger radii must converge to the eventual cluster centre in less than 1Myr -- the upper limit to the observed stellar age spreads in YMCs \citep{L14PPVI}. This converging gas flow should be imprinted in the observed gas kinematics. As a large-area, blind, spectroscopic, dense gas Galactic plane survey with angular resolution corresponding to parsec-scale physical resolutions at the expected distances, HOPS is ideally suited to finding such YMC precursor clouds in the earliest evolutionary stages. Below we describe the method we used to search for YMC progenitor clouds in the HOPS data.

The fundamental property for a gas cloud to be included as a potential YMC precursor is that it has sufficient mass to form a stellar cluster of 10$^4$\,M$_\odot$. To derive the gas mass of the HOPS $\nhthree$ sources we used data from the ATLASGAL survey \citep{schuller09}, providing a map of \textgreater \ 400 square degrees of the inner Galaxy at 870~$\mu$m, with an angular resolution of $\sim$ 19$^{\prime\prime}$. To match the sources in the HOPS catalogue with regions in the ATLASGAL map, we smoothed the full ATLASGAL map to the angular resolution and pixel scale of the HOPS data. We then applied the 2D HOPS masks (described in full detail in Paper II) to the ATLASGAL map and extracted the data within each HOPS mask as an individual map. As the spatial coverage of the two surveys are different, we were unable to obtain a masked dust map for every HOPS source. Of the 687 sources in the HOPS catalogue, we obtained 605 corresponding dust maps.

We then estimated the mass of each HOPS source through,
\[
M = \frac{d^2}{\kappa_{\nu} B_{\nu}(T)} \int I_{\nu} d\Omega = \frac{d^2 F_{\nu}}{\kappa_{\nu} B_{\nu}(T)},
\]
where $M$ is the mass, $B_{\nu}$(T) is the Planck function, $T$ is the dust temperature, $\kappa_{\nu}$ is the dust opacity, $F_{\nu}$ is the integrated flux and $d$ is the distance. To estimate the dust opacity ($\kappa_{\nu}$), we use the following relation \citep{battersby11},

\[
\kappa_{\nu} = 0.04 \rm ~ cm^{2} ~ g^{-1} \left(\frac{\nu}{505 ~ GHz}\right)^{1.75},
\]

where $\nu$ is the frequency. Note that this contains the assumption that the gas-to-dust-ratio is 100.

As we do not know the dust temperature for all of the sources, we assume a uniform value of 15~K in all cases. Though the true dust temperature is likely to be non-uniform and vary between the regions, 15~K is consistent with typical temperatures seen in molecular clouds and $\nhthree$-derived gas temperatures above, and in other studies \citep[e.g.][]{wienen12}. Assuming the dust temperature for each source equals the HOPS $\nhthree$-derived gas temperature derived for that source does not affect the conclusions below. 

The distance to each HOPS source was calculated from the $\nhone$ V$_{\rm LSR}$ following the procedure outlined in \citet{urquhart14}, but adopting the updated Galactic model described in \citet{reid14} which uses the most recent maser parallax measurements to constrain galactic rotation. We then couple these distance estimates with the estimated geometric radii for each catalogued source to determine their physical radii in parsecs.

\begin{figure}
\includegraphics[width=0.45\textwidth]{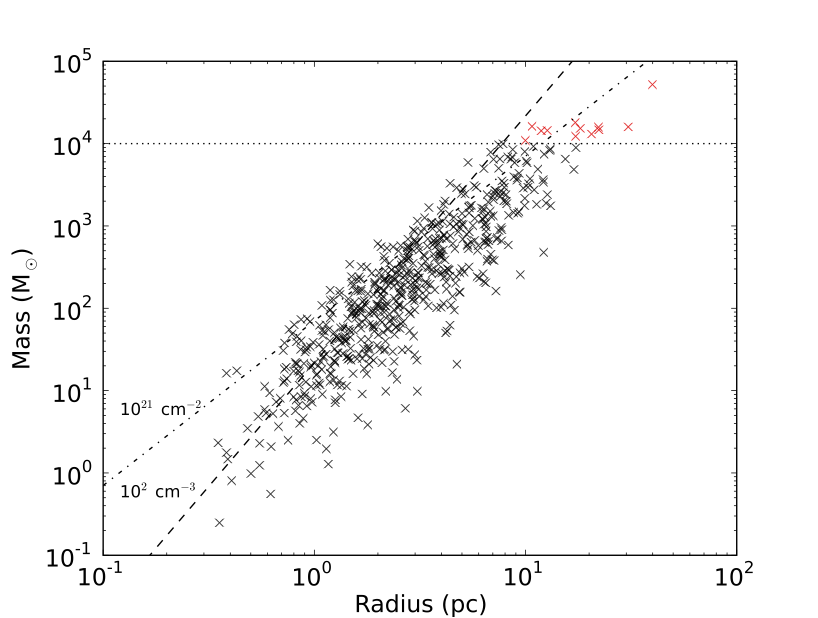}
\caption{Mass vs radius for the 605 HOPS $\nhthree$ sources which could be matched with the ATLASGAL catalogue. The dotted horizontal line shows the 10$^{4}$~M$_{\odot}$ mass threshold. We find a total of 12 sources (marked as red points) above this threshold. The dashed line shows a constant volume density of 10$^2$\,cm$^{-3}$. The dot-dashed line shows a constant column density of 10$^{21}$\,cm$^{-2}$. }
\label{fig:mass_radius}
\end{figure}

With mass and radius estimates for the HOPS $\nhthree$ sources, we then seek to identify which of these may be potential YMC precursor candidates. By definition, the stellar mass of YMCs is $\gtrsim~10^{4}$\,M$_{\odot}$. Our first criteria is therefore to select HOPS sources with a large enough mass reservoir to form a YMC, i.e. M~$>10^{4}$~M$_{\odot}$. Figure~\ref{fig:mass_radius} shows the mass vs. radius for the 605 HOPS $\nhthree$ sources which could be matched with the ATLASGAL catalogue. The dotted horizontal line shows the 10$^{4}$~M$_{\odot}$ mass threshold. We find a total of 12 sources (marked as red points) above this threshold.

\begin{table*}
  \begin{tabular}{cccccccc}
    \hline
    Source	&	M 	&	R & FWHM & d$_{\textrm{near}}$ & d$_{\textrm{far}}$ & d$_{\textrm{adopted}}$ ($\Delta$d) & Probability\\
    & 	10$^{4}$ M$_\odot$	&	pc  & km s$^{-1}$ & kpc & kpc & kpc & \\ \hline
	G003.432$-$0.351 & 1.2 & 17 & 3.08 & $^{*}$ & $^{*}$ & 21.1 (3.7) & 0.59\\
	G316.752$+$0.004 & 1.5 & 18 & 2.95 & 2.6 & 9.8 & 9.8 (0.8) & 0.54\\
	G330.881$-$0.371 & 1.1 & 9 & 3.5 & 3.9 & 11.0 & 11.0 (0.4) & 0.7\\
	G338.464$+$0.034 & 1.6 & 10 & 11.8 & 3.1 & 12.7 & 12.6 (0.5) & 0.74\\
	G341.224$-$0.274 & 1.3 & 20 & 1.89 & 3.6 & 12.5 & 12.4 (0.5) & 0.79\\
	G350.170$+$0.070 & 1.8 & 17 & 5.66 & 5.8 & 11.0 & 10.4 (0.4) & 0.72\\
	G357.555$-$0.323 & 1.5 & 22 & 1.48 & $^{*}$ & $^{*}$ & 16.8 (2.3) & 0.79\\ \hline \hline
  \end{tabular}
  \caption{Properties of HOPS sources identified as candidate YMC precursor gas clouds. The estimated mass (M), radius (R), FWHM, near distance (d$_{\textrm{near}}$), far distance (d$_{\textrm{far}}$), adopted distance (d$_{\textrm{adopted}}$) with associated error ($\Delta$d), and the probability that the source is at the adopted distance are given for each source. $^{*}$A reliable estimate of near and far distance limits could not be obtained for these sources due to their close projected distance to the Galactic Centre.}
       \label{tab:YMC_precursor_candidate_properties}
\end{table*}

Many of the sources are found projected within a few degrees of the Galactic Centre. Inspection of the dust continuum emission and the $\nhone$ integrated intensity, velocity field and the velocity dispersion towards G001.374$+$0.112, G002.826$+$0.048, G003.145$+$0.3014, G003.340$+$0.396 and G358.894$-$0.290, show that these sources all have very broad line profiles, with FWHM well in excess of $>$15\,$\kms$. Inspection of the full HOPS $\nhthree$ data cubes show these sources are clearly associated with prominent features in the Galactic Centre such as the `1.3 degree cloud', `Clump 2' and `Sgr C' (see $\S$~\ref{sec:galactic_centre}). They have likely have only been identified as separate sources in Paper~II as the emission linking them to the main structures falls below the HOPS sensitivity threshold. The kinematic models used to determine distances from the $V_{\rm LSR}$ are not reliable for regions so close to the Galactic Centre. In addition, the 15\,K dust temperature is likely to be an underestimate \citep{ginsburg16}. Placing these sources at the correct distance of $\sim$8.5\,kpc and using a lower limit to the dust temperature of 20\,K, the mass of these sources drops below 10$^4$\,M$_\odot$.  We conclude that these are unlikely to be YMC precursor clouds.

\begin{figure*}
\includegraphics[width=1.0\textwidth, angle=0, trim= 0 0 0 0, clip]{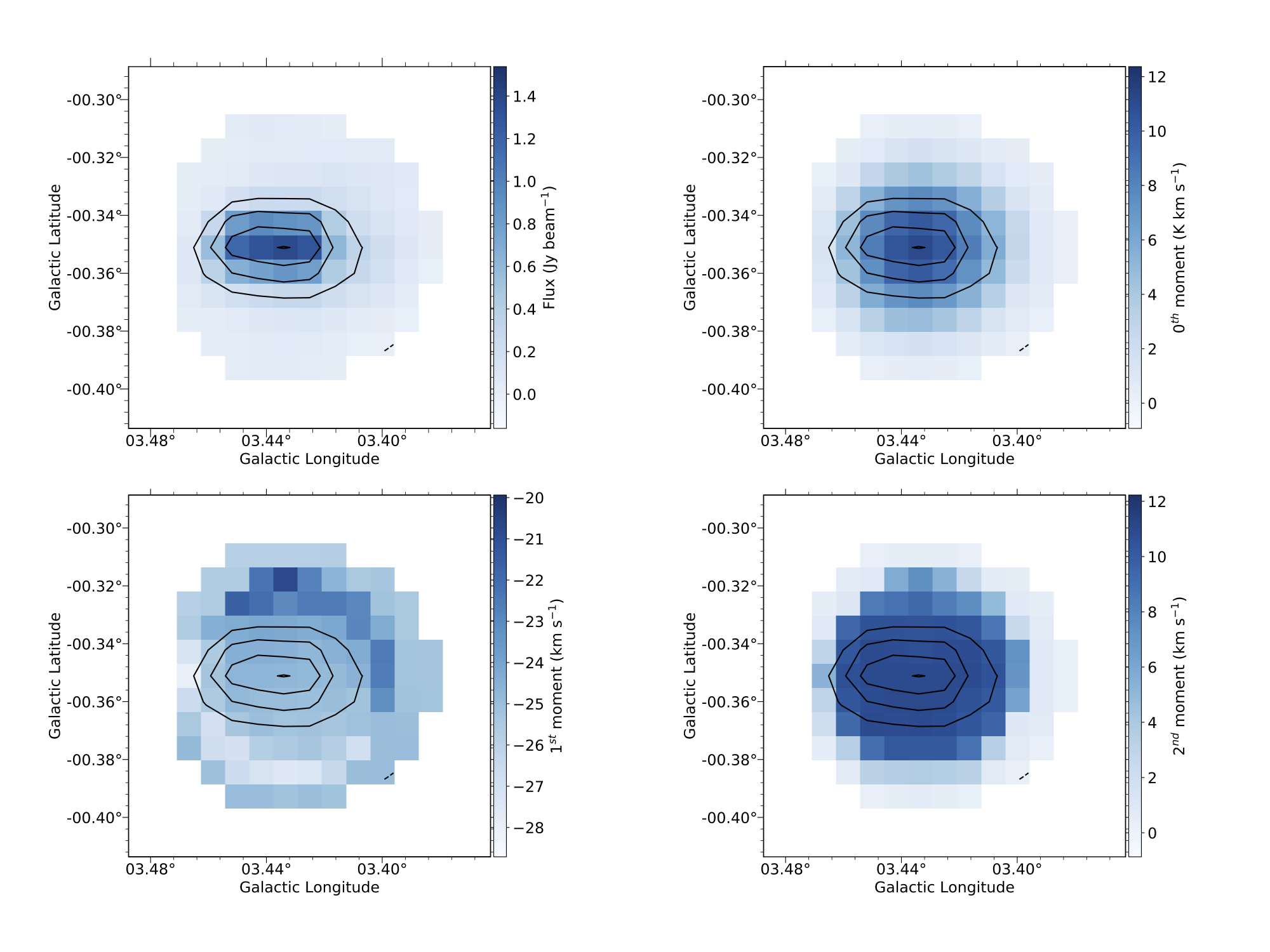} \\
\caption{$\nhone$ peak intensity [top left], integrated intensity (zeroth moment) [top right], velocity-weighted intensity (1st moment) [bottom left], and intensity-weighted velocity dispersion (2nd moment) [bottom right] maps of HOPS YMC progenitor gas cloud candidate, G003.432$-$0.351 \citep{purcell2012}. The contours in all panels show the peak intensity distribution in steps of 20\% of the maximum intensity in the map. }
\label{fig:ymc_candidate_5}
\end{figure*}

\begin{figure*}
\includegraphics[width=0.4\textwidth, angle=0, trim= 0 0 0 0, clip]{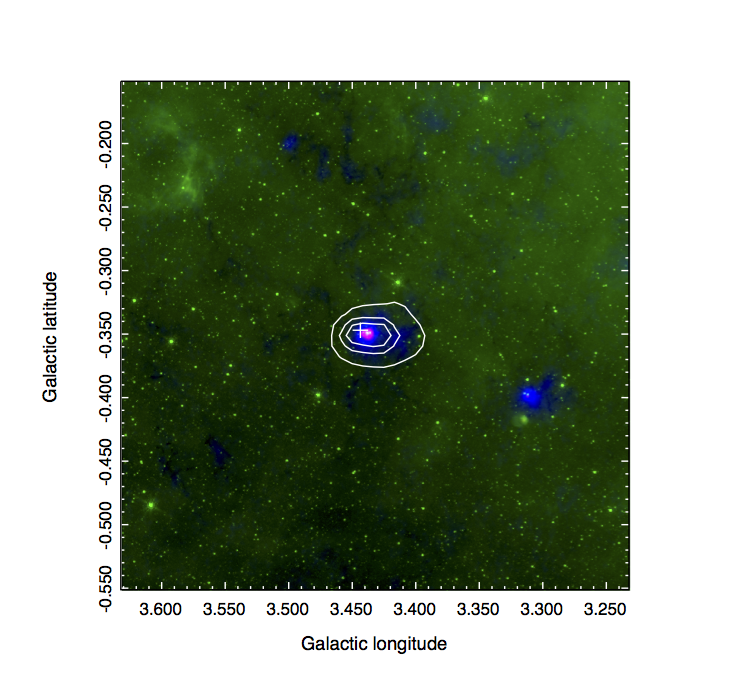} \\
\caption{Three-colour images of young massive cluster progenitor gas cloud candidate G003.432$-$0.351 [Red, HIGAL 70\,$\mu$m \citep{molinari_2010}; green, 8\,$\mu$m GLIMPSE \citep{churchwell09}; blue, ATLASGAL 850\,$\mu$ emission \citep{schuller09}]. Crosses are methanol masers \citep{caswell10,caswell11} and plus symbols are HOPS water masers \citep{walsh2011, walsh14}. White contours show the $\nhone$ integrated intensity emission.}
\label{fig:ymc_3col_G003}
\end{figure*}

\begin{figure*}
\includegraphics[width=1.0\textwidth, angle=0, trim= 0 0 0 0, clip]{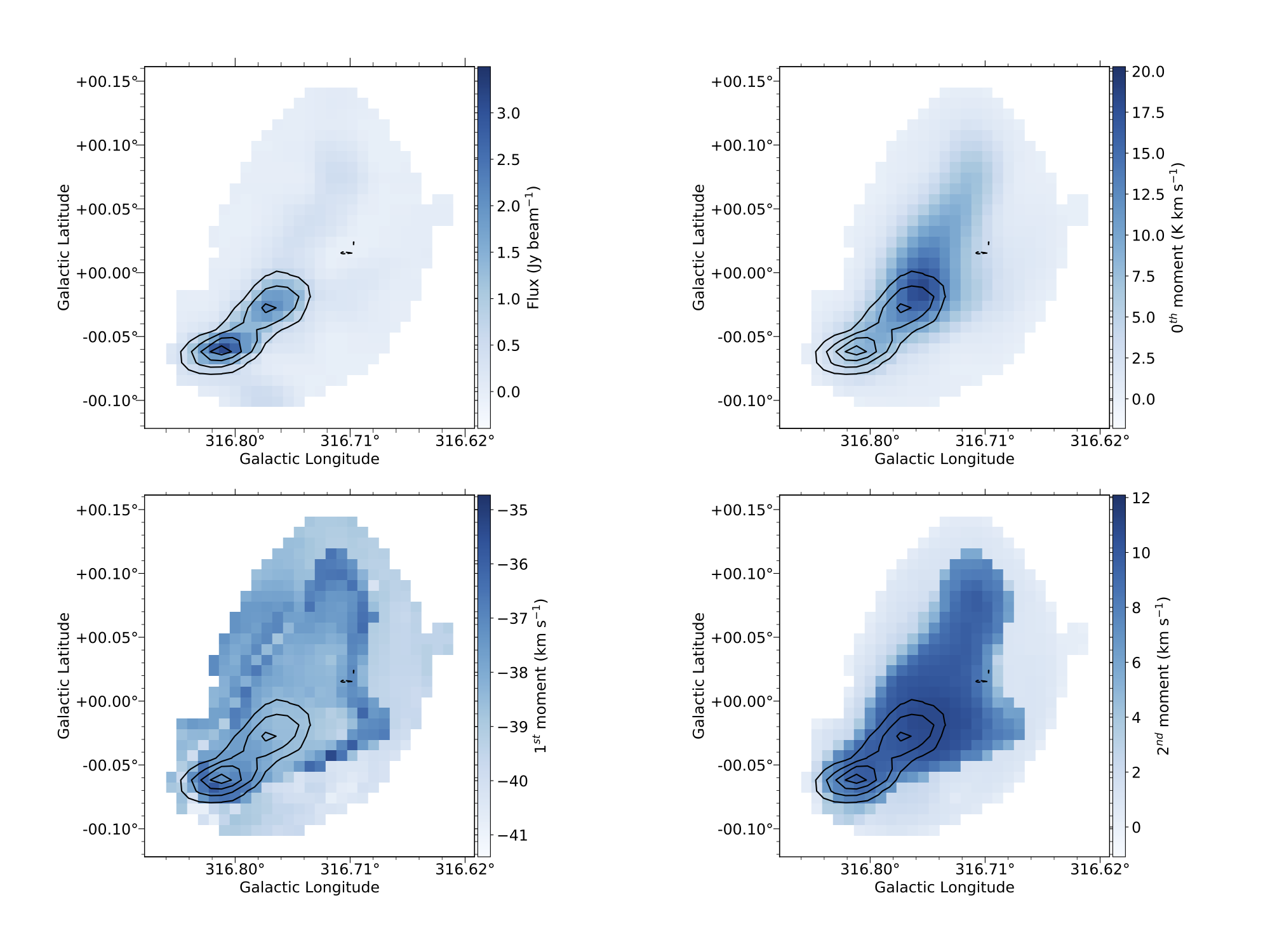} \\
\caption{$\nhone$ peak intensity [top left], integrated intensity (zeroth moment) [top right], velocity-weighted intensity (1st moment) [bottom left], and intensity-weighted velocity dispersion (2nd moment) [bottom right] maps of HOPS YMC progenitor gas cloud candidate, G316.579$+$0.054 \citep{purcell2012}. The contours in all panels show the peak intensity distribution in steps of 20\% of the maximum intensity in the map.}
\label{fig:ymc_candidate_6}
\end{figure*}

\begin{figure*}
\includegraphics[width=0.4\textwidth, angle=0, trim= 0 0 0 0, clip]{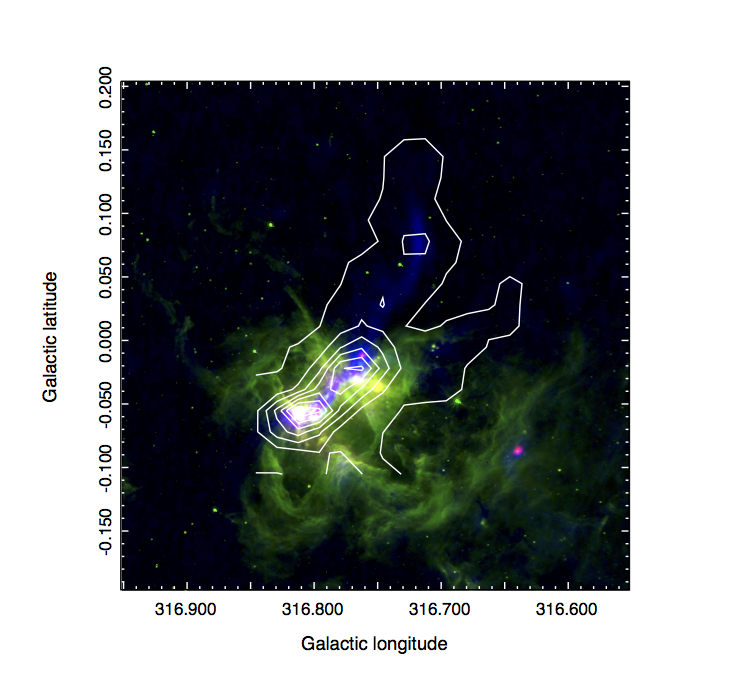} \\
\caption{Three-colour images of young massive cluster progenitor gas cloud candidate G316.579$+$0.054 [Red, HIGAL 70\,$\mu$m \citep{molinari_2010}; green, 8\,$\mu$m GLIMPSE \citep{churchwell09}; blue, ATLASGAL 850\,$\mu$ emission \citep{schuller09}]. Crosses are methanol masers \citep{caswell10,caswell11} and plus symbols are HOPS water masers \citep{walsh2011, walsh14}. White contours show the $\nhone$ integrated intensity emission.}
\label{fig:ymc_3col_G316}
\end{figure*}

\begin{figure*}
\includegraphics[width=1.0\textwidth, angle=0, trim= 0 0 0 0, clip]{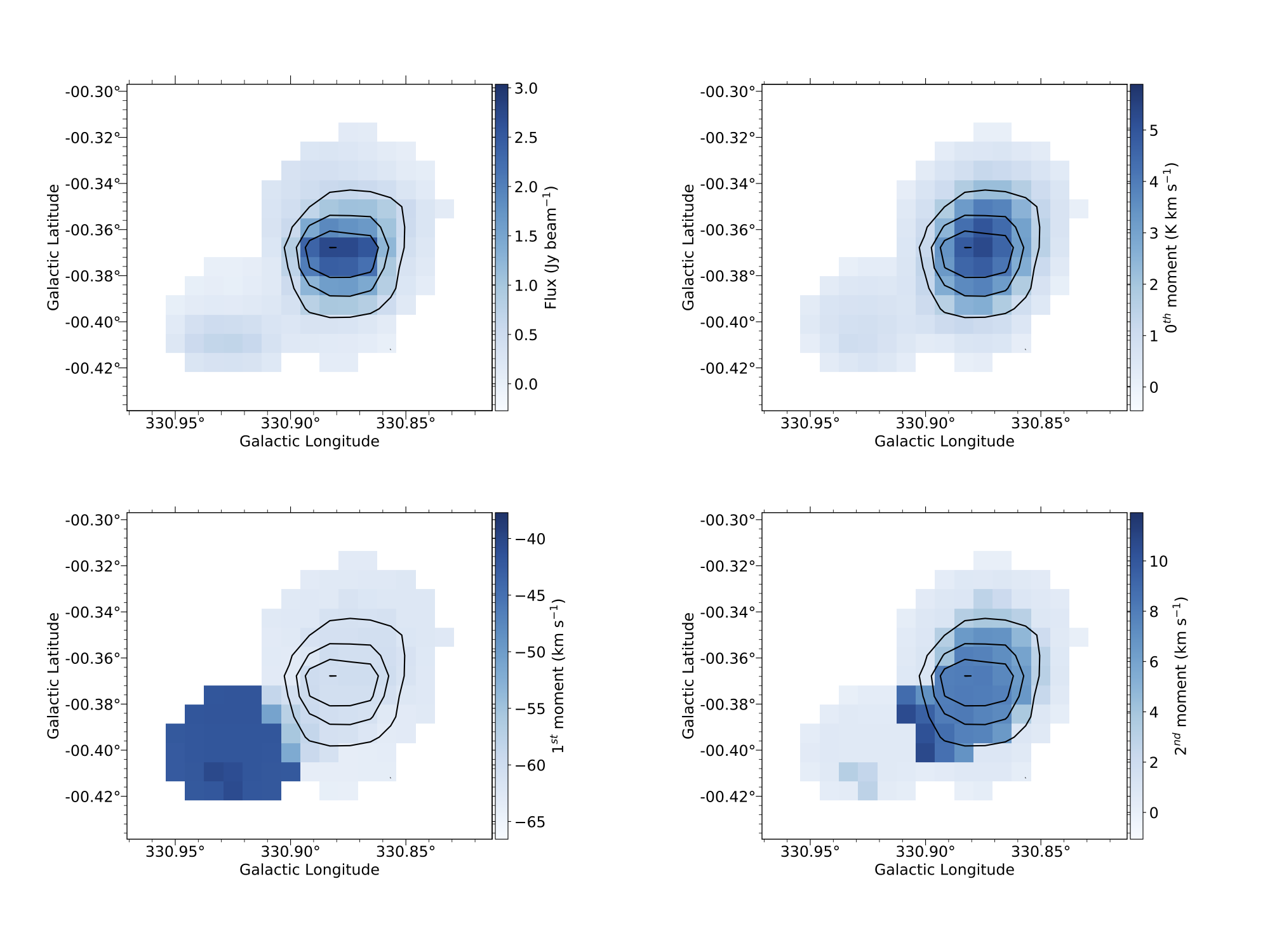} \\
\caption{$\nhone$ peak intensity [top left], integrated intensity (zeroth moment) [top right], velocity-weighted intensity (1st moment) [bottom left], and intensity-weighted velocity dispersion (2nd moment) [bottom right] maps of HOPS YMC progenitor gas cloud candidate, G330.881$-$0.371 \citep{purcell2012}. The contours in all panels show the peak intensity distribution in steps of 20\% of the maximum intensity in the map.}
\label{fig:ymc_candidate_7}
\end{figure*}

\begin{figure*}
\includegraphics[width=0.4\textwidth, angle=0, trim= 0 0 0 0, clip]{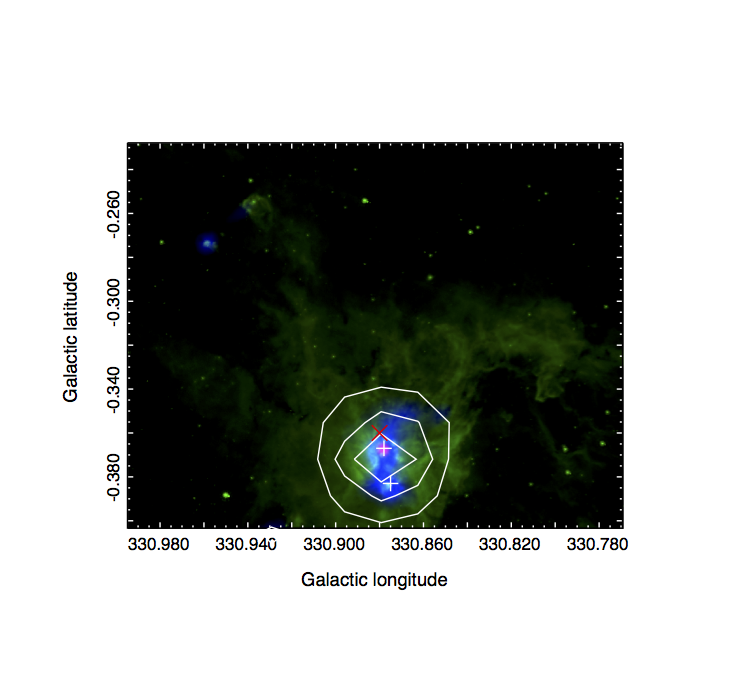} \\
\caption{Three-colour images of young massive cluster progenitor gas cloud candidate G330.881$-$0.371 [Red, HIGAL 70\,$\mu$m \citep{molinari_2010}; green, 8\,$\mu$m GLIMPSE \citep{churchwell09}; blue, ATLASGAL 850\,$\mu$ emission \citep{schuller09}]. Crosses are methanol masers \citep{caswell10,caswell11} and plus symbols are HOPS water masers \citep{walsh2011, walsh14}. White contours show the $\nhone$ integrated intensity emission.}
\label{fig:ymc_3col_G330}
\end{figure*}

\begin{figure*}
\includegraphics[width=1.0\textwidth, angle=0, trim= 0 0 0 0, clip]{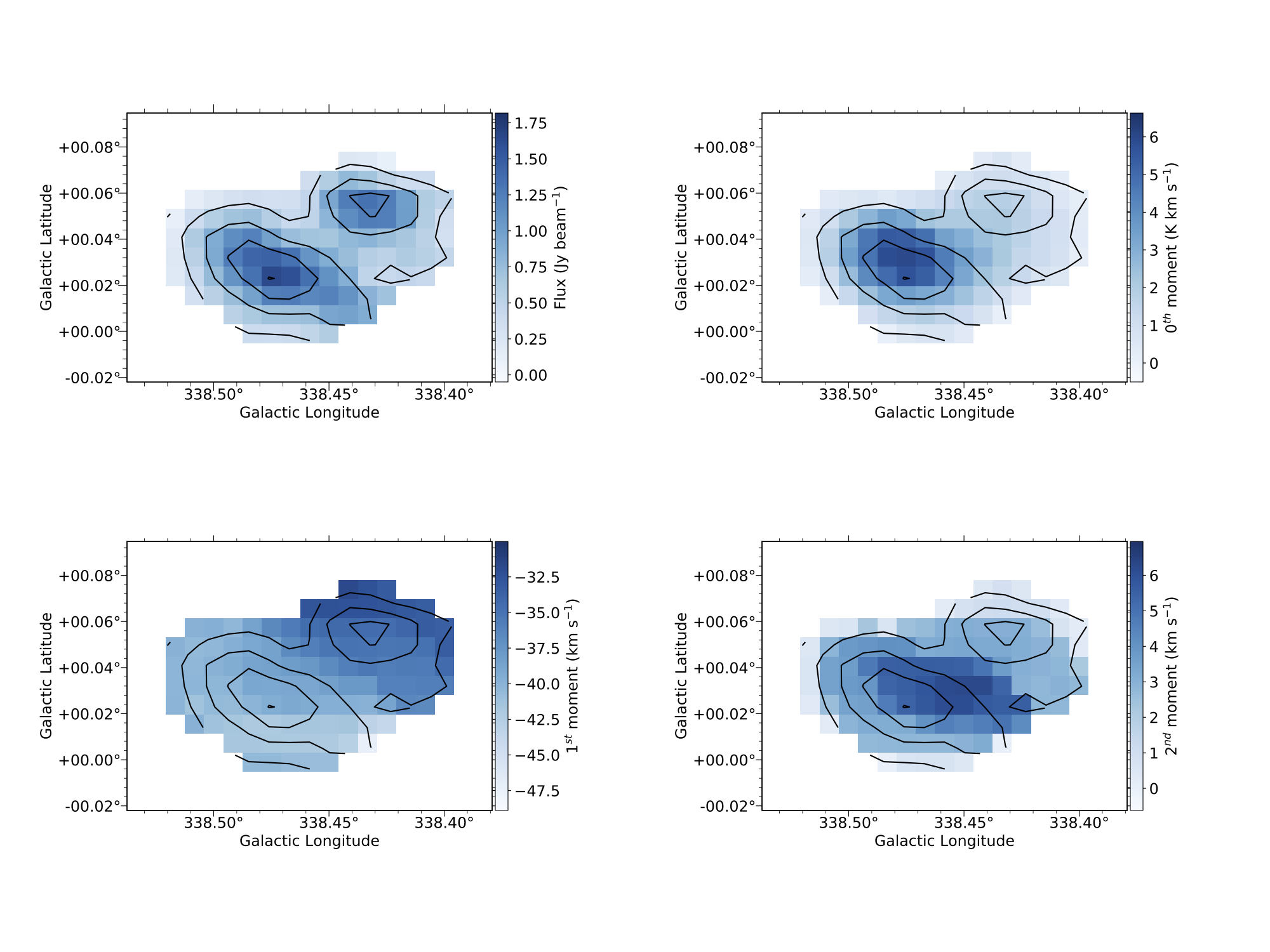} \\
\caption{$\nhone$ peak intensity [top left], integrated intensity (zeroth moment) [top right], velocity-weighted intensity (1st moment) [bottom left], and intensity-weighted velocity dispersion (2nd moment) [bottom right] maps of HOPS YMC progenitor gas cloud candidate, G338.464$+$0.034 \citep{purcell2012}. The contours in all panels show the peak intensity distribution in steps of 20\% of the maximum intensity in the map.}
\label{fig:ymc_candidate_8}
\end{figure*}

\begin{figure*}
\includegraphics[width=0.4\textwidth, angle=0, trim= 0 0 0 0, clip]{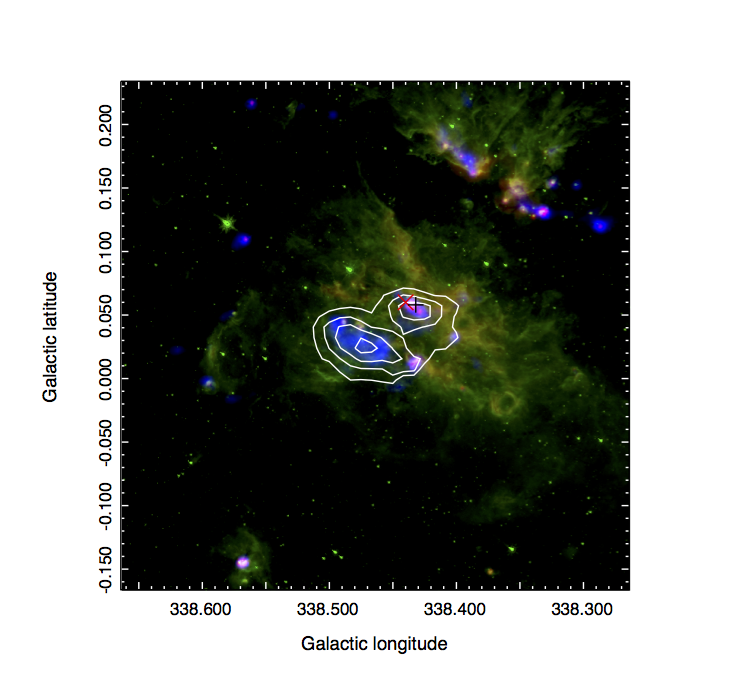} \\
\caption{Three-colour images of young massive cluster progenitor gas cloud candidate G338.464$+$0.034 [Red, HIGAL 70\,$\mu$m \citep{molinari_2010}; green, 8\,$\mu$m GLIMPSE \citep{churchwell09}; blue, ATLASGAL 850\,$\mu$ emission \citep{schuller09}]. Crosses are methanol masers \citep{caswell10,caswell11} and plus symbols are HOPS water masers \citep{walsh2011, walsh14}. White contours show the $\nhone$ integrated intensity emission.}
\label{fig:ymc_3col_G338}
\end{figure*}

\begin{figure*}
\includegraphics[width=1.0\textwidth, angle=0, trim= 0 0 0 0, clip]{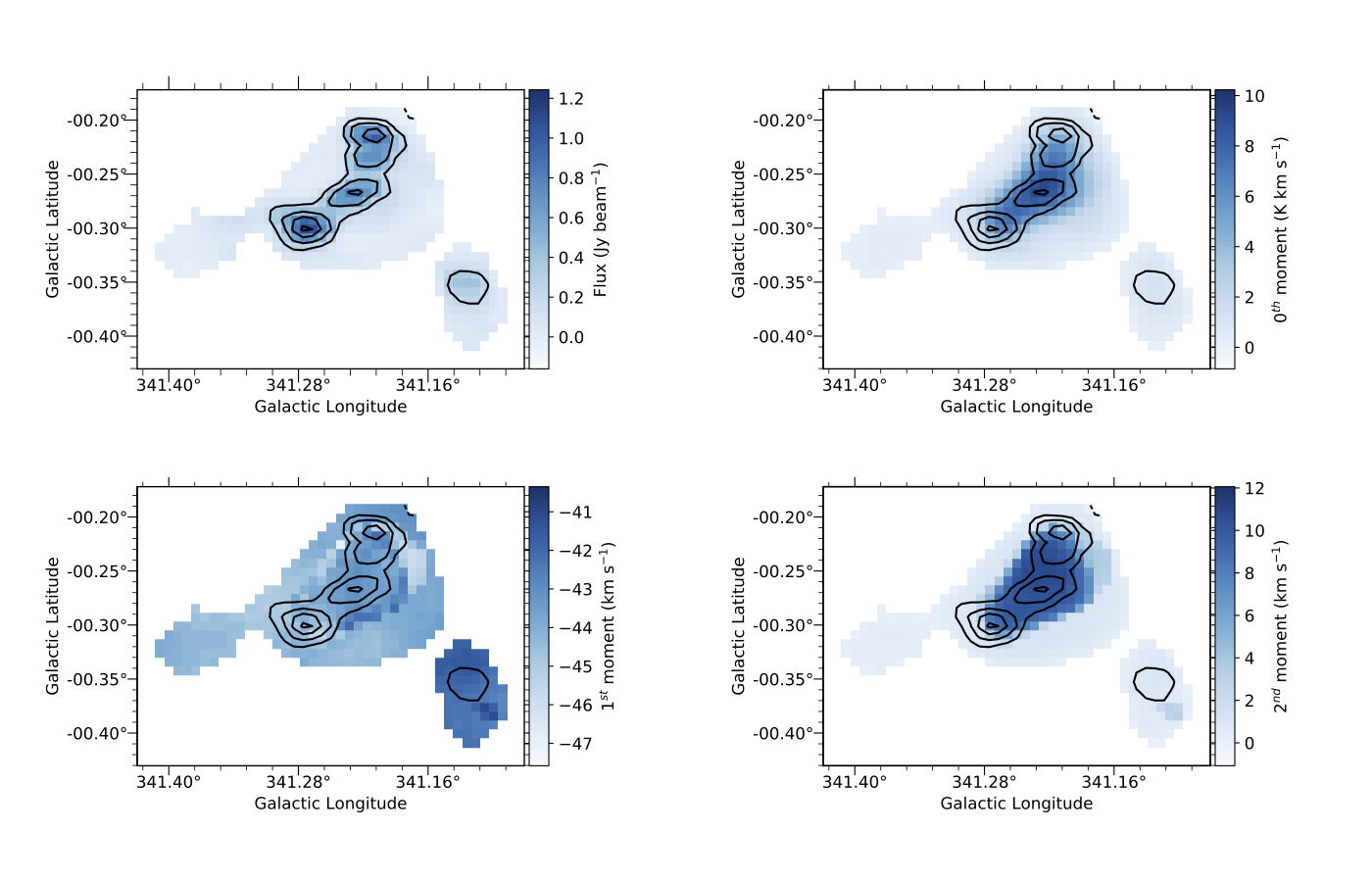} \\
\caption{$\nhone$ peak intensity [top left], integrated intensity (zeroth moment) [top right], velocity-weighted intensity (1st moment) [bottom left], and intensity-weighted velocity dispersion (2nd moment) [bottom right] maps of HOPS YMC progenitor gas cloud candidate, G341.224$-$0.274 \citep{purcell2012}. The contours in all panels show the peak intensity distribution in steps of 20\% of the maximum intensity in the map.}
\label{fig:ymc_candidate_9}
\end{figure*}

\begin{figure*}
\includegraphics[width=0.4\textwidth, angle=0, trim= 0 0 0 0, clip]{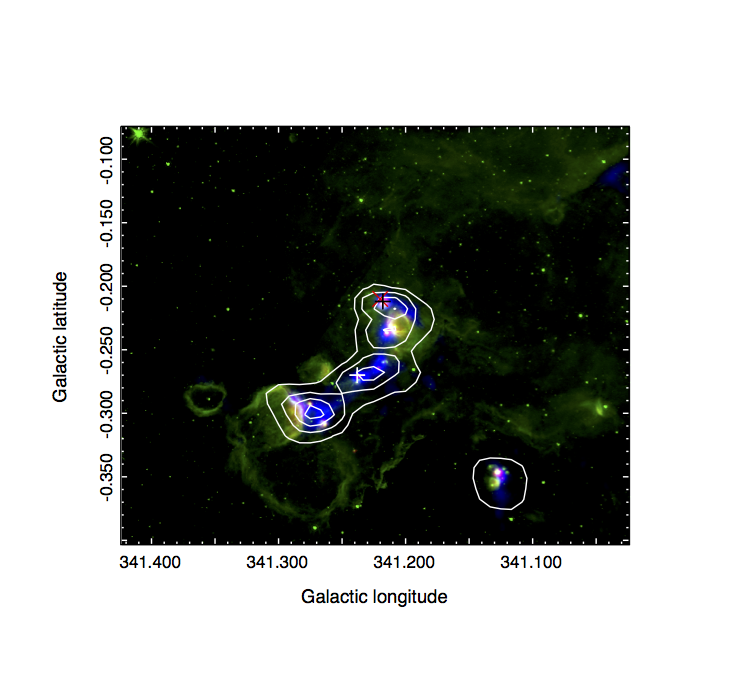} \\
\caption{Three-colour images of young massive cluster progenitor gas cloud candidate G341.224$-$0.274  [Red, HIGAL 70\,$\mu$m \citep{molinari_2010}; green, 8\,$\mu$m GLIMPSE \citep{churchwell09}; blue, ATLASGAL 850\,$\mu$ emission \citep{schuller09}]. Crosses are methanol masers \citep{caswell10,caswell11} and plus symbols are HOPS water masers \citep{walsh2011, walsh14}. White contours show the $\nhone$ integrated intensity emission.}
\label{fig:ymc_3col_G341}
\end{figure*}

\begin{figure*}
\includegraphics[width=1.0\textwidth, angle=0, trim= 0 0 0 0, clip]{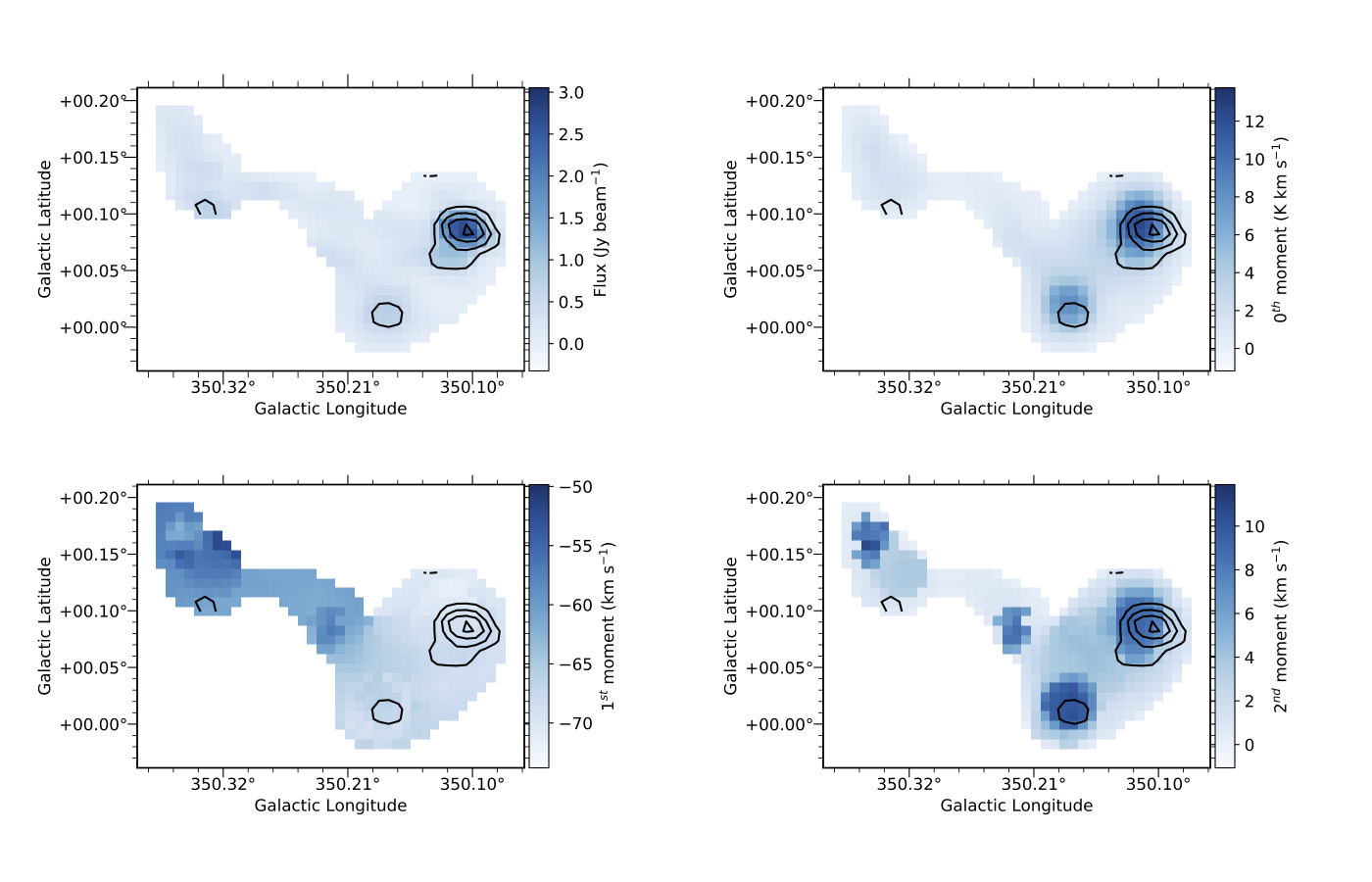} \\
\caption{$\nhone$ peak intensity [top left], integrated intensity (zeroth moment) [top right], velocity-weighted intensity (1st moment) [bottom left], and intensity-weighted velocity dispersion (2nd moment) [bottom right] maps of HOPS YMC progenitor gas cloud candidate, G350.170$+$0.070 \citep{purcell2012}. The contours in all panels show the peak intensity distribution in steps of 20\% of the maximum intensity in the map.}
\label{fig:ymc_candidate_10}
\end{figure*}

\begin{figure*}
\includegraphics[width=0.4\textwidth, angle=0, trim= 0 0 0 0, clip]{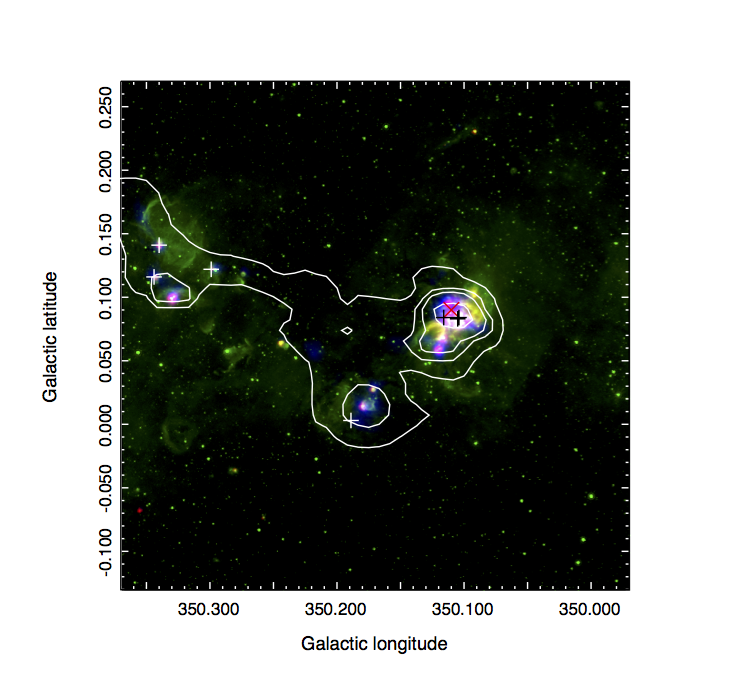} \\
\caption{Three-colour images of young massive cluster progenitor gas cloud candidate G350.170$+$0.070 [Red, HIGAL 70\,$\mu$m \citep{molinari_2010}; green, 8\,$\mu$m GLIMPSE \citep{churchwell09}; blue, ATLASGAL 850\,$\mu$ emission \citep{schuller09}]. Crosses are methanol masers \citep{caswell10,caswell11} and plus symbols are HOPS water masers \citep{walsh2011, walsh14}. White contours show the $\nhone$ integrated intensity emission.}
\label{fig:ymc_3col_G350}
\end{figure*}

\begin{figure*}
\includegraphics[width=1.0\textwidth, angle=0, trim= 0 0 0 0, clip]{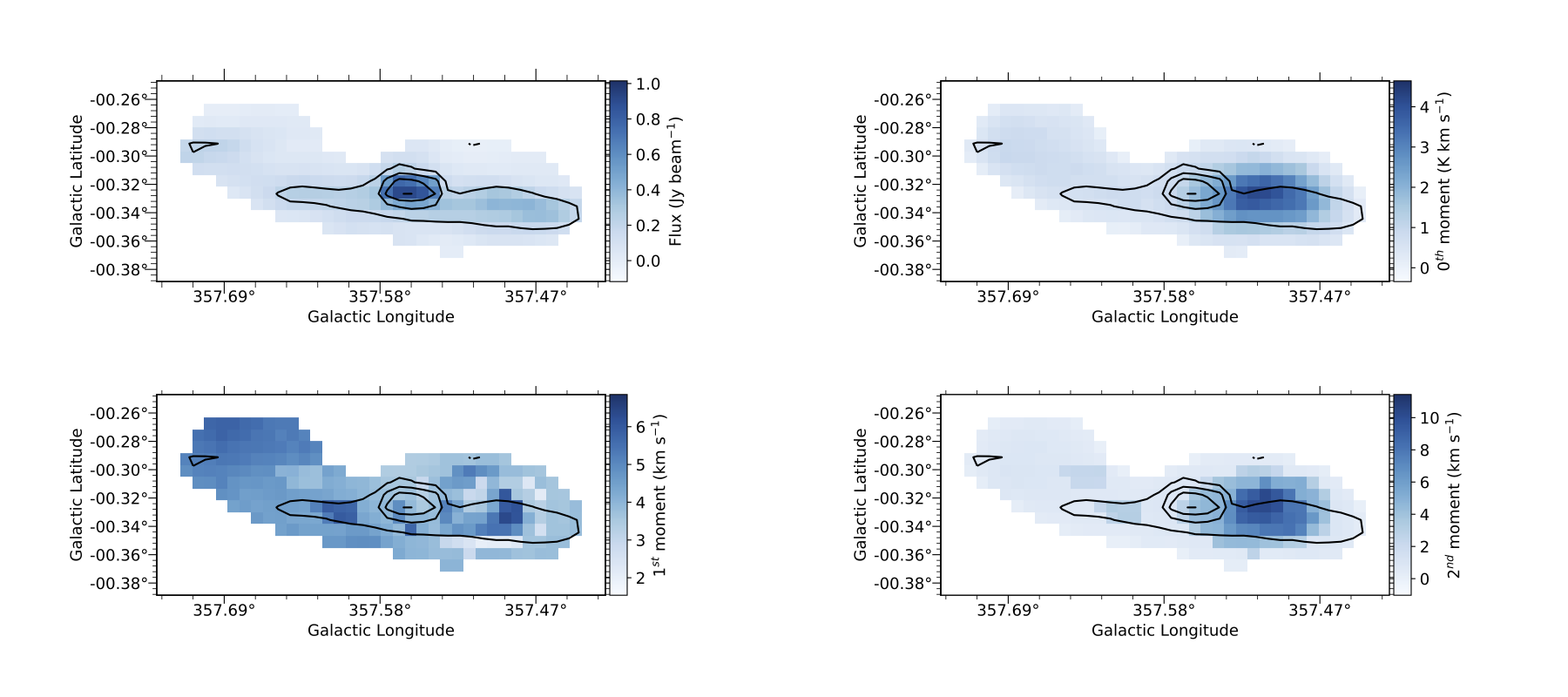} \\
\caption{$\nhone$ peak intensity [top left], integrated intensity (zeroth moment) [top right], velocity-weighted intensity (1st moment) [bottom left], and intensity-weighted velocity dispersion (2nd moment) [bottom right] maps of HOPS YMC progenitor gas cloud candidate, G357.555$-$0.323 \citep{purcell2012}. The contours in all panels show the peak intensity distribution in steps of 20\% of the maximum intensity in the map.}
\label{fig:ymc_candidate_11}
\end{figure*}

\begin{figure*}
\includegraphics[width=0.4\textwidth, angle=0, trim= 0 0 0 0, clip]{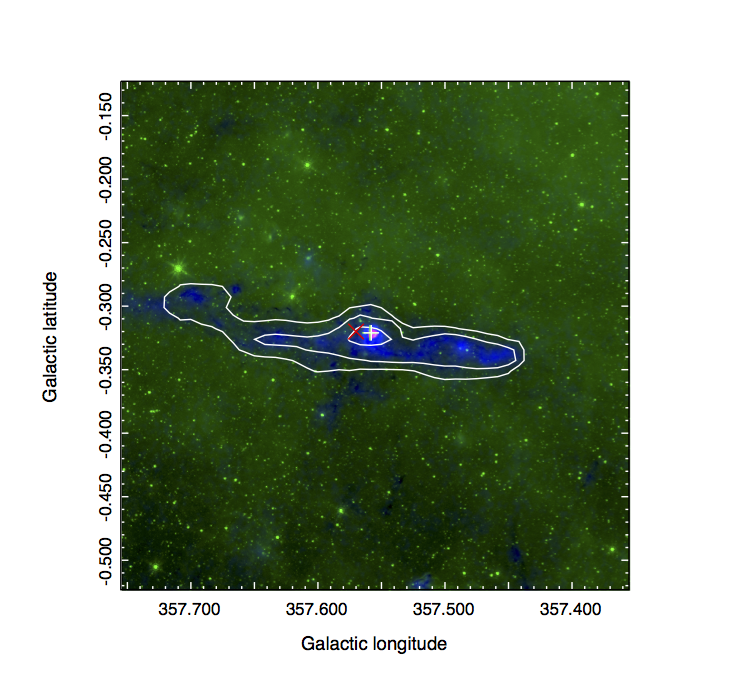} \\
\caption{Three-colour images of young massive cluster progenitor gas cloud candidate G357.555$-$0.323 [Red, HIGAL 70\,$\mu$m \citep{molinari_2010}; green, 8\,$\mu$m GLIMPSE \citep{churchwell09}; blue, ATLASGAL 850\,$\mu$ emission \citep{schuller09}]. Crosses are methanol masers \citep{caswell10,caswell11} and plus symbols are HOPS water masers \citep{walsh2011, walsh14}. White contours show the $\nhone$ integrated intensity emission.}
\label{fig:ymc_3col_G357}
\end{figure*}

Table~\ref{tab:YMC_precursor_candidate_properties} summarises the properties of the remaining 7 sources,  Figures~\ref{fig:ymc_candidate_5}, \ref{fig:ymc_candidate_6}, \ref{fig:ymc_candidate_7}, \ref{fig:ymc_candidate_8}, \ref{fig:ymc_candidate_9}, \ref{fig:ymc_candidate_10} and \ref{fig:ymc_candidate_11} show their dust continuum emission and the $\nhone$ integrated intensity, velocity field and the velocity dispersion. Figures~\ref{fig:ymc_3col_G003}, \ref{fig:ymc_3col_G316}, \ref{fig:ymc_3col_G330}, \ref{fig:ymc_3col_G338}, \ref{fig:ymc_3col_G341}, \ref{fig:ymc_3col_G350} and \ref{fig:ymc_3col_G357} shows the mid-IR images for the 7 sources. Below, we investigate each in turn and weigh up the likelihood that each may be a YMC progenitor gas cloud.

Although G003.432$-$0.351 lies close in projection to the Galactic Centre, the much narrower velocity dispersion is consistent with a source outside the Galactic Centre. The large mass for this source comes primarily because the adopted distance is very large, 21.1\,kpc, putting it at the outer reaches on the far side of the Galaxy. Inspection of mid-IR images (Figure~\ref{fig:ymc_3col_G003}) shows the $\nhone$ and dust emission line up very well with IR-dark gas clouds, suggesting the source lies on this side of the Galactic Centre (d$<$8.5\,pc) in front of the extended Galactic mid-IR emission. We note that the probability that the source lies at the far distance is rather low (0.59) and conclude that the distance is factors of several overestimated. The likely mass is therefore much less than 10$^4$\,M$_\odot$, and this source is not a YMC progenitor cloud.

The story for G316.752$+$0.004 seems to be similar. From the distance likelihood estimation, it is only marginally more probable that the source lies at the far distance of 9.8\,kpc, rather than the near distance of 2.6\,kpc. The mid-IR images also show absorption features with a similar morphology as the dust and $\nhthree$ emission, which would suggest this source is at the near distance. Indeed, previous studies have concluded this source is at the near distance \citep[see][and references therein]{L07A}. In which case, the mass in Table~\ref{tab:YMC_precursor_candidate_properties} is overestimated by a factor $\sim$13. We conclude this source is unlikely to be a YMC progenitor cloud.

The dust continuum and $\nhone$ maps of G330.881$-$0.371 show two peaks offset by $\sim$5$-$10\,pc. These are clearly at two very different velocities. The brighter component is at V$_{\rm LSR} = -63$\,$\kms$, and the weaker at $-41$\,$\kms$. We conclude the gas at the different velocities is not physically associated. We focus on the brightest component as this is the one which dominates the dust and $\nhthree$ emission, and for which the distance has been calculated. Masking the weaker component, the mass drops, by $\sim$10\% to $\sim10^4$\,M$_\odot$. Although the mid-IR images show absorption with similar morphology as the dust emission, the nebulosity may be associated with the region, weakening the argument that this absorption places the source at the near distance. The brighter gas and dust component is associated with an embedded 70\,$\mu$m source and maser emission, showing that star formation is underway, although at an early evolutionary stage. Whether or not this source is a YMC progenitor candidate depends primarily on correctly resolving the distance ambiguity. But even if it does lie at the far distance, more detailed calculations of the mass (e.g. including the actual distribution of dust temperature) are required to see if the mass reservoir is large enough for the cloud to produce a YMC. 
	
The dust continuum and $\nhone$ maps of G338.464$+$0.034 show two, distinct peaks of roughly equal intensity, offset by $\sim$5$-$10\,pc.  The mid-IR images show significant nebulosity, although it is not clear if this is physically associated with the dense gas. The northern gas component has an embedded 70\,$\mu$m source and maser emission showing that star formation is underway. The probability that this source lies at the far distance is high (0.74), and there is no evidence of mid-IR absorption features with similar morphology to the $\nhone$ and dust emission. At the peak of the $\nhone$ integrated intensity and dust emission the two gas clumps are offset in velocity by $\sim$6\,$\kms$. Although there are clear signs of a velocity gradient along the major axis of the two clumps, suggesting that they may be physically associated, the angular resolution of the HOPS data is too coarse to see whether this velocity gradient is unambiguously contiguous from one clump to the other. As such, we cannot rule out that these may be physically unassociated dense clumps which are serendipitously projected along our line of sight. However, if they are physically associated, it is intriguing that the velocity gradient is sufficient to bring the two clumps together in $\sim$1\,Myr -- the maximum age spread observed for YMCs. This is an interesting candidate to follow up as a potential YMC progenitor candidate.

Both the dust and $\nhthree$ emission for G341.224$-$0.274 show multiple components spread over a projected radius of $>$20\,pc. The probability that this source is at the far distance of 0.79 is the most robust for all the sources. Figure~\ref{fig:ymc_3col_G341} shows that this is an area of intense star formation activity. The $\nhone$ data cubes show there is almost no change in velocity across the main ridge of source which contains most of the mass. Unless there are convergent gas motions of $>$10\,$\kms$ purely in the plane of the sky, which seems unlikely, there is no way the gas in this source can condense to a radius of 0.1\,pc within 1\,Myr. We therefore conclude this is unlikely to be a YMC progenitor.

The dust continuum and $\nhone$ maps of G350.170$+$0.070 show two, distinct peaks of roughly equal intensity, offset by $\sim$5$-$10\,pc, and surrounding lower intensity, filamentary emission, all linked contiguously in the $\nhone$ velocity cube. Figure~\ref{fig:ymc_3col_G350} shows that this is an area of intense star formation activity across the complex. There is a $\sim$5\,$\kms$ velocity gradient across the two brightest components, but there is no way to tell if this is due to convergent motions or not. Further analysis of the physical and kinematic substructure of this source are needed to determine whether it has the potential to form a YMC.

For G357.555$-$0.323, both the dust and $\nhone$ integrated intensity emission are dominated by a single component. However, although the dust and dense gas have a similar, elongated morphology, the peaks are offset from each other by a projected distance of a few pc. The $\nhone$ data cubes show there is almost no change in velocity across the main ridge of the source which contains most of the mass. Similar to G341.224$-$0.274, unless there are convergent gas motions of $>$10\,$\kms$ purely in the plane of the sky, which seems unlikely, there is no way the gas in this source can condense to a radius of 0.1\,pc within 1\,Myr. We therefore conclude this is unlikely to be a YMC progenitor.

In summary, our search to find YMC progenitor candidate clouds using the HOPS data has proved inconclusive. The main limiting factor in determining whether or not most of the sources are genuine YMC progenitor candidates is the uncertainty in the distance. Although several sources may lie at the far kinematic distance, and therefore have sufficient mass to form a YMC, more detailed investigation is required to prove this. The most promising candidates are G330.881$-$0.371, G338.464$+$0.034 and G350.170$+$0.070. If these do lie at the far kinematic distance, and all the gas is physically associated, the $\nhone$ velocity structure suggests the magnitude of the gas motion is sufficient to bring the mass to a radius of $<$1\,pc in under 1\,Myr. However, the major caveat for this to occur is that these gas flows must be convergent. Further analysis of the kinematics of the dense gas and surrounding lower density envelopes from existing survey data \citep[e.g.][]{jackson13, beuther15, beuther16, jordan15, rathborne16, rigby16, bihr16}, and fragmentation at higher resolution \citep[e.g.][]{L06, beuther09, L11, henshaw16_frag, henshaw17} will form the basis of a subsequent paper.

Regardless of whether these turn out to be YMC progenitor clouds or not, one thing is clear -- all the potential candidates we found are already forming stars. Despite relaxing our search criteria from those in \citet{bressert12} to find younger sources, and despite HOPS having sufficient sensitivity and resolution to find quiescent progenitor clouds, we did not find one. We conclude that the starless phase for YMC progenitor gas clouds in the disk of the Galaxy is either non existent, or so short it is effectively not observable given the small number of YMCs expected to be forming at any time in the Milky Way.  
	
%------

%------

%------

%------

%------

%------

%------

%==============================================================================
\section{Galactic Centre gas kinematic properties}
\label{sec:galactic_centre}

\begin{figure*}
\includegraphics[width=1.05\textwidth, angle=0, trim= 0 0 0 0, clip]{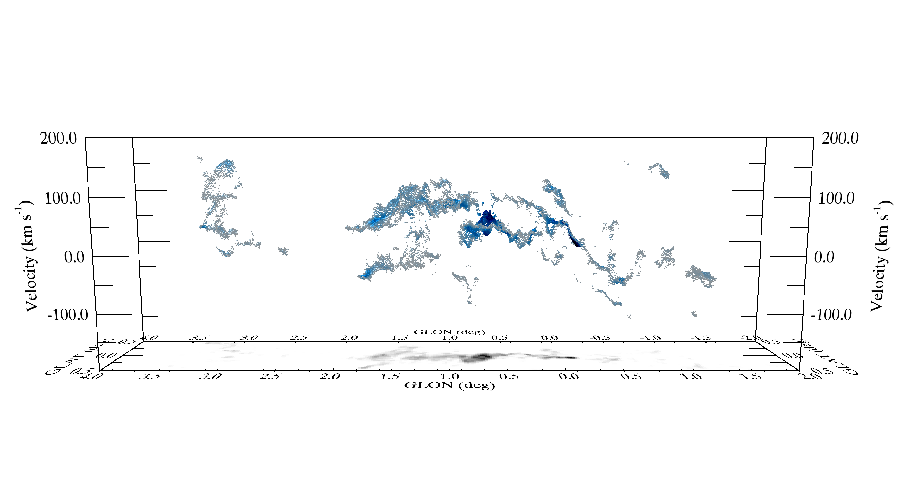} \\
\includegraphics[width=1.05\textwidth, angle=0, trim= 0 0 0 0, clip]{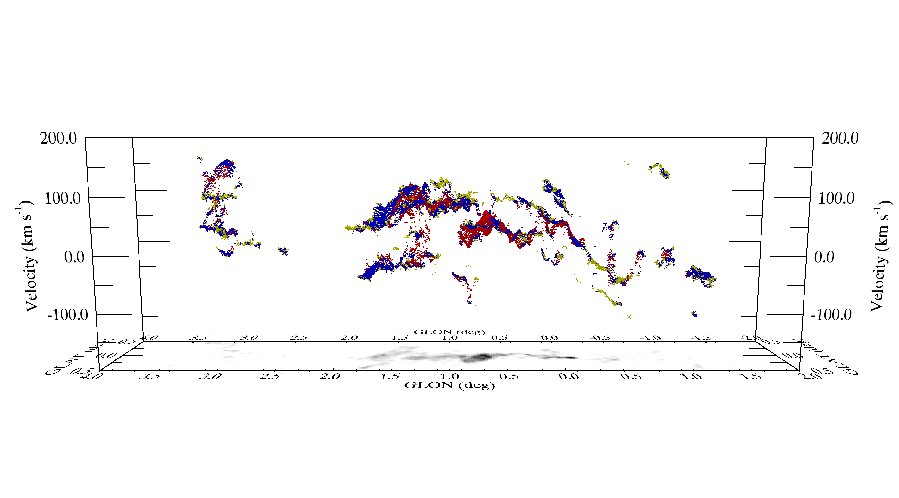} \\
\caption{Position-position-velocity (PPV) map of the HOPS $\nhthree$(1,1) {\sc scouse} \citep{henshaw16} fit results for gas detected in the inner 500~pc of the Galaxy (see text for details). The grey scale on the bottom surface of each panel is the integrated intensity of the $\nhthree$(1,1) emission. [Top] The colour and size of the symbols correspond to the peak brightness temperature of the $\nhthree$(1,1) emission at that PPV location, ranging on a linear scale from small, grey dots at the lowest peak brightness temperature (0.1 K) to large, blue dots at the highest peak brightness temperature (5.8 K). [Bottom] The size of the symbols is equivalent to those in the top panel. The colour of each pixel however, now corresponds to the velocity dispersion of the emission separated into three ranges: \textbf{Yellow}: $3.2\,{\rm km\,s^{-1}}<\sigma<14.5\,{\rm km\,s^{-1}}$; \textbf{Blue}: $14.5\,{\rm km\,s^{-1}}<\sigma<19.5\,{\rm km\,s^{-1}}$; \textbf{Red}: $19.5\,{\rm km\,s^{-1}}<\sigma<47.8\,{\rm km\,s^{-1}}$. }
\label{fig:gc_nh3_11}
\end{figure*}

\begin{figure*}
\includegraphics[width=1.05\textwidth, angle=0, trim= 0 0 0 0, clip]{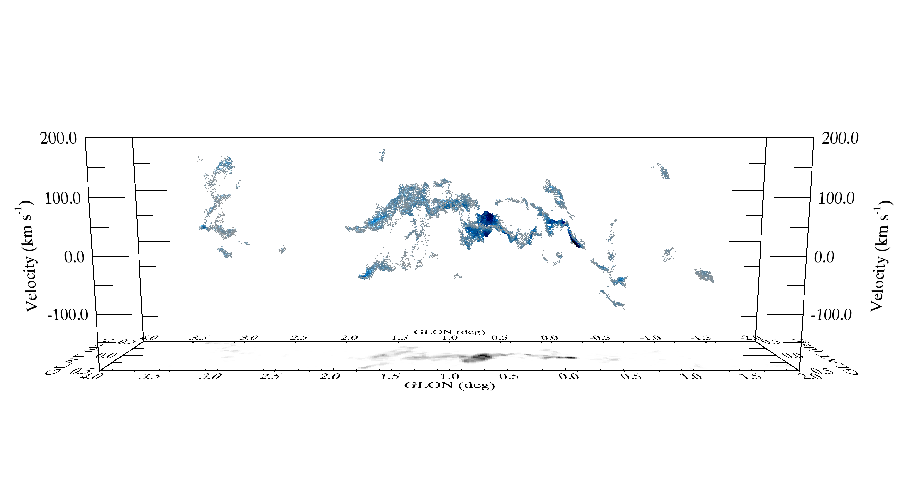} \\
\includegraphics[width=1.05\textwidth, angle=0, trim= 0 0 0 0, clip]{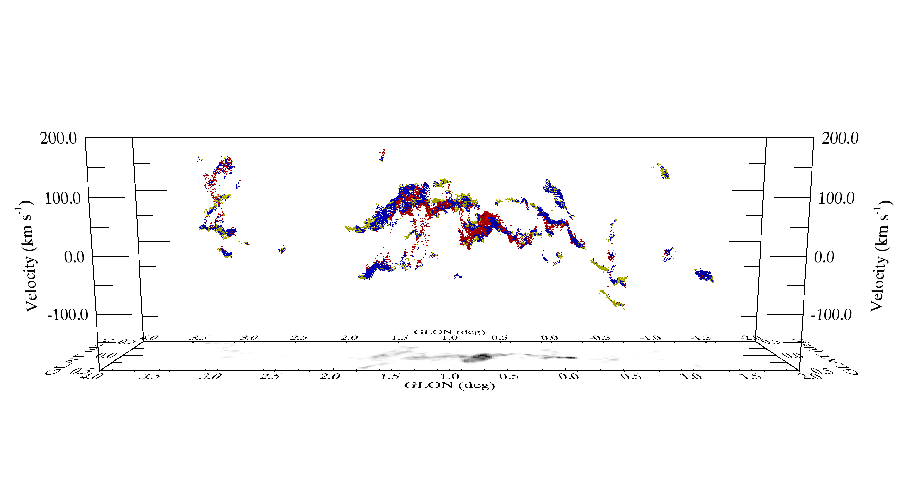} \\
\caption{Equivalent to Figure~\ref{fig:gc_nh3_11} but for the HOPS $\nhthree$(2,2) {\sc scouse} fit results. The colour and size of each pixel in the top panel scales linearly from small, grey dots at lowest peak brightness temperature (0.1 K) to large, blue dots at the highest peak brightness temperature (4.0 K). The colour of each pixel in the bottom panel corresponds to the following three ranges: \textbf{Yellow}: $2.2\,{\rm km\,s^{-1}}<\sigma<12.7\,{\rm km\,s^{-1}}$; \textbf{Blue}: $12.7\,{\rm km\,s^{-1}}<\sigma<19.9\,{\rm km\,s^{-1}}$; \textbf{Red}: $19.9\,{\rm km\,s^{-1}}<\sigma<53.9\,{\rm km\,s^{-1}}$. }
\label{fig:gc_nh3_22}
\end{figure*}

\begin{figure*}
\includegraphics[width=1.05\textwidth, angle=0, trim= 0 0 0 0, clip]{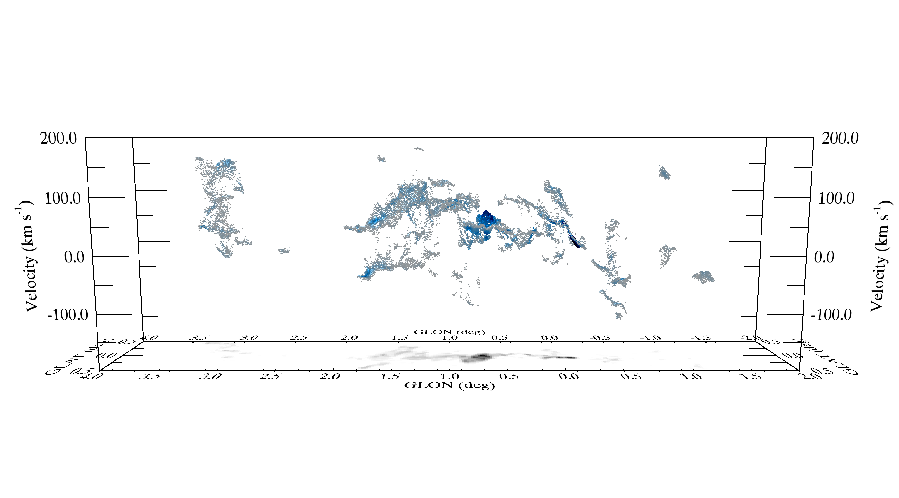} \\
\includegraphics[width=1.05\textwidth, angle=0, trim= 0 0 0 0, clip]{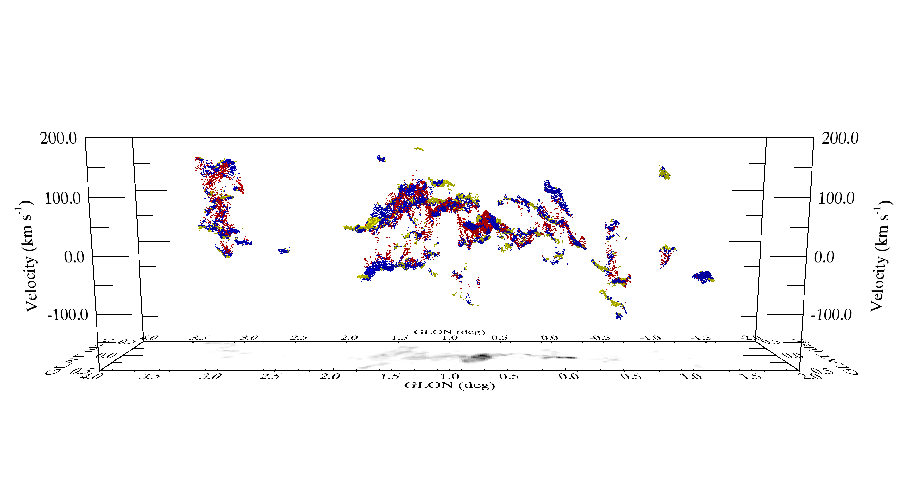} \\
\caption{Equivalent to Figure~\ref{fig:gc_nh3_11} but for the HOPS $\nhthree$(3,3) {\sc scouse} fit results. The colour and size of each pixel in the top panel scales linearly from small, grey dots at lowest peak brightness temperature (0.1 K) to large, blue dots at the highest peak brightness temperature (6.8 K). The colour of each pixel in the bottom panel corresponds to the following three ranges: \textbf{Yellow}: $2.1\,{\rm km\,s^{-1}}<\sigma<10.2\,{\rm km\,s^{-1}}$; \textbf{Blue}: $10.2\,{\rm km\,s^{-1}}<\sigma<19.0\,{\rm km\,s^{-1}}$; \textbf{Red}: $19.0\,{\rm km\,s^{-1}}<\sigma<62.6\,{\rm km\,s^{-1}}$. }
\label{fig:gc_nh3_33}
\end{figure*}

\begin{figure*}
\includegraphics[width=1.05\textwidth, angle=0, trim= 0 0 0 0, clip]{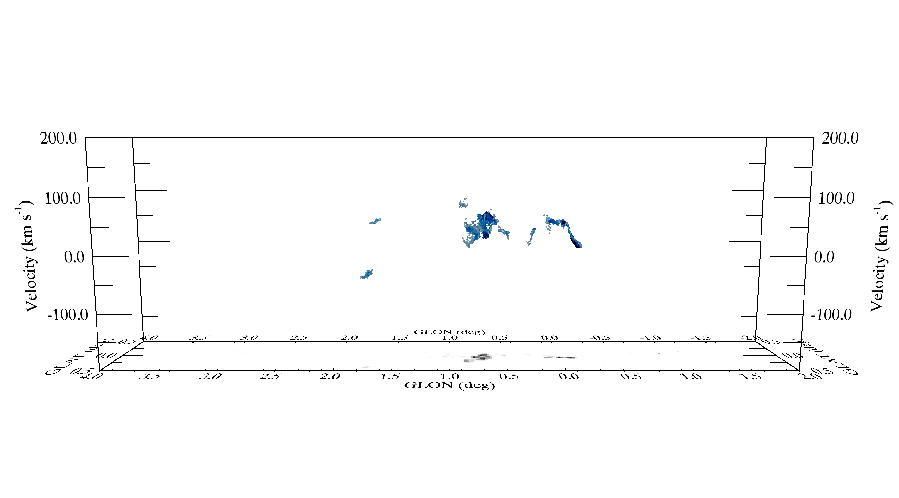} \\
\includegraphics[width=1.05\textwidth, angle=0, trim= 0 0 0 0, clip]{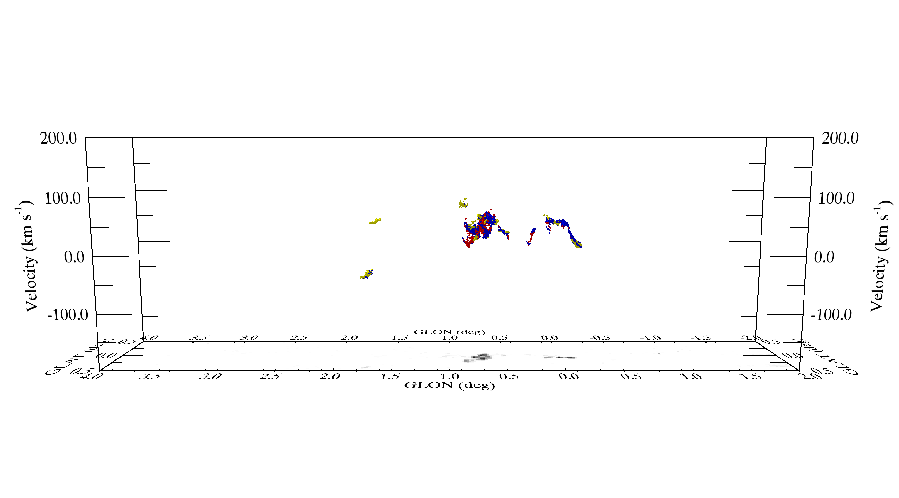} \\
\caption{Equivalent to Figure~\ref{fig:gc_nh3_11} but for the HOPS $\nhthree$(6,6) {\sc scouse} fit results. The colour and size of each pixel in the top panel scales linearly from small, grey dots at lowest peak brightness temperature (0.1 K) to large, blue dots at the highest peak brightness temperature (0.8 K). The colour of each pixel in the bottom panel corresponds to the following three ranges: \textbf{Yellow}: $3.7\,{\rm km\,s^{-1}}<\sigma<9.6\,{\rm km\,s^{-1}}$; \textbf{Blue}: $9.6\,{\rm km\,s^{-1}}<\sigma<17.7\,{\rm km\,s^{-1}}$; \textbf{Red}: $17.7\,{\rm km\,s^{-1}}<\sigma<32.5\,{\rm km\,s^{-1}}$. 
}
\label{fig:gc_nh3_66}
\end{figure*}

Figures~\ref{fig:gc_nh3_11},~\ref{fig:gc_nh3_22},~\ref{fig:gc_nh3_33} and \ref{fig:gc_nh3_66} show the position-position-velocity (PPV) diagrams of the {\sc scouse} $\nhthree$ fit results for gas detected in the inner 500\,pc of the Galaxy (see $\S$~\ref{sub:CMZ_fitting} for details). In the various sub plots, the size and colour of the symbols show the variation in either peak brightness temperature or velocity dispersion. The grey-scale on the bottom surface of each panel shows the integrated intensity emission for that transition.

The detected PPV structure, variation in peak brightness temperature and variation in velocity dispersion of the independently-fit $\nhone$, (2,2) and (3,3) emission are very similar, giving further confidence in the robustness of the {\sc scouse} fit results. The integrated intensity of the emission from these transitions corresponds well to the high columsn density HiGAL gas distribution \citep[][Battersby et al., in prep]{molinari11}. 

In contrast, the $\nhthree$(6,6) emission is detected over a much smaller spatial extent than the other $\nhthree$ transitions, being almost exclusively confined to the inner 150\,pc ($|l|<1^\circ$). The brightest $\nhthree$(6,6) emission is associated with the well-known star formation regions Sgr B2 ($l\sim0.65^\circ$, $b\sim0^\circ$, V$_{\rm LSR}\sim65\,\kms$), the ``20$\kms$" and ``50$\kms$'' clouds ($l\sim0^\circ$, $b\sim0^\circ$, V$_{\rm LSR}\sim20-50\,\kms$). Slightly weaker $\nhthree$(6,6) emission is detected towards a few locations along the ``dust ridge" ($l\sim0.1-0.5^\circ$, $b\sim0-0.2^\circ$, V$_{\rm LSR}\sim30\,\kms$). The only $\nhthree$(6,6) emission detected outside the inner 100\,pc is at the high longitude end of both velocity components of the ``1.3 degree" cloud complex. 

As we view the Galactic Centre through the disk of the Galaxy, there is a large column of gas along the line sight which is unassociated with this region. This can lead to (sometimes severe) contamination when trying to isolate emission from gas in the Galactic Centre \citep[for example, see Figure 4 in][]{bally88} making it difficult to interpret the kinematics, especially at velocities close to V$_{\rm LSR} \sim 0\,\kms$. The advantage of using transitions of $\nhthree$ with a higher critical density than CO is immediately obvious in Figures~\ref{fig:gc_nh3_11},~\ref{fig:gc_nh3_22},~\ref{fig:gc_nh3_33} and \ref{fig:gc_nh3_66}. The $\nhthree$ emission cleanly picks out the high density Galactic Centre gas and none of the intervening, lower density, CO-bright molecular clouds in the disk.

A direct comparison between the $\nhthree$ PPV diagrams in Figures~\ref{fig:gc_nh3_11},~\ref{fig:gc_nh3_22},~\ref{fig:gc_nh3_33} and \ref{fig:gc_nh3_66} and other observations of high critical density molecular line transitions over the same region is difficult, because previous papers have tended to use moment maps and channel maps to investigate the kinematics \citep[e.g.][]{jones12, ott_14}. However, the general trends in $\nhthree$ velocity structure appear qualitatively similar to those of other dense gas tracers in the literature.

\begin{figure*}
\includegraphics[width=1.05\textwidth, angle=0, trim= 0 0 0 0, clip]{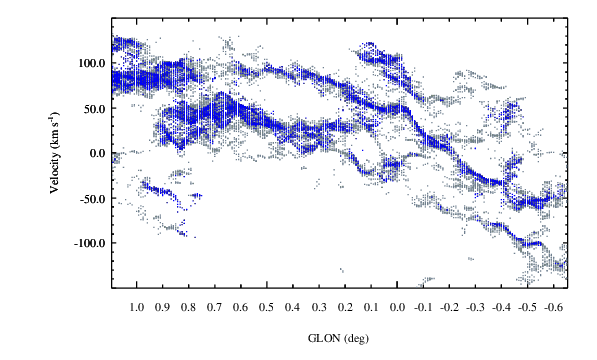} \\
\caption{Comparison of the \citet{henshaw16} fits to the \citet{jones12} Mopra 3mm emission (grey) and the HOPS $\nhone$ emission (blue). In the overlapping spatial coverage range between the two datasets (longitudes of $-0.7^\circ \lesssim l \lesssim1^\circ$) the $\nhthree$ PPV structure closely matches that of the HNCO, N$_2$H$^+$ and HNC, despite the factor $\sim$2 lower angular resolution. This gives confidence that the {\sc scouse} fits to these high density tracers are robustly tracing the global gas kinematics.
}
\label{fig:gc_hops_jones_comp}
\end{figure*}

Figure~\ref{fig:gc_hops_jones_comp} shows a comparison of the \citet{henshaw16} fits to the \citet{jones12} Mopra 3mm HNCO emission in grey and the HOPS $\nhone$ emission in blue. In the overlapping spatial coverage range (between longitudes of $-0.7^\circ \lesssim l \lesssim1^\circ$) the $\nhthree$ PPV structure closely matches that of the HNCO, despite the factor $\sim$2 lower angular resolution. This gives further confidence that the {\sc scouse} fits to these high density tracers are robustly tracing the global gas kinematics.

\begin{figure*}
\includegraphics[width=1.05\textwidth, angle=0]{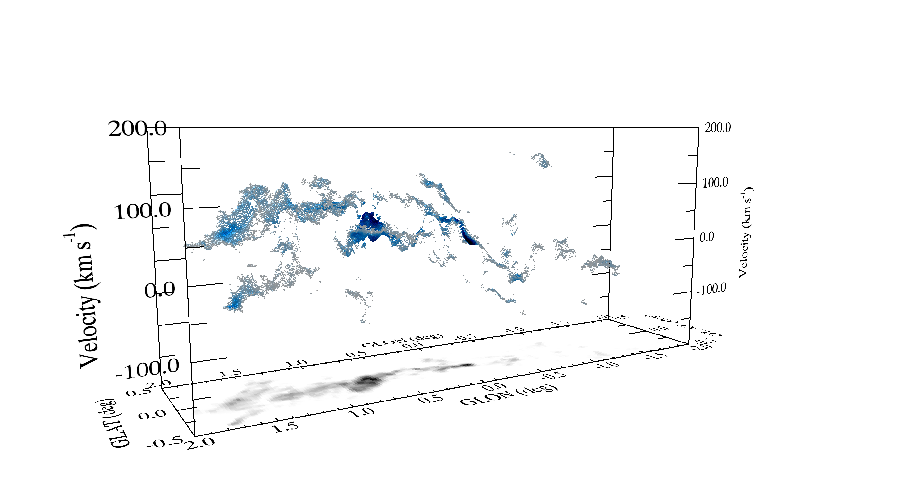} 
\caption{Equivalent to the top panel of Figure~\ref{fig:gc_nh3_11}, but zoomed into $-2\fdg0<l<2\fdg0$ and rotated to accentuate oscillatory patterns observed in the gas kinematic structure.}
\label{fig:gc_nh3_wiggles}
\end{figure*}

The general trends in velocity structure -- such as the two streams with large velocity gradients in the inner 100\,pc and the large velocity dispersion towards Sgr B2 --  are similar to those reported by other authors \citep[see][and references therein]{KDL15, henshaw16}. One important advantage of fitting independent profiles to every velocity component in the data cube is to minimise blending in velocity space prevalent in moment maps or when creating PV diagrams by collapsing the data cube over one of the two position axes (e.g. summing along latitude to create a longitude-velocity diagram). By fitting (multiple) Gaussian components to every pixel and viewing in the full 3D PPV space, it is possible to pick out finer kinematic details than in moment maps and 2D PV diagrams. \citet{henshaw16} reported seeing regular oscillatory patterns, or ``wiggles" in the two coherent gas streams in the inner 100\,pc. As shown in Figure~\ref{fig:gc_nh3_wiggles}, the same oscillatory pattern is seen in the $\nhthree$ data. The detection of these oscillations with an independent data set, at a different angular resolution and using a different molecular species to trace the gas, confirms these are robust kinematic features. We discuss the potential origin of these features as gravitational instabilities in the gas streams in \citet{henshaw16_wiggles}.

Figure~\ref{fig:gc_nh3_wiggles} shows oscillatory patterns in the kinematic structure of both velocity components of the gas in the 1.3 degree cloud. Although the oscillation length is similar to those in the inner 100\,pc, the gas in these clouds does not appear to be self gravitating \citep{kruijssen14}. Therefore, the origin of these features is unlikely to be related to gravitational instabilities. We are currently investigating what mechanisms may give rise to such features.

\begin{figure*}
\includegraphics[width=1.05\textwidth, angle=0]{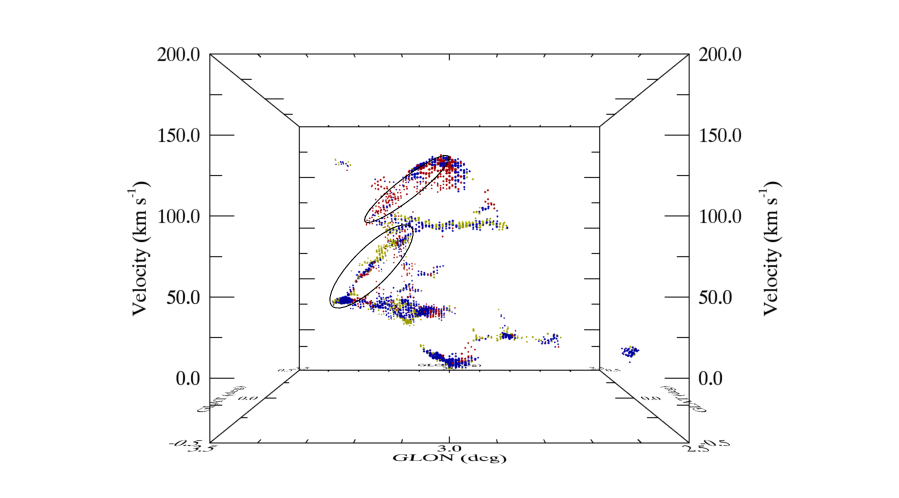} \\
\includegraphics[width=1.05\textwidth, angle=0]{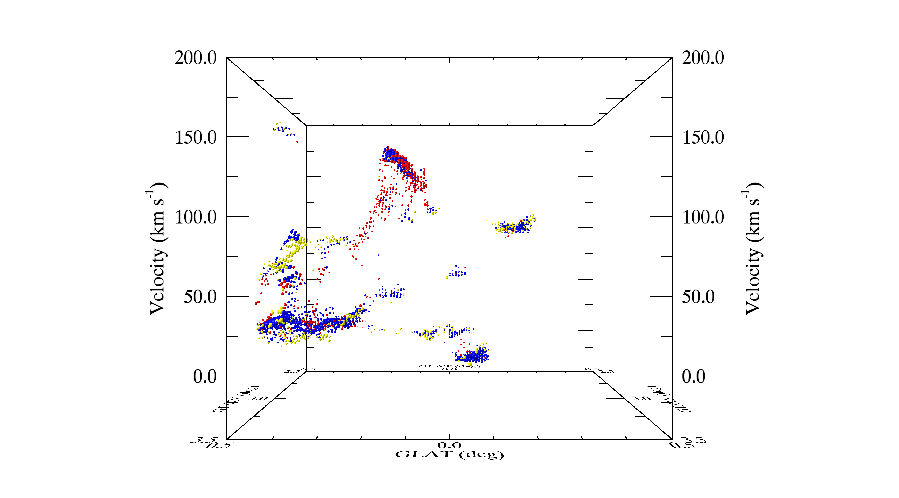}
\caption{Position-position-velocity (PPV) diagram of the HOPS $\nhthree$(1,1) fit results for gas detected towards the region with $l>3\fdg0$, known as Clump 2. The size and colour of each pixel is equivalent to that presented in the bottom panel of Figure~\ref{fig:gc_nh3_11}. The top panel and bottom panels depict the $\{l,\,v_{\rm LSR}\}$ and $\{b,\,v_{\rm LSR}\}$ projections, respectively. The black ellipses in the top panel highlight the kinematic components referred to in the text with steep velocity gradients. }
\label{fig:gc_nh3_bc2}
\end{figure*}

The only region which does not show oscillatory motions in the kinematic structure is the gas at $l>3^\circ$, known as Clump 2. Interestingly, this is also the region with the most complicated velocity structure. Figure~\ref{fig:gc_nh3_bc2} shows the PPV structure of this region from two different perspectives. There are clearly several coherent kinematic components, but without knowing their (relative) line of sight positions, it is not possible to infer if they are physically connected, or simply projected and overlapping in PPV space. Most of the coherent velocity components have a constant velocity along their full extent. The exception to this is are the two features highlighted by ellipses in the top panel of Figure~\ref{fig:gc_nh3_bc2}. These have very steep velocity gradients, both rising by $\sim$50\,$\kms$ over a projected length of 30\,pc (assuming a distance of 8\,kpc). Understanding the origin of the kinematic complexity of Clump 2 remains a challenge.

We now turn to the measured velocity dispersion across the region. As discussed in $\S$~\ref{sub:CMZ_fitting}, the lower $\nhthree$(J,K) transitions can be affected by opacity broadening. However, as shown in Figures~\ref{fig:gc_nh3_11},~\ref{fig:gc_nh3_22},~\ref{fig:gc_nh3_33} and \ref{fig:gc_nh3_66}, the trends in velocity dispersion are the same for all the $\nhthree$ transitions. While the absolute velocity dispersion of the $\nhone$ and (2,2) may be overestimated, the relative velocity dispersion of gas over the region should be robust against opacity-broadening.

The highest velocity dispersions are observed towards the Sgr~B2 and 20 \& 50\,$\kms$ clouds. These clouds contain some of the most active embedded star formation activity in the Galactic Centre region \citep{lu15,lu17}. It is therefore unclear if these clouds had intrinsically large velocity dispersions before forming stars, or whether the affect of stellar feedback may have been responsible for increasing their velocity dispersions from their pre-star-formation state. 

The remaining regions all contain PPV components with a large spread in the measured velocity dispersion. However, there are some broad trends in the relative numbers of low and high velocity dispersion components as a function of longitude. Excluding the gas associated with known star forming regions, the gas in the inner 100\,pc has a noticeably higher concentration of smaller velocity dispersion components than the gas outside of the inner 100\,pc. Conversely, the largest concentration of high velocity dispersion components (again excluding the gas associated with known star forming regions) is in the centre of the 1.3 degree cloud. The gas at $l\sim3^\circ$ is dominated by components with an intermediate velocity dispersion. We note that this result is very different than the velocity dispersion that would be inferred for the $l\sim3^\circ$ cloud from moment analysis. The multiple, overlapping velocity components, including some with very large projected velocity gradients, would lead to moment analysis overestimating the intrinsic velocity dispersion of this gas by factors of several.

Finally, we note that this new level of kinematic precision may help understand which mechanism(s) are responsible for driving gas from the disk of the Milky Way towards the Galactic Centre. Several gas transport mechanisms have been proposed in the literature. In those which can be loosely described as ``direct-infall", gas in the disk feels a torque from the bar, loses angular momentum and falls directly towards the Galactic Centre along the leading edge of the bar, and $x_1$ and $x_2$ orbits play a major role in the gas transport \citep[e.g.][]{sofue95, morris_serabyn96, bally10}. In other scenarios, gas within the bar's inner Linblad resonance ($\sim$1\,kpc for the Milky Way) is driven inwards by angular momentum transport induced by acoustic instabilities \citep{montenegro_yuan_elmegreen_1999, kruijssen14, kk15, kkc17}. Approaching the galactic centre, the galaxy's rotation curve turns from flat to solid body and the inward transport slows down, leading to a build up of gas. This is predicted to occur at Galactocentric radii approaching $\sim100\,$pc in the Milky Way \citep{kk15}. These different transport mechanisms may not be mutually exclusive and may both operate in the same galaxy with varying influences at different galactocentric radii \citep[][]{kruijssen17}. 

With the detailed kinematic information above, it may be possible to test predictions of different gas transport mechanisms. For example, we note that the observed trend of increasing velocity dispersion towards a (projected) Galactocentric radius of $\sim$150\,pc is qualitatively similar to that predicted by the acoustic instability model of \citet{kk15} for the Milky Way. Further detailed comparisons with predictions from these and other models/simulations \citep[][Ridley et al. (sub)]{emsellem15} which can also be tuned to match conditions in the centre of the Millky Way, will help explain the origin of the observed kinematic features, and ultimately what determines the long-term mass flows and energy cycles in the Galactic Centre.

%------------------

%------------------

%------------------

%------------------

%------------------

%------------------

%------------------

%==========================================================
\section{Conclusions}
\label{sec:conclusions}

We have described the automated spectral-line fitting pipelines developed to determine the properties of spectral line emission for HOPS data cubes. We then used this to determine the characteristic molecular gas properties in the HOPS sample from the $\nhthree$ emission. 

We find that HOPS $\nhthree$ sources in the disk of the Milky Way associated with methanol and/or water maser emission are warmer and have larger line widths than $\nhthree$ sources with no maser emission. We conclude the HOPS $\nhthree$-only sources in the disk represent an interesting sample of gas clouds at the earliest evolutionary phases of high-mass star formation.

Our search to find YMC progenitor candidate clouds using the HOPS data has proved inconclusive. The main limiting factor in determining whether or not most of the sources are genuine YMC progenitor candidates is the uncertainty in the distance. The most promising candidates are G330.881$-$0.371, G338.464$+$0.034 and G350.170$+$0.070. If these do lie at the far kinematic distance, and all the gas is physically associated, the $\nhone$ velocity structure suggests the magnitude of the gas motion is sufficient to bring the mass to a radius of $<$1\,pc in under 1\,Myr. However, the major caveat for this to occur is that these gas flows must be convergent. Further analysis of these clouds will form the basis of a subsequent paper.

Regardless of whether these turn out to be YMC progenitor clouds or not, one thing is clear -- all the potential candidates we found are already forming stars. Despite relaxing our search criteria from those in \citet{bressert12} to find younger sources, and despite HOPS having sufficient sensitivity and resolution to find quiescent progenitor clouds, we did not find one. We conclude that the starless phase for YMC progenitor gas clouds in the disk of the Galaxy is either non existent, or so short it is effectively not observable given the small number of YMCs expected to be forming at any time in the Milky Way.

We used the {\sc scouse} algorithm of \citet{henshaw16} to fit the complex $\nhthree$ spectra in the Galactic Centre. We confirm the findings of \citet{henshaw16_wiggles} that there are regular velocity oscillations in the inner 100\,pc gas streams. We find there are also similar oscillatory patterns in the kinematic structure of gas in the 1.3 degree cloud. Although the oscillation length is similar to those in the inner 100 pc, the gas in these clouds does not appear to be self gravitating, so the origin of these features is unlikely to be related to gravitational instabilities. We are currently investigating what mechanisms may give rise to such features.

The only region which does not show oscillatory motions in the kinematic structure is the gas at $l\sim3^\circ$, known as Clump 2 -- the region with the most complicated velocity structure. Most of the coherent velocity components have a constant velocity along their full extent. However, two kinematic features have very steep velocity gradients, both rising by $\sim$50\,$\kms$ over a projected length of 30\,pc (assuming a distance of 8\,kpc). Understanding the origin of the kinematic complexity of Clump 2 remains a challenge. The velocity dispersion and latitude extent of the dense gas as a function of longitude in the Galactic Centre qualitatively matches the predictions of \citet{kk15}. 

Finally, we provide the output catalogues, fit results, maps of the derived properties and all the remaining unpublished data as a resource for future studies through the HOPS website (http://www.hops.org.au) and online appendices.

%----------------------------------------------------------------------
\section{Acknowledgements}
DJB was supported by NASA contract NAS8-03060.

%================================================================
\bibliography{hops_p3}

%================================================================
\section{Appendix 1: Additional HOPS data}
\label{appendix1}

\begin{figure*}
\begin{center}
\includegraphics[width=0.96\textwidth, angle=0]{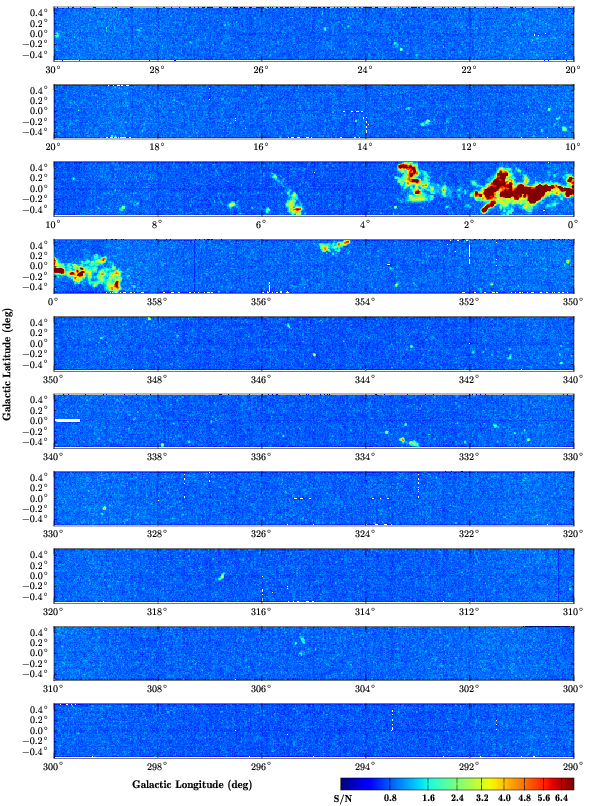} 
\caption{Signal-to-noise map of the HOPS $\nhthree$(3,3) emission across the 100 square degrees of the survey. The colour bar in the bottom-right corner provides the conversion of colour in the map to S/N (see Paper~II for details of the S/N calculation).}
\label{fig:nhthree_33_sn_map}
\end{center}
\end{figure*}

\begin{figure*}
\begin{center}
\includegraphics[width=0.96\textwidth, angle=0]{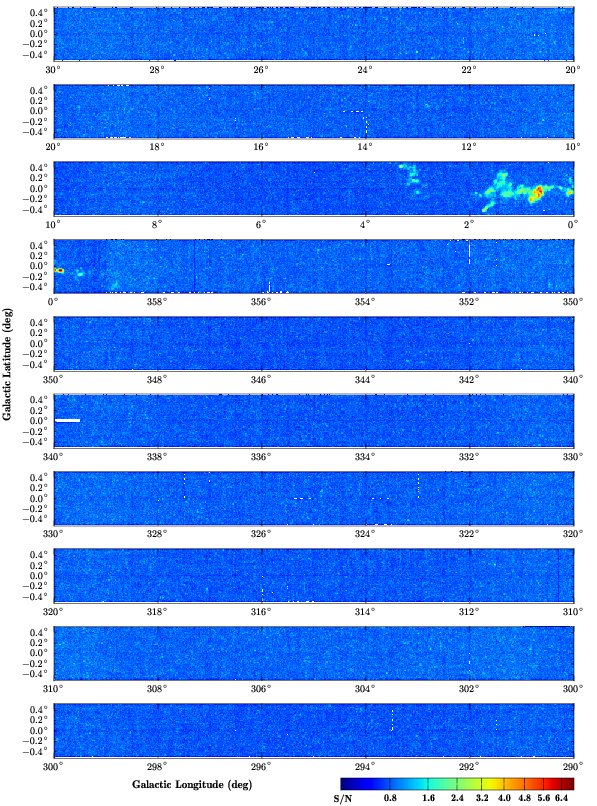} 
\caption{Signal-to-noise map of the HOPS $\nhthree$(6,6)  emission across the 100 square degrees of the survey. The colour bar in the bottom-right corner provides the conversion of colour in the map to S/N (see Paper~II for details of the S/N calculation).}
\label{fig:nhthree_66_sn_map}
\end{center}
\end{figure*}

\begin{figure*}
\begin{center}
\includegraphics[width=0.96\textwidth, angle=0]{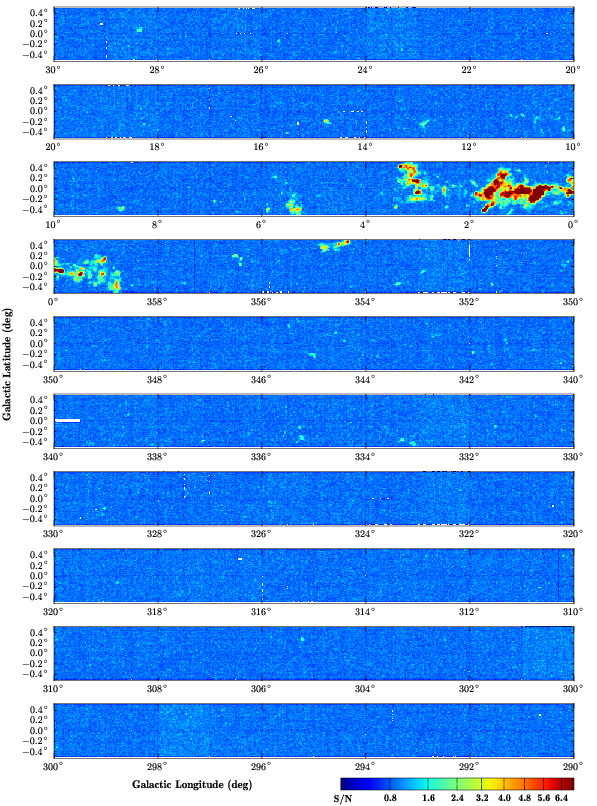} 
\caption{Signal-to-noise map of the HOPS HC$_3$N emission across the 100 square degrees of the survey. The colour bar in the bottom-right corner provides the conversion of colour in the map to S/N (see Paper~II for details of the S/N calculation).}
\label{fig:nhthree_HC3N_sn_map}
\end{center}
\end{figure*}

\begin{figure*}
\begin{center}
\includegraphics[width=0.96\textwidth, angle=0]{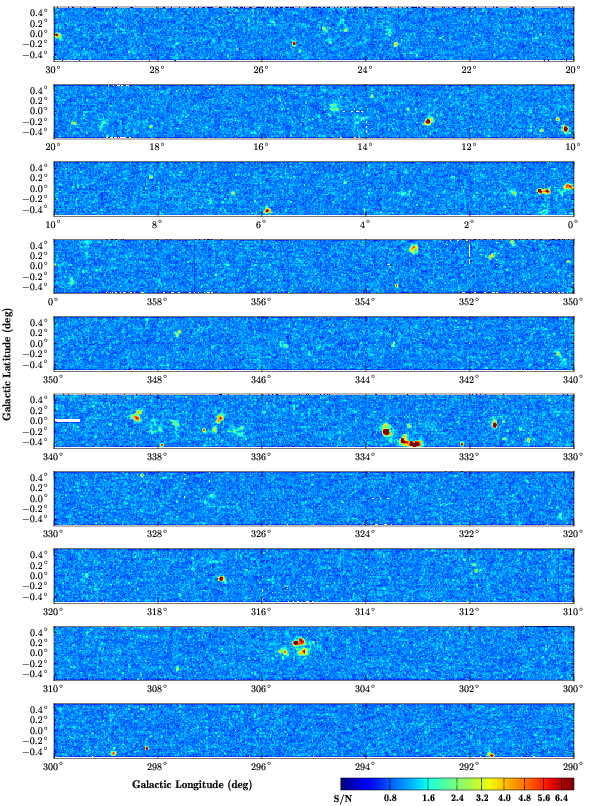} 
\caption{Signal-to-noise (S/N) map of the HOPS H69$\alpha$ emission across the 100 square degrees of the survey. The colour bar in the bottom-right corner provides the conversion of colour in the map to S/N (see Paper~II for details of the S/N calculation).}
\label{fig:nhthree_H69a_sn_map}
\end{center}
\end{figure*}

Here we make available the remaining  HOPS data cubes. In Paper II we described the procedure we used to measured the signal to noise of the \nhone\ and (2,2) data cubes and provided maps of the peak signal to noise ratio. In Figures~\ref{fig:nhthree_33_sn_map}, \ref{fig:nhthree_66_sn_map}, \ref{fig:nhthree_HC3N_sn_map} and \ref{fig:nhthree_H69a_sn_map} we show similar maps illustrating the spatial extent of the $\nhthree$(3,3),  $\nhthree$(6,6), HC$_3$N and H69$\alpha$ emission\footnote{Note the H69$\alpha$ map is the same data as published in Figure~1 of \citet{longmore13}.} detected across the 100 square degrees of the survey. 

$\nhthree$(3,3), $\nhthree$(6,6) and HC$_3$N emission traces reservoirs of dense molecular gas. The maps of these transitions (Figures~\ref{fig:nhthree_33_sn_map}, \ref{fig:nhthree_66_sn_map}, \ref{fig:nhthree_HC3N_sn_map}) are dominated by emission within $|l|<4^\circ$, corresponding to gas in the inner few hundred pc of the Galaxy (the Central Molecular Zone, CMZ). As noted by \citet{longmore13}, the H69$\alpha$ emission, which predominantly traces young HII regions, is much more uniformly distributed. This intriguing lack of correspondence between the large dense gas reservoir and star formation activity tracers in the CMZ, and the implications of this for our understanding of star formation in other environments, is currently under active investigation \citep{longmore13, kruijssen14, kk15, kkc17}.

While the $\nhthree$(6,6) emission is only seen towards the CMZ, the $\nhthree$(3,3) and HC$_3$N emission is detected at many other locations throughout the inner Galaxy, sometime coincident with H69$\alpha$ emission. Further inspection shows these locations are sites of well-known, bright star formation regions.

In addition to the H$_2$O masers \citep[presented in][]{walsh2008, walsh2011, walsh14}, the $\nhone$ and $\nhthree$(2,2) data (presented in Paper~II), and the $\nhthree$(3,3),  $\nhthree$(6,6), HC$_3$N and H69$\alpha$ data presented above, HOPS also detected emission from several CH$_3$OH transitions, CS, $\nhthree$(2,1) and the H65$\alpha$ and H62$\alpha$ radio recombination lines. However, emission from the CH$_3$OH, CS and $\nhthree$(2,1) transitions is only detected towards a small number of regions -- Sgr B2, the $l=2^\circ$ cloud complex, and the G305 star formation complex. The other recombination lines show a similar spatial extent as the H69$\alpha$ emission, but suffer from interference and other imaging artefacts.

We have made all this data, and the fits results described above, available to the community through the HOPS website -- http://www.hops.org.au.

%------------------------------------------------------------
\clearpage
\newpage

%================================================================
\section{Appendix 2: Online Tables}
\label{appendix2}

\begin{landscape} 
\footnotesize
{\bf Table 2:} Output from the automated spectral line fitting procedure described in $\S$~\ref{sec:spectral-line_fitting}. Columns 1 and 2 give the Galactic longitude and latitude of the region taken from the source name in Paper~II. Columns 3 to 11 list the best-fit $\nhone$ and $\nhtwo$ parameters at the peak pixel towards each region. For each transition, T$_{\rm B}$ is the peak brightness temperature of the main component, V$_{\rm LSR}$ is the line-centre velocity and $\Delta V$ is the linewidth.  Columns 6 and 7 show the opacity and excitation temperature of the $\nhone$ transition. The subscript `m' denotes the main hyperfine component. Column 8 shows the $\nhtwo$ integrated intensity. A dash denotes no reliable fit was obtained for that parameter. 
\begin{center} 
\begin{tabular}{|c|c|c|c|c|c|c|c|c|c|c|} 
\hline \hline 
 l             & b     & T$_{\rm B,m}^{11}$ & V$_{\rm LSR}^{11}$ & $\Delta$V$^{11}$  & $\tau_{m}^{11}$ & T$_{\rm ex}^{11}$ & $\int$ T$_{\rm B}^{22}$dV & V$_{\rm LSR}^{22}$ & $\Delta$V$^{22}$  & T$_{\rm B}^{22}$ \\ 
 ($^\circ$) & ($^\circ$) & (K) & (kms$^{-1}$) & (kms$^{-1}$) & & (K) & (K.kms$^{-1}$) & (kms$^{-1}$) & (kms$^{-1}$ &  (K)  \\ \hline 
0.223 & -0.473 & 2.87 $\pm$ 0.39 & 16.90 $\pm$ 0.07 & 1.96 $\pm$ 0.15 & 1.65 $\pm$ 0.49 & 4.46 & 0.65 $\pm$ 0.20 & 16.71 $\pm$ 0.17 & 1.08 $\pm$ 0.40 & 0.57 \\  
0.238 & -0.189 & 0.74 $\pm$ 0.61 & 76.60 $\pm$ 1.55 & 6.75 $\pm$ 3.78 & 3.37 $\pm$ 3.81 & 2.94 & 2.05 $\pm$ 0.45 & 79.13 $\pm$ 0.75 & 6.34 $\pm$ 1.39 & 0.30 \\  
0.507 & -0.321 & 0.31 $\pm$ 0.07 & -64.00 $\pm$ 0.85 & 6.69 $\pm$ 1.40 & 0.10 $\pm$ 2.74 & 5.83 & 2.47 $\pm$ 0.69 & -64.48 $\pm$ 3.40 & 24.64 $\pm$ 7.92 & 0.09 \\  
0.658 & -0.213 & 0.51 $\pm$ 0.37 & -50.00 $\pm$ 0.67 & 2.60 $\pm$ 1.28 & 1.27 $\pm$ 3.13 & 3.13 & 0.44 $\pm$ 0.16 & -59.87 $\pm$ 0.22 & 1.20 $\pm$ 0.47 & 0.35 \\  
0.860 & -0.067 & 11.30 $\pm$ 1.13 & 12.80 $\pm$ 0.26 & 15.00 $\pm$ 0.90 & 1.92 $\pm$ 0.25 & 8.61 & 126.40 $\pm$ 2.11 & 12.67 $\pm$ 0.25 & 31.51 $\pm$ 0.64 & 3.77 \\  
0.970 & 0.065 & 0.58 $\pm$ 0.61 & -40.60 $\pm$ 0.39 & 1.58 $\pm$ 0.62 & 0.63 $\pm$ 5.14 & 3.66 & 0.56 $\pm$ 0.16 & -23.85 $\pm$ 0.20 & 1.14 $\pm$ 0.34 & 0.46 \\
\hline 
\end{tabular} 
\end{center} 
\label{tmp} 
\end{landscape} 

\newpage 
 \clearpage 

\begin{landscape} 
\footnotesize
{\bf Table 2:} Continued. 
\begin{center} 
\begin{tabular}{|c|c|c|c|c|c|c|c|c|c|c|} 
\hline \hline 
 l             & b     & T$_{\rm B,m}^{11}$ & V$_{\rm LSR}^{11}$ & $\Delta$V$^{11}$  & $\tau_{m}^{11}$ & T$_{\rm ex}^{11}$ & $\int$ T$_{\rm B}^{22}$dV & V$_{\rm LSR}^{22}$ & $\Delta$V$^{22}$  & T$_{\rm B}^{22}$ \\ 
 ($^\circ$) & ($^\circ$) & (K) & (kms$^{-1}$) & (kms$^{-1}$) & & (K) & (K.kms$^{-1}$) & (kms$^{-1}$) & (kms$^{-1}$ &  (K)  \\ \hline 
3.340 & 0.396 & 0.41 $\pm$ 0.27 & 163.00 $\pm$ 1.99 & 14.00 $\pm$ 5.61 & 1.71 $\pm$ 1.55 & 2.96 & 4.48 $\pm$ 0.78 & 163.97 $\pm$ 2.66 & 29.15 $\pm$ 6.34 & 0.14 \\  
3.432 & -0.351 & 3.88 $\pm$ 0.45 & -25.50 $\pm$ 0.09 & 3.08 $\pm$ 0.19 & 2.75 $\pm$ 0.53 & 4.14 & 2.40 $\pm$ 0.35 & -24.71 $\pm$ 0.22 & 3.25 $\pm$ 0.71 & 0.69 \\  
3.448 & 0.008 & 2.25 $\pm$ 0.48 & 5.11 $\pm$ 0.06 & 1.26 $\pm$ 0.17 & 0.61 $\pm$ 0.58 & 6.42 & -2.16 $\pm$ 0.90 & 16.93 $\pm$ 0.67 & 21.46 $\pm$ 12.20 & -0.09 \\  
3.460 & -0.212 & 0.97 $\pm$ 0.24 & -23.50 $\pm$ 0.13 & 2.59 $\pm$ 0.39 & 0.31 $\pm$ 0.59 & 5.83 & -1.09 $\pm$ 0.34 & -43.29 $\pm$ 1.03 & 6.20 $\pm$ 1.87 & -0.17 \\  
3.680 & -0.131 & 1.24 $\pm$ 0.66 & 4.57 $\pm$ 0.14 & 0.84 $\pm$ 0.23 & 1.79 $\pm$ 2.33 & 3.42 & -0.37 $\pm$ 0.13 & 38.20 $\pm$ 0.14 & 0.73 $\pm$ 0.30 & -0.48 \\  
3.736 & -0.076 & 0.78 $\pm$ 0.22 & 4.12 $\pm$ 0.20 & 1.95 $\pm$ 0.72 & 0.10 $\pm$ 2.24 & 10.49 & 0.66 $\pm$ 0.25 & -29.98 $\pm$ 0.57 & 2.54 $\pm$ 0.84 & 0.25 \\
\hline 
\end{tabular} 
\end{center} 
\label{tmp} 
\end{landscape} 

\newpage 
 \clearpage 

\begin{landscape} 
\footnotesize
{\bf Table 2:} Continued. 
\begin{center} 
\begin{tabular}{|c|c|c|c|c|c|c|c|c|c|c|} 
\hline \hline 
 l             & b     & T$_{\rm B,m}^{11}$ & V$_{\rm LSR}^{11}$ & $\Delta$V$^{11}$  & $\tau_{m}^{11}$ & T$_{\rm ex}^{11}$ & $\int$ T$_{\rm B}^{22}$dV & V$_{\rm LSR}^{22}$ & $\Delta$V$^{22}$  & T$_{\rm B}^{22}$ \\ 
 ($^\circ$) & ($^\circ$) & (K) & (kms$^{-1}$) & (kms$^{-1}$) & & (K) & (K.kms$^{-1}$) & (kms$^{-1}$) & (kms$^{-1}$ &  (K)  \\ \hline 
8.349 & 0.497 & 0.22 $\pm$ 0.09 & -183.00 $\pm$ 0.66 & 3.24 $\pm$ 1.00 & 0.10 $\pm$ 2.65 & 4.96 & -3.60 $\pm$ 0.93 & -183.37 $\pm$ 2.70 & 23.05 $\pm$ 7.12 & -0.15 \\  
8.621 & -0.346 & 3.64 $\pm$ 0.36 & 38.70 $\pm$ 0.05 & 2.01 $\pm$ 0.12 & 1.77 $\pm$ 0.35 & 4.78 & 1.57 $\pm$ 0.23 & 38.53 $\pm$ 0.15 & 2.27 $\pm$ 0.43 & 0.65 \\  
8.642 & -0.143 & 0.30 $\pm$ 0.08 & 31.70 $\pm$ 1.15 & 5.54 $\pm$ 1.50 & 0.10 $\pm$ 0.66 & 5.71 & 1.75 $\pm$ 0.60 & 32.59 $\pm$ 1.91 & 11.15 $\pm$ 4.91 & 0.15 \\  
8.831 & -0.036 & 3.00 $\pm$ 0.49 & 0.24 $\pm$ 0.08 & 1.69 $\pm$ 0.16 & 2.82 $\pm$ 0.75 & 3.79 & 1.22 $\pm$ 0.20 & 0.49 $\pm$ 0.14 & 1.73 $\pm$ 0.37 & 0.67 \\  
9.159 & 0.048 & 1.19 $\pm$ 0.47 & 30.60 $\pm$ 0.10 & 1.31 $\pm$ 0.24 & 1.15 $\pm$ 1.28 & 3.76 & 0.26 $\pm$ 0.13 & 30.02 $\pm$ 0.31 & 1.14 $\pm$ 0.49 & 0.21 \\  
9.237 & -0.190 & 3.52 $\pm$ 0.54 & 42.00 $\pm$ 0.06 & 1.57 $\pm$ 0.16 & 2.18 $\pm$ 0.62 & 4.34 & 2.23 $\pm$ 0.34 & 41.77 $\pm$ 0.25 & 2.90 $\pm$ 0.55 & 0.72 \\
\hline 
\end{tabular} 
\end{center} 
\label{tmp} 
\end{landscape} 

\newpage 
 \clearpage 

\begin{landscape} 
\footnotesize
{\bf Table 2:} Continued. 
\begin{center} 
\begin{tabular}{|c|c|c|c|c|c|c|c|c|c|c|} 
\hline \hline 
 l             & b     & T$_{\rm B,m}^{11}$ & V$_{\rm LSR}^{11}$ & $\Delta$V$^{11}$  & $\tau_{m}^{11}$ & T$_{\rm ex}^{11}$ & $\int$ T$_{\rm B}^{22}$dV & V$_{\rm LSR}^{22}$ & $\Delta$V$^{22}$  & T$_{\rm B}^{22}$ \\ 
 ($^\circ$) & ($^\circ$) & (K) & (kms$^{-1}$) & (kms$^{-1}$) & & (K) & (K.kms$^{-1}$) & (kms$^{-1}$) & (kms$^{-1}$ &  (K)  \\ \hline 
13.672 & -0.121 & 0.52 $\pm$ 0.13 & 57.00 $\pm$ 0.12 & 0.71 $\pm$ 0.28 & 0.10 $\pm$ 1.12 & 7.96 & 0.53 $\pm$ 0.18 & 37.91 $\pm$ 0.28 & 1.59 $\pm$ 0.72 & 0.32 \\  
13.738 & -0.081 & 0.70 $\pm$ 0.28 & 49.70 $\pm$ 0.79 & 6.33 $\pm$ 1.25 & 3.27 $\pm$ 1.87 & 2.94 & 0.29 $\pm$ 0.11 & 38.71 $\pm$ 0.11 & 0.60 $\pm$ 0.26 & 0.45 \\  
13.861 & 0.186 & 1.79 $\pm$ 1.18 & 45.90 $\pm$ 0.27 & 1.16 $\pm$ 0.36 & 4.37 $\pm$ 4.50 & 3.13 & - & - & - & - \\  
13.914 & -0.287 & 1.02 $\pm$ 0.33 & 39.50 $\pm$ 0.18 & 1.12 $\pm$ 0.53 & 0.10 $\pm$ 0.44 & 12.93 & -1.16 $\pm$ 0.54 & 28.86 $\pm$ 1.15 & 5.50 $\pm$ 3.72 & -0.20 \\  
13.932 & -0.452 & 1.62 $\pm$ 0.34 & 22.80 $\pm$ 0.09 & 1.93 $\pm$ 0.23 & 0.63 $\pm$ 0.56 & 5.30 & 1.24 $\pm$ 0.30 & 22.93 $\pm$ 0.30 & 2.07 $\pm$ 0.66 & 0.56 \\  
14.010 & -0.144 & 1.53 $\pm$ 0.43 & 40.50 $\pm$ 0.10 & 1.93 $\pm$ 0.30 & 0.58 $\pm$ 0.72 & 5.36 & 1.71 $\pm$ 0.34 & 39.86 $\pm$ 0.43 & 4.15 $\pm$ 0.83 & 0.39 \\
\hline 
\end{tabular} 
\end{center} 
\label{tmp} 
\end{landscape} 

\newpage 
 \clearpage 

\begin{landscape} 
\footnotesize
{\bf Table 2:} Continued. 
\begin{center} 
\begin{tabular}{|c|c|c|c|c|c|c|c|c|c|c|} 
\hline \hline 
 l             & b     & T$_{\rm B,m}^{11}$ & V$_{\rm LSR}^{11}$ & $\Delta$V$^{11}$  & $\tau_{m}^{11}$ & T$_{\rm ex}^{11}$ & $\int$ T$_{\rm B}^{22}$dV & V$_{\rm LSR}^{22}$ & $\Delta$V$^{22}$  & T$_{\rm B}^{22}$ \\ 
 ($^\circ$) & ($^\circ$) & (K) & (kms$^{-1}$) & (kms$^{-1}$) & & (K) & (K.kms$^{-1}$) & (kms$^{-1}$) & (kms$^{-1}$ &  (K)  \\ \hline 
18.080 & 0.042 & 3.54 $\pm$ 0.84 & 50.90 $\pm$ 0.33 & 2.50 $\pm$ 0.36 & 30.00 $\pm$ 16.40 & 2.84 & 1.14 $\pm$ 0.72 & 60.19 $\pm$ 17.78 & 32.31 $\pm$ 20.12 & 0.03 \\  
18.122 & 0.359 & 1.46 $\pm$ 0.76 & 22.40 $\pm$ 0.11 & 0.89 $\pm$ 0.26 & 3.51 $\pm$ 2.87 & 3.14 & -0.83 $\pm$ 0.29 & 41.65 $\pm$ 1.15 & 6.63 $\pm$ 2.62 & -0.12 \\  
18.231 & -0.272 & 3.95 $\pm$ 0.43 & 68.00 $\pm$ 0.06 & 2.24 $\pm$ 0.15 & 2.23 $\pm$ 0.43 & 4.50 & 1.96 $\pm$ 0.22 & 67.52 $\pm$ 0.14 & 2.52 $\pm$ 0.31 & 0.73 \\  
18.249 & -0.478 & 0.31 $\pm$ 0.13 & 48.80 $\pm$ 0.82 & 5.36 $\pm$ 1.28 & 0.12 $\pm$ 19.10 & 5.23 & -0.31 $\pm$ 0.13 & 85.47 $\pm$ 0.14 & 0.70 $\pm$ 0.42 & -0.41 \\  
18.302 & -0.381 & 1.60 $\pm$ 0.49 & 32.30 $\pm$ 0.15 & 1.98 $\pm$ 0.26 & 3.20 $\pm$ 1.69 & 3.23 & 1.38 $\pm$ 0.25 & 31.84 $\pm$ 0.35 & 3.68 $\pm$ 0.67 & 0.35 \\  
18.330 & -0.389 & 1.66 $\pm$ 1.55 & 32.30 $\pm$ 0.25 & 1.04 $\pm$ 0.48 & 3.25 $\pm$ 4.60 & 3.24 & -0.38 $\pm$ 0.21 & 39.59 $\pm$ 0.44 & 1.42 $\pm$ 0.63 & -0.25 \\
\hline 
\end{tabular} 
\end{center} 
\label{tmp} 
\end{landscape} 

\newpage 
 \clearpage 

\begin{landscape} 
\footnotesize
{\bf Table 2:} Continued. 
\begin{center} 
\begin{tabular}{|c|c|c|c|c|c|c|c|c|c|c|} 
\hline \hline 
 l             & b     & T$_{\rm B,m}^{11}$ & V$_{\rm LSR}^{11}$ & $\Delta$V$^{11}$  & $\tau_{m}^{11}$ & T$_{\rm ex}^{11}$ & $\int$ T$_{\rm B}^{22}$dV & V$_{\rm LSR}^{22}$ & $\Delta$V$^{22}$  & T$_{\rm B}^{22}$ \\ 
 ($^\circ$) & ($^\circ$) & (K) & (kms$^{-1}$) & (kms$^{-1}$) & & (K) & (K.kms$^{-1}$) & (kms$^{-1}$) & (kms$^{-1}$ &  (K)  \\ \hline 
23.241 & -0.480 & 0.61 $\pm$ 0.14 & 61.10 $\pm$ 0.14 & 1.32 $\pm$ 0.38 & 0.10 $\pm$ 18.60 & 8.57 & 0.33 $\pm$ 0.12 & 66.50 $\pm$ 0.13 & 0.58 $\pm$ 0.20 & 0.53 \\  
23.265 & 0.087 & 1.28 $\pm$ 0.53 & 77.40 $\pm$ 0.14 & 1.39 $\pm$ 0.31 & 1.25 $\pm$ 1.36 & 3.75 & -0.69 $\pm$ 0.20 & 62.37 $\pm$ 0.10 & 0.82 $\pm$ 0.32 & -0.80 \\  
23.355 & -0.224 & 1.67 $\pm$ 0.26 & 99.40 $\pm$ 0.23 & 4.97 $\pm$ 0.40 & 1.89 $\pm$ 0.57 & 3.61 & 2.92 $\pm$ 0.43 & 100.30 $\pm$ 0.55 & 6.92 $\pm$ 1.13 & 0.40 \\  
23.365 & 0.105 & 0.53 $\pm$ 0.09 & 83.40 $\pm$ 0.27 & 3.06 $\pm$ 0.54 & 0.10 $\pm$ 2.47 & 8.05 & - & - & - & - \\  
23.410 & 0.446 & 2.69 $\pm$ 0.76 & 107.00 $\pm$ 0.13 & 1.91 $\pm$ 0.26 & 3.56 $\pm$ 1.59 & 3.48 & 1.22 $\pm$ 0.32 & 108.70 $\pm$ 0.41 & 2.74 $\pm$ 0.67 & 0.42 \\  
23.415 & -0.102 & 1.50 $\pm$ 0.68 & 52.60 $\pm$ 0.14 & 1.17 $\pm$ 0.27 & 2.09 $\pm$ 1.97 & 3.44 & 1.36 $\pm$ 0.73 & 54.02 $\pm$ 3.86 & 14.08 $\pm$ 10.14 & 0.09 \\
\hline 
\end{tabular} 
\end{center} 
\label{tmp} 
\end{landscape} 

\newpage 
 \clearpage 

\begin{landscape} 
\footnotesize
{\bf Table 2:} Continued. 
\begin{center} 
\begin{tabular}{|c|c|c|c|c|c|c|c|c|c|c|} 
\hline \hline 
 l             & b     & T$_{\rm B,m}^{11}$ & V$_{\rm LSR}^{11}$ & $\Delta$V$^{11}$  & $\tau_{m}^{11}$ & T$_{\rm ex}^{11}$ & $\int$ T$_{\rm B}^{22}$dV & V$_{\rm LSR}^{22}$ & $\Delta$V$^{22}$  & T$_{\rm B}^{22}$ \\ 
 ($^\circ$) & ($^\circ$) & (K) & (kms$^{-1}$) & (kms$^{-1}$) & & (K) & (K.kms$^{-1}$) & (kms$^{-1}$) & (kms$^{-1}$ &  (K)  \\ \hline 
25.735 & 0.223 & 1.38 $\pm$ 0.76 & 110.00 $\pm$ 0.50 & 3.59 $\pm$ 0.79 & 4.90 $\pm$ 4.08 & 3.01 & 1.27 $\pm$ 0.38 & 106.95 $\pm$ 0.49 & 3.79 $\pm$ 1.54 & 0.32 \\  
25.799 & -0.037 & 0.86 $\pm$ 0.52 & 91.10 $\pm$ 0.14 & 0.85 $\pm$ 0.25 & 0.96 $\pm$ 2.00 & 3.63 & -0.84 $\pm$ 0.22 & 101.14 $\pm$ 0.27 & 2.00 $\pm$ 0.65 & -0.39 \\  
25.828 & -0.177 & 1.36 $\pm$ 0.38 & 93.20 $\pm$ 0.16 & 2.28 $\pm$ 0.36 & 1.90 $\pm$ 1.03 & 3.44 & 2.16 $\pm$ 0.36 & 93.23 $\pm$ 0.41 & 4.89 $\pm$ 0.93 & 0.41 \\  
26.274 & 0.148 & - & - & - & - & - & -0.37 $\pm$ 0.26 & 92.06 $\pm$ 0.72 & 1.74 $\pm$ 0.87 & -0.20 \\  
26.277 & -0.068 & 0.41 $\pm$ 0.22 & 97.70 $\pm$ 1.54 & 10.10 $\pm$ 3.20 & 1.85 $\pm$ 1.46 & 2.95 & 0.28 $\pm$ 0.07 & 94.80 $\pm$ 0.05 & 0.43 $\pm$ 2.99 & 0.61 \\  
26.523 & 0.284 & 0.73 $\pm$ 0.24 & 101.00 $\pm$ 0.28 & 4.57 $\pm$ 0.65 & 0.27 $\pm$ 0.89 & 5.42 & 2.39 $\pm$ 0.78 & 103.08 $\pm$ 2.90 & 16.62 $\pm$ 7.22 & 0.14 \\
\hline 
\end{tabular} 
\end{center} 
\label{tmp} 
\end{landscape} 

\newpage 
 \clearpage 

\begin{landscape} 
\footnotesize
{\bf Table 2:} Continued. 
\begin{center} 
\begin{tabular}{|c|c|c|c|c|c|c|c|c|c|c|} 
\hline \hline 
 l             & b     & T$_{\rm B,m}^{11}$ & V$_{\rm LSR}^{11}$ & $\Delta$V$^{11}$  & $\tau_{m}^{11}$ & T$_{\rm ex}^{11}$ & $\int$ T$_{\rm B}^{22}$dV & V$_{\rm LSR}^{22}$ & $\Delta$V$^{22}$  & T$_{\rm B}^{22}$ \\ 
 ($^\circ$) & ($^\circ$) & (K) & (kms$^{-1}$) & (kms$^{-1}$) & & (K) & (K.kms$^{-1}$) & (kms$^{-1}$) & (kms$^{-1}$ &  (K)  \\ \hline 
29.304 & -0.318 & 0.95 $\pm$ 0.35 & 96.40 $\pm$ 0.08 & 0.71 $\pm$ 0.11 & 1.55 $\pm$ 1.70 & 3.34 & 0.27 $\pm$ 0.08 & 99.07 $\pm$ 0.04 & 0.43 $\pm$ 1.01 & 0.59 \\  
29.394 & -0.100 & 0.39 $\pm$ 0.21 & 103.00 $\pm$ 0.92 & 5.72 $\pm$ 1.35 & 1.04 $\pm$ 1.67 & 3.10 & 0.38 $\pm$ 0.13 & 103.12 $\pm$ 0.14 & 0.79 $\pm$ 0.33 & 0.45 \\  
29.419 & 0.361 & 0.46 $\pm$ 0.16 & 0.23 $\pm$ 0.14 & 0.98 $\pm$ 0.50 & 0.10 $\pm$ 3.04 & 7.28 & -0.73 $\pm$ 0.28 & -18.42 $\pm$ 1.01 & 4.85 $\pm$ 2.20 & -0.14 \\  
29.552 & 0.185 & 0.84 $\pm$ 0.15 & 79.10 $\pm$ 0.11 & 1.27 $\pm$ 0.28 & 0.10 $\pm$ 6.53 & 11.06 & -0.66 $\pm$ 0.26 & 98.47 $\pm$ 0.80 & 3.66 $\pm$ 1.52 & -0.17 \\  
29.904 & -0.225 & 0.34 $\pm$ 0.12 & 95.30 $\pm$ 0.33 & 1.77 $\pm$ 0.66 & 0.10 $\pm$ 1.12 & 6.08 & 3.06 $\pm$ 1.08 & 103.03 $\pm$ 2.83 & 49.34 $\pm$ 17.12 & 0.06 \\  
29.939 & -0.031 & 1.96 $\pm$ 0.26 & 101.00 $\pm$ 0.11 & 3.37 $\pm$ 0.32 & 1.16 $\pm$ 0.40 & 4.41 & 1.65 $\pm$ 0.24 & 100.48 $\pm$ 0.14 & 2.18 $\pm$ 0.43 & 0.71 \\
\hline 
\end{tabular} 
\end{center} 
\label{tmp} 
\end{landscape} 

\newpage 
 \clearpage 

\begin{landscape} 
\footnotesize
{\bf Table 2:} Continued. 
\begin{center} 
\begin{tabular}{|c|c|c|c|c|c|c|c|c|c|c|} 
\hline \hline 
 l             & b     & T$_{\rm B,m}^{11}$ & V$_{\rm LSR}^{11}$ & $\Delta$V$^{11}$  & $\tau_{m}^{11}$ & T$_{\rm ex}^{11}$ & $\int$ T$_{\rm B}^{22}$dV & V$_{\rm LSR}^{22}$ & $\Delta$V$^{22}$  & T$_{\rm B}^{22}$ \\ 
 ($^\circ$) & ($^\circ$) & (K) & (kms$^{-1}$) & (kms$^{-1}$) & & (K) & (K.kms$^{-1}$) & (kms$^{-1}$) & (kms$^{-1}$ &  (K)  \\ \hline 
309.379 & -0.122 & 0.78 $\pm$ 0.33 & -50.80 $\pm$ 0.21 & 2.23 $\pm$ 0.42 & 0.29 $\pm$ 1.51 & 5.38 & 0.92 $\pm$ 0.24 & -50.57 $\pm$ 0.35 & 2.55 $\pm$ 0.72 & 0.34 \\  
309.908 & 0.322 & 0.91 $\pm$ 0.12 & -59.80 $\pm$ 0.20 & 3.06 $\pm$ 0.49 & 0.10 $\pm$ 0.37 & 11.81 & -0.02 $\pm$ 4.15 & -58.41 $\pm$ 16.44 & 2.16 $\pm$ 47.13 & -0.01 \\  
309.916 & 0.484 & - & - & - & - & - & 0.55 $\pm$ 0.21 & -43.18 $\pm$ 0.19 & 1.01 $\pm$ 0.37 & 0.51 \\  
310.016 & 0.384 & 1.58 $\pm$ 0.57 & -41.60 $\pm$ 0.27 & 2.95 $\pm$ 0.68 & 3.26 $\pm$ 1.82 & 3.21 & -0.40 $\pm$ 0.16 & -46.88 $\pm$ 0.17 & 0.83 $\pm$ 0.31 & -0.46 \\  
310.218 & 0.382 & 0.62 $\pm$ 0.17 & -32.70 $\pm$ 0.13 & 0.71 $\pm$ 0.57 & 0.10 $\pm$ 19.10 & 8.90 & - & - & - & - \\  
310.390 & -0.301 & 1.61 $\pm$ 0.76 & -52.90 $\pm$ 0.14 & 1.15 $\pm$ 0.28 & 2.81 $\pm$ 2.26 & 3.30 & 0.04 $\pm$ 0.11 & -24.61 $\pm$ 0.52 & 0.43 $\pm$ 2.66 & 0.09 \\
\hline 
\end{tabular} 
\end{center} 
\label{tmp} 
\end{landscape} 

\newpage 
 \clearpage 

\begin{landscape} 
\footnotesize
{\bf Table 2:} Continued. 
\begin{center} 
\begin{tabular}{|c|c|c|c|c|c|c|c|c|c|c|} 
\hline \hline 
 l             & b     & T$_{\rm B,m}^{11}$ & V$_{\rm LSR}^{11}$ & $\Delta$V$^{11}$  & $\tau_{m}^{11}$ & T$_{\rm ex}^{11}$ & $\int$ T$_{\rm B}^{22}$dV & V$_{\rm LSR}^{22}$ & $\Delta$V$^{22}$  & T$_{\rm B}^{22}$ \\ 
 ($^\circ$) & ($^\circ$) & (K) & (kms$^{-1}$) & (kms$^{-1}$) & & (K) & (K.kms$^{-1}$) & (kms$^{-1}$) & (kms$^{-1}$ &  (K)  \\ \hline 
320.227 & -0.282 & 0.58 $\pm$ 0.12 & -66.80 $\pm$ 0.27 & 2.25 $\pm$ 0.57 & 0.10 $\pm$ 0.78 & 8.54 & 1.04 $\pm$ 0.27 & -67.36 $\pm$ 0.31 & 2.20 $\pm$ 0.56 & 0.44 \\  
320.237 & 0.403 & 1.96 $\pm$ 0.76 & -33.10 $\pm$ 0.15 & 1.74 $\pm$ 0.31 & 3.57 $\pm$ 2.14 & 3.27 & 1.06 $\pm$ 0.20 & -33.40 $\pm$ 0.13 & 1.32 $\pm$ 0.28 & 0.75 \\  
320.322 & -0.438 & -0.45 $\pm$ 0.18 & -18.20 $\pm$ 0.18 & 0.71 $\pm$ 0.31 & 0.10 $\pm$ 2.87 & -1.73 & -0.36 $\pm$ 0.30 & -77.97 $\pm$ 1.28 & 3.04 $\pm$ 1.54 & -0.11 \\  
320.325 & -0.189 & 0.39 $\pm$ 0.08 & -10.40 $\pm$ 0.41 & 4.07 $\pm$ 0.96 & 0.10 $\pm$ 1.22 & 6.64 & 0.45 $\pm$ 0.38 & -37.20 $\pm$ 1.06 & 6.62 $\pm$ 4.18 & 0.06 \\  
320.414 & 0.118 & 0.96 $\pm$ 0.50 & -6.78 $\pm$ 0.18 & 1.73 $\pm$ 0.54 & 1.29 $\pm$ 1.65 & 3.47 & 1.10 $\pm$ 0.35 & -30.78 $\pm$ 0.54 & 3.26 $\pm$ 1.20 & 0.32 \\  
320.886 & -0.385 & 1.11 $\pm$ 0.44 & -45.10 $\pm$ 0.21 & 2.09 $\pm$ 0.44 & 2.27 $\pm$ 1.68 & 3.21 & 3.99 $\pm$ 0.89 & -43.33 $\pm$ 2.73 & 24.71 $\pm$ 6.29 & 0.15 \\
\hline 
\end{tabular} 
\end{center} 
\label{tmp} 
\end{landscape} 

\newpage 
 \clearpage 

\begin{landscape} 
\footnotesize
{\bf Table 2:} Continued. 
\begin{center} 
\begin{tabular}{|c|c|c|c|c|c|c|c|c|c|c|} 
\hline \hline 
 l             & b     & T$_{\rm B,m}^{11}$ & V$_{\rm LSR}^{11}$ & $\Delta$V$^{11}$  & $\tau_{m}^{11}$ & T$_{\rm ex}^{11}$ & $\int$ T$_{\rm B}^{22}$dV & V$_{\rm LSR}^{22}$ & $\Delta$V$^{22}$  & T$_{\rm B}^{22}$ \\ 
 ($^\circ$) & ($^\circ$) & (K) & (kms$^{-1}$) & (kms$^{-1}$) & & (K) & (K.kms$^{-1}$) & (kms$^{-1}$) & (kms$^{-1}$ &  (K)  \\ \hline 
329.179 & -0.301 & 1.16 $\pm$ 0.29 & -50.50 $\pm$ 0.31 & 4.09 $\pm$ 0.51 & 2.05 $\pm$ 1.04 & 3.29 & 0.97 $\pm$ 0.22 & -51.41 $\pm$ 0.22 & 1.79 $\pm$ 0.47 & 0.51 \\  
329.243 & 0.487 & -0.45 $\pm$ 0.15 & -93.20 $\pm$ 0.26 & 1.58 $\pm$ 0.66 & 0.10 $\pm$ 1.52 & -1.75 & -0.90 $\pm$ 1.13 & -80.67 $\pm$ 17.01 & 59.15 $\pm$ 46.44 & -0.01 \\  
329.343 & 0.139 & 1.45 $\pm$ 0.28 & -108.00 $\pm$ 0.15 & 3.29 $\pm$ 0.35 & 1.61 $\pm$ 0.69 & 3.63 & 2.19 $\pm$ 0.34 & -108.17 $\pm$ 0.34 & 5.06 $\pm$ 1.06 & 0.41 \\  
329.381 & 0.071 & 0.28 $\pm$ 0.10 & -115.00 $\pm$ 0.38 & 1.93 $\pm$ 0.62 & 0.10 $\pm$ 1.17 & 5.51 & 1.29 $\pm$ 0.46 & -105.87 $\pm$ 2.60 & 13.40 $\pm$ 4.56 & 0.09 \\  
329.422 & -0.168 & 1.05 $\pm$ 0.25 & -77.20 $\pm$ 0.15 & 2.35 $\pm$ 0.32 & 1.47 $\pm$ 0.87 & 3.44 & 1.12 $\pm$ 0.23 & -76.68 $\pm$ 0.34 & 3.04 $\pm$ 0.58 & 0.34 \\  
329.440 & 0.486 & 2.57 $\pm$ 0.47 & -68.40 $\pm$ 0.10 & 1.98 $\pm$ 0.23 & 2.61 $\pm$ 0.80 & 3.71 & 0.72 $\pm$ 0.23 & -69.05 $\pm$ 0.26 & 1.88 $\pm$ 0.86 & 0.36 \\
\hline 
\end{tabular} 
\end{center} 
\label{tmp} 
\end{landscape} 

\newpage 
 \clearpage 

\begin{landscape} 
\footnotesize
{\bf Table 2:} Continued. 
\begin{center} 
\begin{tabular}{|c|c|c|c|c|c|c|c|c|c|c|} 
\hline \hline 
 l             & b     & T$_{\rm B,m}^{11}$ & V$_{\rm LSR}^{11}$ & $\Delta$V$^{11}$  & $\tau_{m}^{11}$ & T$_{\rm ex}^{11}$ & $\int$ T$_{\rm B}^{22}$dV & V$_{\rm LSR}^{22}$ & $\Delta$V$^{22}$  & T$_{\rm B}^{22}$ \\ 
 ($^\circ$) & ($^\circ$) & (K) & (kms$^{-1}$) & (kms$^{-1}$) & & (K) & (K.kms$^{-1}$) & (kms$^{-1}$) & (kms$^{-1}$ &  (K)  \\ \hline 
332.975 & -0.035 & 0.42 $\pm$ 0.21 & -67.80 $\pm$ 0.58 & 4.62 $\pm$ 1.08 & 0.75 $\pm$ 1.24 & 3.28 & -0.53 $\pm$ 0.15 & -46.74 $\pm$ 0.17 & 1.25 $\pm$ 0.35 & -0.40 \\  
333.029 & -0.039 & 0.77 $\pm$ 0.40 & -42.40 $\pm$ 0.11 & 0.71 $\pm$ 0.16 & 0.27 $\pm$ 4.19 & 5.59 & -0.30 $\pm$ 0.13 & -69.00 $\pm$ 0.31 & 1.22 $\pm$ 0.51 & -0.23 \\  
333.183 & -0.405 & 4.82 $\pm$ 0.35 & -51.10 $\pm$ 0.05 & 2.62 $\pm$ 0.11 & 1.86 $\pm$ 0.27 & 5.32 & 4.12 $\pm$ 0.26 & -51.16 $\pm$ 0.09 & 2.84 $\pm$ 0.22 & 1.36 \\  
333.218 & -0.066 & 1.21 $\pm$ 0.24 & -88.80 $\pm$ 0.18 & 3.60 $\pm$ 0.36 & 1.08 $\pm$ 0.64 & 3.85 & 2.03 $\pm$ 0.33 & -88.49 $\pm$ 0.44 & 5.33 $\pm$ 0.94 & 0.36 \\  
333.479 & -0.480 & 1.22 $\pm$ 0.39 & -55.40 $\pm$ 0.18 & 1.98 $\pm$ 0.35 & 1.84 $\pm$ 1.21 & 3.39 & 0.93 $\pm$ 0.23 & -55.02 $\pm$ 0.31 & 2.43 $\pm$ 0.59 & 0.36 \\  
333.511 & -0.224 & 2.70 $\pm$ 0.50 & -49.80 $\pm$ 0.11 & 2.25 $\pm$ 0.28 & 2.11 $\pm$ 0.72 & 4.00 & - & - & - & - \\
\hline 
\end{tabular} 
\end{center} 
\label{tmp} 
\end{landscape} 

\newpage 
 \clearpage 

\begin{landscape} 
\footnotesize
{\bf Table 2:} Continued. 
\begin{center} 
\begin{tabular}{|c|c|c|c|c|c|c|c|c|c|c|} 
\hline \hline 
 l             & b     & T$_{\rm B,m}^{11}$ & V$_{\rm LSR}^{11}$ & $\Delta$V$^{11}$  & $\tau_{m}^{11}$ & T$_{\rm ex}^{11}$ & $\int$ T$_{\rm B}^{22}$dV & V$_{\rm LSR}^{22}$ & $\Delta$V$^{22}$  & T$_{\rm B}^{22}$ \\ 
 ($^\circ$) & ($^\circ$) & (K) & (kms$^{-1}$) & (kms$^{-1}$) & & (K) & (K.kms$^{-1}$) & (kms$^{-1}$) & (kms$^{-1}$ &  (K)  \\ \hline 
336.985 & -0.017 & 1.19 $\pm$ 0.26 & -122.00 $\pm$ 0.28 & 4.81 $\pm$ 0.54 & 0.98 $\pm$ 0.66 & 3.94 & 1.71 $\pm$ 0.36 & -120.25 $\pm$ 0.52 & 4.64 $\pm$ 1.14 & 0.35 \\  
337.126 & 0.010 & 1.87 $\pm$ 1.07 & -114.00 $\pm$ 0.26 & 2.31 $\pm$ 0.37 & 3.49 $\pm$ 3.47 & 3.26 & -1.15 $\pm$ 0.28 & -130.57 $\pm$ 0.16 & 1.27 $\pm$ 0.35 & -0.85 \\  
337.139 & -0.381 & 3.28 $\pm$ 0.65 & -40.90 $\pm$ 0.11 & 1.98 $\pm$ 0.27 & 2.33 $\pm$ 0.91 & 4.13 & 1.89 $\pm$ 0.34 & -41.19 $\pm$ 0.19 & 2.20 $\pm$ 0.51 & 0.81 \\  
337.217 & -0.050 & 1.02 $\pm$ 0.46 & -68.50 $\pm$ 0.21 & 2.46 $\pm$ 0.57 & 0.81 $\pm$ 1.22 & 3.99 & 2.02 $\pm$ 0.51 & -67.44 $\pm$ 0.68 & 5.61 $\pm$ 1.62 & 0.34 \\  
337.266 & -0.387 & 5.21 $\pm$ 1.21 & -40.50 $\pm$ 0.38 & 2.57 $\pm$ 0.38 & 30.00 $\pm$ 23.40 & 2.90 & 0.19 $\pm$ 0.10 & -14.54 $\pm$ 0.27 & 0.43 $\pm$ 0.15 & 0.43 \\  
337.306 & 0.020 & 0.71 $\pm$ 0.27 & -107.00 $\pm$ 0.29 & 1.46 $\pm$ 0.88 & 0.10 $\pm$ 1.34 & 9.85 & 0.52 $\pm$ 0.16 & -128.16 $\pm$ 0.17 & 1.08 $\pm$ 0.34 & 0.46 \\
\hline 
\end{tabular} 
\end{center} 
\label{tmp} 
\end{landscape} 

\newpage 
 \clearpage 

\begin{landscape} 
\footnotesize
{\bf Table 2:} Continued. 
\begin{center} 
\begin{tabular}{|c|c|c|c|c|c|c|c|c|c|c|} 
\hline \hline 
 l             & b     & T$_{\rm B,m}^{11}$ & V$_{\rm LSR}^{11}$ & $\Delta$V$^{11}$  & $\tau_{m}^{11}$ & T$_{\rm ex}^{11}$ & $\int$ T$_{\rm B}^{22}$dV & V$_{\rm LSR}^{22}$ & $\Delta$V$^{22}$  & T$_{\rm B}^{22}$ \\ 
 ($^\circ$) & ($^\circ$) & (K) & (kms$^{-1}$) & (kms$^{-1}$) & & (K) & (K.kms$^{-1}$) & (kms$^{-1}$) & (kms$^{-1}$ &  (K)  \\ \hline 
340.246 & -0.198 & 1.71 $\pm$ 0.27 & -46.80 $\pm$ 0.17 & 4.15 $\pm$ 0.36 & 1.25 $\pm$ 0.52 & 4.09 & 2.03 $\pm$ 0.50 & -46.90 $\pm$ 0.26 & 3.10 $\pm$ 1.30 & 0.61 \\  
340.302 & -0.376 & 4.34 $\pm$ 0.44 & -51.00 $\pm$ 0.06 & 2.31 $\pm$ 0.14 & 1.93 $\pm$ 0.38 & 4.97 & 3.27 $\pm$ 0.34 & -51.43 $\pm$ 0.15 & 3.06 $\pm$ 0.38 & 1.00 \\  
340.775 & -0.124 & 1.53 $\pm$ 0.49 & -40.10 $\pm$ 0.09 & 1.43 $\pm$ 0.27 & 0.56 $\pm$ 0.83 & 5.44 & -2.93 $\pm$ 0.93 & -55.76 $\pm$ 2.86 & 19.77 $\pm$ 8.54 & -0.14 \\  
340.776 & -0.096 & 1.71 $\pm$ 0.59 & -101.00 $\pm$ 0.21 & 1.82 $\pm$ 0.41 & 3.22 $\pm$ 1.71 & 3.26 & 0.33 $\pm$ 0.15 & -84.07 $\pm$ 0.25 & 0.97 $\pm$ 0.43 & 0.33 \\  
340.930 & -0.235 & 2.15 $\pm$ 0.68 & -45.60 $\pm$ 0.13 & 1.54 $\pm$ 0.32 & 2.42 $\pm$ 1.31 & 3.61 & -0.34 $\pm$ 0.11 & -30.58 $\pm$ 0.23 & 0.45 $\pm$ 2.82 & -0.70 \\  
340.952 & -0.355 & 0.70 $\pm$ 0.29 & -44.20 $\pm$ 0.29 & 3.18 $\pm$ 0.54 & 0.80 $\pm$ 1.19 & 3.60 & 4.82 $\pm$ 0.94 & -47.80 $\pm$ 2.67 & 26.29 $\pm$ 5.65 & 0.17 \\
\hline 
\end{tabular} 
\end{center} 
\label{tmp} 
\end{landscape} 

\newpage 
 \clearpage 

\begin{landscape} 
\footnotesize
{\bf Table 2:} Continued. 
\begin{center} 
\begin{tabular}{|c|c|c|c|c|c|c|c|c|c|c|} 
\hline \hline 
 l             & b     & T$_{\rm B,m}^{11}$ & V$_{\rm LSR}^{11}$ & $\Delta$V$^{11}$  & $\tau_{m}^{11}$ & T$_{\rm ex}^{11}$ & $\int$ T$_{\rm B}^{22}$dV & V$_{\rm LSR}^{22}$ & $\Delta$V$^{22}$  & T$_{\rm B}^{22}$ \\ 
 ($^\circ$) & ($^\circ$) & (K) & (kms$^{-1}$) & (kms$^{-1}$) & & (K) & (K.kms$^{-1}$) & (kms$^{-1}$) & (kms$^{-1}$ &  (K)  \\ \hline 
345.588 & -0.210 & 0.54 $\pm$ 0.51 & -122.00 $\pm$ 1.31 & 6.21 $\pm$ 2.23 & 5.34 $\pm$ 6.76 & 2.83 & - & - & - & - \\  
345.641 & 0.351 & 1.03 $\pm$ 0.23 & -17.00 $\pm$ 0.14 & 1.29 $\pm$ 0.33 & 0.10 $\pm$ 3.48 & 13.03 & -0.55 $\pm$ 0.21 & 5.95 $\pm$ 0.20 & 0.96 $\pm$ 0.36 & -0.54 \\  
346.257 & 0.110 & 1.56 $\pm$ 0.38 & -98.20 $\pm$ 0.10 & 1.77 $\pm$ 0.23 & 2.42 $\pm$ 1.03 & 3.37 & 0.79 $\pm$ 0.26 & -98.87 $\pm$ 0.89 & 5.49 $\pm$ 1.96 & 0.14 \\  
346.300 & 0.126 & 0.91 $\pm$ 0.37 & -28.50 $\pm$ 0.14 & 1.35 $\pm$ 0.32 & 0.88 $\pm$ 1.12 & 3.75 & 0.40 $\pm$ 0.12 & -57.09 $\pm$ 0.11 & 0.62 $\pm$ 0.17 & 0.61 \\  
346.476 & 0.225 & 2.65 $\pm$ 0.66 & -17.10 $\pm$ 0.11 & 1.71 $\pm$ 0.18 & 4.85 $\pm$ 1.82 & 3.27 & -0.26 $\pm$ 0.14 & -8.32 $\pm$ 0.29 & 0.95 $\pm$ 0.60 & -0.26 \\  
346.602 & 0.049 & 0.78 $\pm$ 0.41 & -89.10 $\pm$ 0.26 & 1.89 $\pm$ 0.56 & 3.24 $\pm$ 2.70 & 2.97 & -1.73 $\pm$ 0.54 & -82.54 $\pm$ 2.02 & 12.53 $\pm$ 4.75 & -0.13 \\
\hline 
\end{tabular} 
\end{center} 
\label{tmp} 
\end{landscape} 

\newpage 
 \clearpage 

\begin{landscape} 
\footnotesize
{\bf Table 2:} Continued. 
\begin{center} 
\begin{tabular}{|c|c|c|c|c|c|c|c|c|c|c|} 
\hline \hline 
 l             & b     & T$_{\rm B,m}^{11}$ & V$_{\rm LSR}^{11}$ & $\Delta$V$^{11}$  & $\tau_{m}^{11}$ & T$_{\rm ex}^{11}$ & $\int$ T$_{\rm B}^{22}$dV & V$_{\rm LSR}^{22}$ & $\Delta$V$^{22}$  & T$_{\rm B}^{22}$ \\ 
 ($^\circ$) & ($^\circ$) & (K) & (kms$^{-1}$) & (kms$^{-1}$) & & (K) & (K.kms$^{-1}$) & (kms$^{-1}$) & (kms$^{-1}$ &  (K)  \\ \hline 
352.315 & -0.442 & 0.57 $\pm$ 0.10 & -10.90 $\pm$ 0.28 & 3.46 $\pm$ 0.77 & 0.10 $\pm$ 3.34 & 8.39 & 0.31 $\pm$ 0.10 & -32.22 $\pm$ 0.05 & 0.43 $\pm$ 5.91 & 0.69 \\  
352.604 & -0.213 & 1.07 $\pm$ 0.42 & -85.10 $\pm$ 0.20 & 2.16 $\pm$ 0.65 & 1.13 $\pm$ 1.18 & 3.67 & - & - & - & - \\  
352.652 & -0.343 & 0.60 $\pm$ 0.25 & -20.00 $\pm$ 0.18 & 0.90 $\pm$ 0.41 & 0.10 $\pm$ 3.36 & 8.69 & -0.17 $\pm$ 0.10 & 16.64 $\pm$ 0.19 & 0.43 $\pm$ 0.22 & -0.39 \\  
352.863 & -0.196 & 0.80 $\pm$ 0.24 & -57.90 $\pm$ 0.22 & 3.75 $\pm$ 0.55 & 0.46 $\pm$ 0.76 & 4.49 & 2.00 $\pm$ 0.47 & -56.57 $\pm$ 0.73 & 6.48 $\pm$ 1.80 & 0.29 \\  
352.864 & 0.436 & 1.02 $\pm$ 0.93 & 16.40 $\pm$ 0.23 & 1.03 $\pm$ 0.59 & 2.67 $\pm$ 4.34 & 3.11 & -1.00 $\pm$ 0.29 & -16.81 $\pm$ 0.29 & 1.85 $\pm$ 0.57 & -0.51 \\  
352.915 & -0.102 & 1.79 $\pm$ 0.48 & -17.80 $\pm$ 0.07 & 0.71 $\pm$ 0.93 & 0.47 $\pm$ 0.91 & 6.53 & -0.69 $\pm$ 0.22 & -20.46 $\pm$ 0.31 & 1.39 $\pm$ 0.47 & -0.47 \\
\hline 
\end{tabular} 
\end{center} 
\label{tmp} 
\end{landscape} 

\newpage 
 \clearpage 

\begin{landscape} 
\footnotesize
{\bf Table 2:} Continued. 
\begin{center} 
\begin{tabular}{|c|c|c|c|c|c|c|c|c|c|c|} 
\hline \hline 
 l             & b     & T$_{\rm B,m}^{11}$ & V$_{\rm LSR}^{11}$ & $\Delta$V$^{11}$  & $\tau_{m}^{11}$ & T$_{\rm ex}^{11}$ & $\int$ T$_{\rm B}^{22}$dV & V$_{\rm LSR}^{22}$ & $\Delta$V$^{22}$  & T$_{\rm B}^{22}$ \\ 
 ($^\circ$) & ($^\circ$) & (K) & (kms$^{-1}$) & (kms$^{-1}$) & & (K) & (K.kms$^{-1}$) & (kms$^{-1}$) & (kms$^{-1}$ &  (K)  \\ \hline 
358.094 & -0.143 & 0.59 $\pm$ 0.10 & 7.01 $\pm$ 0.31 & 3.33 $\pm$ 0.66 & 0.10 $\pm$ 1.10 & 8.59 & 0.96 $\pm$ 0.34 & 7.25 $\pm$ 0.54 & 2.74 $\pm$ 0.99 & 0.33 \\  
358.251 & -0.436 & 0.91 $\pm$ 0.30 & 3.99 $\pm$ 0.17 & 2.17 $\pm$ 0.37 & 1.02 $\pm$ 1.00 & 3.62 & 0.54 $\pm$ 0.19 & 28.00 $\pm$ 0.50 & 2.58 $\pm$ 0.75 & 0.20 \\  
358.486 & -0.398 & 2.69 $\pm$ 0.36 & -4.67 $\pm$ 0.06 & 2.02 $\pm$ 0.17 & 1.46 $\pm$ 0.44 & 4.57 & 1.60 $\pm$ 0.22 & -5.08 $\pm$ 0.17 & 2.42 $\pm$ 0.38 & 0.62 \\  
358.653 & -0.487 & - & - & - & - & - & 1.18 $\pm$ 0.37 & -43.95 $\pm$ 0.71 & 5.04 $\pm$ 2.38 & 0.22 \\  
358.707 & -0.135 & 0.44 $\pm$ 0.16 & 9.31 $\pm$ 0.80 & 7.51 $\pm$ 1.35 & 0.92 $\pm$ 0.90 & 3.20 & 2.34 $\pm$ 0.96 & -18.02 $\pm$ 15.15 & 59.15 $\pm$ 10.97 & 0.04 \\  
358.762 & -0.041 & 0.48 $\pm$ 0.25 & -189.00 $\pm$ 0.55 & 4.54 $\pm$ 1.05 & 1.60 $\pm$ 1.77 & 3.03 & -1.05 $\pm$ 0.51 & -182.84 $\pm$ 2.71 & 11.43 $\pm$ 5.41 & -0.09 \\
\hline 
\end{tabular} 
\end{center} 
\label{tmp} 
\end{landscape} 

%------------------------------------------------------------
%\clearpage
%\newpage

\setcounter{table}{2}

\begin{table*}
\caption{Derived gas properties for sources with robust $\nhone$ and $\nhtwo$ detections (see $\S$~\ref{sec:peak_spectra_fit_results}). The rotational temperature, total $\nhthree$ column density,  kinetic temperature and the non-thermal and thermal contributions to the line width are listed in columns 3 to 7, respectively, for the peak $\nhone$ pixel.
}
\begin{tabular}{|c|c|c|c|c|c|c|}

\hline \hline
  \multicolumn{1}{|c|}{l} &
  \multicolumn{1}{c|}{b} &
  \multicolumn{1}{c|}{T$_{\rm rot}$} &
  \multicolumn{1}{c|}{N$_{\rm NH3}^{\rm tot}$} &
  \multicolumn{1}{c|}{T$_{\rm kin}$} &
  \multicolumn{1}{c|}{$\Delta$V$_{\rm nt}$} &
  \multicolumn{1}{c|}{$\Delta$V$_{\rm th}$} \\
  \multicolumn{1}{|c|}{($^\circ$)} &
  \multicolumn{1}{c|}{($^\circ$)} &
  \multicolumn{1}{c|}{(K)} &
  \multicolumn{1}{c|}{(10$^{14}$cm$^{-2}$)} &
  \multicolumn{1}{c|}{(K)} &
  \multicolumn{1}{c|}{($\kms$)} &
  \multicolumn{1}{c|}{($\kms$)} \\
\hline
  29.9386 & -0.03095 & 19 & 1.69 & 22 & 3.13 & 0.24\\
  27.77443 & 0.06791 & 21 & 0.69 & 26 & 2.37 & 0.26\\
  27.27837 & 0.15187 & 42 & 0.45 & 93 & 2.32 & 0.50\\
  24.78522 & 0.09488 & 17 & 2.22 & 20 & 3.58 & 0.23\\
  24.47508 & 0.47827 & 34 & 0.33 & 59 & 3.04 & 0.39\\
  24.42476 & -0.21978 & 19 & 1.86 & 23 & 3.27 & 0.25\\
\hline\end{tabular}
\label{tab:nh3_derived_vals}
\end{table*}

\end{document}